\documentclass[11pt,a4paper]{article}
\usepackage{packages}
\usepackage{jheppub}
\pdfoutput=1
\newcommand{\condcomment}[2]{\ifthenelse{#1}{#2}{}} 
\newboolean{includefeyn}
\setboolean{includefeyn}{true}

\numberwithin{equation}{section}

\preprint{IPPP/14/23, DCPT/14/46}

\title{Constrained Dirac gluino mediation}
 
\author{Daniel Busbridge}

\affiliation{Institute for Particle Physics Phenomenology,\\Durham University, South Road\\Durham, DH1 3LE, UK}

\emailAdd{d.w.busbridge@durham.ac.uk}

\abstract{\small 
We perform a comparison study of the Constrained Minimal Supersymmetric Standard Model and Constrained General Gauge Mediation with and without a heavy Dirac gluino. These extremely simple models have very few free parameters and exhibit the characteristic features of \emph{supersoftness} and \emph{supersafeness}. We determine the characteristic low energy spectra, the production cross sections of key processes at the Large Hadron Collider and the degree of fine tuning for a representative range of parameters for each model.
}

\keywords{Dirac gauginos, SUSY breaking, Supersymmetry}

\arxivnumber{1408.4605}

\begin{document}
\begin{fmffile}{supersoft}

%%%% Additional FeynMF commands for SUSY stuff %%%%

\fmfcmd{vardef cross_bar (expr p, len, ang) = ((-len/2,0)--(len/2,0)) rotated (ang + angle direction length(p)/2 of p) shifted point length(p)/2 of p enddef;} % Defines a cross
\fmfcmd{style_def gaugino expr p =  cdraw (wiggly p);  cdraw p; enddef;} % Defines a gaugino line in FeynMF
\fmfcmd{style_def gluino expr p =  cdraw (curly p);  cdraw p; enddef;} % Defines a gluino line in FeynMF
\fmfcmd{style_def gaugino_arrow expr p =  draw_gaugino p;  cfill (arrow p); enddef;} % Gaugino with an arrow
\fmfcmd{style_def gluino_arrow expr p =  draw_gluino p;  cfill (arrow p); enddef;} % Gluino with an arrow
\fmfcmd{style_def crossp expr p = cdraw p; ccutdraw cross_bar (p, 4mm, 45); ccutdraw cross_bar (p, 4mm, -45) enddef;} % Defines a plain crossed line for insertions
\fmfcmd{style_def crossd expr p = draw_dashes p; ccutdraw cross_bar (p, 4mm, 45); ccutdraw cross_bar (p, 4mm, -45) enddef;} % Defines a dashed crossed line for instertions

\maketitle

\glsresetall

\section{Introduction}

In light of the data taken by the ATLAS and CMS collaborations during Run I of the \gls{LHC}, many popular \gls{UV} completions of the \gls{MSSM} are now severely challenged as `natural' realisations of microscopic physics: 
\begin{itemize}
	\item The discovery of a particle closely resembling the \gls{SM} Higgs Boson \cite{Collaboration2012a,Collaboration2012} with a mass $\m\ph= 125.9\pm0.4$ GeV \cite{Beringer2012} typically requires one loop corrections to its mass to be the same order as its tree level contribution\footnote
{
We use $c_\tta\equiv\cos(\tta)$, $s_\tta\equiv\sin(\tta)$ and $t_\tta\equiv\tan(\tta)$ throughout.
}
\begin{equation}
\mm\ph2 \simeq \mm\pZ2 \,c_{2\be}^2 + \frac3{2\,\pi^2}\frac{\mm\pt4}{v^2}\left[
\log\left(\frac{\m{\pstn1}\m{\pstn2}}{\mm\pt2}\right)+\frac{X_t^2}{\m{\pstn1}\m{\pstn2}}\left(1-\frac{X_t^2}{12\,\m{\pstn1}\m{\pstn2}}\right)
\right]
\label{eq:1loopmh}
\end{equation}
if it is identified with the lightest \gls{CP} even neutral MSSM scalar. Here $X_t=A_t-\mu\,\cot(\be)$ is the stop mixing parameterk $\be$ is the ArcTan of the \gls{2HDM} vacuum parameters
\begin{equation}
	\vev{\pHu}
	=\frac1{\sqrt2}\,(0\quad v_u)^T,
	\quad
	\vev{\pHd}
	=\frac1{\sqrt2}\,(v_d\quad0)^T,
	\quad
	t_\be
	\equiv\frac{v_u}{v_d}
\end{equation}
and $\mu$ is the Higgsino mass parameter. Without increasing the field content beyond the \gls{MSSM}, raising the Higgs mass this leaves residual tuning in electroweak minimisation conditions.
	\item The non-observation of strongly interacting sparticles by ATLAS (CMS) has put stringent bounds on gluino and squark masses: $\m\pglui\gtrsim 1350$ $(1000-1200)$ GeV, $\m\psq\gtrsim 780$ $(780)$ GeV in simplified models with degenerate squark masses and a neutralino \gls{LOSP} with mass $\m\pneuz=0$ $(\m\pneuz<80)$ GeV \cite{Collaboration2013,CMSCollaboration2014}. If the stops and gluinos are heavy for the entire \gls{RG} flow they lead to a large logarithmic dependence of \gls{EWSB} upon the \gls{UV} parameters.
\end{itemize}

There are many ways of reducing tuning. A well studied example is the \gls{NMSSM} which adds to the \gls{MSSM} a \gls{SM} gauge singlet chiral superfield which acquires a \gls{VEV} to dynamically solve the \emph{$\mu$-problem}. New quartic Higgs interactions are present which raise the tree level Higgs mass so that a smaller fraction is required from radiative corrections. Unfortunately, these interactions are suppressed at large $t_\be$, where the $D$-term contributions to the tree level Higgs mass are maximised.  Typically $\m\pst\sim1$ TeV is required if perturbativity holds all the way to the \gls{GUT} scale \cite{Ellwanger2007,Ellwanger2009}, leaving the \gls{NMSSM} with a \textit{little hierarchy problem}. If one also adds a hypercharge neutral $\SUL$ triplet as is done in the Triplet Extended \gls{NMSSM}, further quartic Higgs interactions can be induced that allow the correct Higgs mass to be achieved at tree level, removing heavy stops as a requirement \cite{Basak2012,Basak2013}. The direct constraints on the gluino and squarks stops still remain however, so this is not a complete solution.

Solutions accounting for the non-observation of \gls{SUSY} are also available: compressed spectra \cite{LeCompte2011,Dreiner2012b,Bhattacherjee2013} softens jet activity, \gls{RPV} reduces the about of \gls{MET} \cite{Barbier2005} and \gls{FGM} \cite{Abdullah2012,Galon2013} can break the squark mass degeneracy, weakening the reduced limits at current experiments. Combining these mechanisms with models that generate natural spectra can give a plausible explanation of \gls{SUSY} non-observation \emph{and} the Higgs mass.

Instead of using a combination of the above, one can use a Dirac gluino to achieve both. A Dirac gluino doesn't enter the scalar \glspl{RGE} at one loop. This is known as \emph{supersoftness}. The Dirac gluino does however give the squarks a one loop threshold correction at the Dirac gluino mass, acting as a low scale messenger for the strong sector of the \gls{SM}. In this situation, larger $\m\pst$ can be more acceptable as the \gls{RG} running will be shorter and so will induce smaller corrections to the parameters involved in \gls{EWSB} \cite{Fox2002}. The electroweak sector will then less \gls{UV} sensitive. Models with Dirac gauginos have been studied in a wide range of scenarios \cite{Fayet1978,Polchinski1982,Hall1991,Fox2002,Nelson2002a,Antoniadis2005,Antoniadis2006,Antoniadis2006a,Hsieh2008,Amigo2008,Choi2008a,Choi2008,Blechman2009,Benakli2008, Belanger2009,Choi2009,Benakli2010a,Chun2010,Benakli2010,Carpenter2010,Kribs2010,Choi2010,Abel2011a,Davies2011a,Benakli2011,Benakli2011a,Heikinheimo2011,Itoyama2011,Kribs2012,Davies2012,Goodsell2012,Benakli2012,Frugiuele2012b,Frugiuele2012a,Bertuzzo2012,Riva2012,Frugiuele2012,Itoyama2013,Abel2013,Kribs2013,Kribs2013a,Banks2013,Csaki2013,Dudas2013,Bertuzzo2014a,Benakli2014,Fox2014,Goodsell2014a,Ipek2014}. The simplest known way of generating a Dirac gluino mass $m_{D3}$ is to generate it at the messenger scale $M$ by integrating out the messenger sector coupled to a source of $D$ term breaking
\begin{equation}
	\de m_{D3}
	=\;\,\condcomment{\boolean{includefeyn}}{
	\parbox{41mm}{\begin{fmfgraph*}(41,24)
	\fmfleft{i} \fmfright{o} 
	\fmf{gluino_arrow,label=$\pglui$,label.side=right,label.dist=12}{i,v1} 
	\fmf{fermion,label=$\tilde{A}_3$,label.side=left,label.dist=12}{o,v2}
	\fmf{crossd,right,tension=.3,label=$D^\prime$}{v2,v1}
	\fmf{fermion}{v1,c1}
	\fmf{fermion}{v2,c2}
	\fmf{crossp,label=$M$}{c1,c2}
	\end{fmfgraph*}}}\;
	=\;\frac{y\,g_3}{16\,\pi^2}\frac{D^\prime}M,
	\label{eq:Dirac_mass_1-loop}
\end{equation}
where $D^\prime$ is the \gls{SUSY} breaking $D$-term \gls{VEV} of a $\U1^\prime$ gauge group in the hidden sector: $\vev{\WW^\prime_\al}=\tta_\al\,D^\prime$ and $M$ is the messenger scale and $y$ is couples vector-like messengers $(\Phi,\Phib)$ to the chiral field $A_3$ containing the right handed component of the Dirac gluino $\pglui_R=(\At_3)^\dagger$
\begin{equation}
	W_\textrm{Mess}=\sqrt2\,y\,\Phib\,A_3\,\Phi+M\,\Phib\,\Phi.
	\label{eq:Dirac_W_Mess}
\end{equation}
This theory is \gls{RG} evolved to the physical Dirac gluino mass where we must switch to an effective theory with the gluino and the sgluons integrated out. This generates one loop threshold corrections for the squarks
\begin{align}
	\de\mm\psq2
	=&\;\,
	\condcomment{\boolean{includefeyn}}{
	\parbox{41mm}{\begin{fmfgraph*}(41,24)
	\fmfleft{i} \fmfright{o} 
	\fmf{dashes_arrow}{i,v1} \fmf{dashes_arrow}{v2,o}
	\fmf{plain_arrow,left,tension=0.,label=${\pq}$,label.dist=8}{v1,v2}
	\fmf{gluino}{v1,v2}
	\end{fmfgraph*}}}
	+\;\,
	\condcomment{\boolean{includefeyn}}{
	\parbox{41mm}{\begin{fmfgraph*}(41,24)
	\fmfleft{i} \fmfright{o} 
	\fmf{dashes_arrow}{i,c} \fmf{dashes_arrow}{c,o}
	\fmf{curly,right,tension=1.}{c,c}
	\end{fmfgraph*}}}
	+\;\,
	\condcomment{\boolean{includefeyn}}{
	\parbox{41mm}{\begin{fmfgraph*}(41,24)
	\fmfleft{i} \fmfright{o} 
	\fmf{dashes_arrow}{i,v1} \fmf{dashes_arrow}{v2,o}
	\fmf{curly,right,tension=.4}{v2,v1}
	\fmf{dashes_arrow}{v1,v2}
	\end{fmfgraph*}}}
	\nonumber\\
	&+\;\,
	\condcomment{\boolean{includefeyn}}{
	\parbox{41mm}{\begin{fmfgraph*}(41,24)
	\fmfleft{i} \fmfright{o} 
	\fmf{dashes_arrow}{i,v1} \fmf{dashes_arrow}{v2,o}
	\fmf{dashes,left,tension=0.,label=${\phi_3}$,label.dist=8}{v1,v2}
	\fmf{dashes_arrow,label=$\psq$}{v1,v2}
	\end{fmfgraph*}}}
	=\;\frac{C_2(\fun,3)\,g_3^2\,m_{D3}^2}{4\,\pi^2} \log\left(\frac{m_{\phi 3}^2}{m_{D3}^2}\right),
	\label{eq:Squark_mass_1-loop}
\end{align}
where $C_2(\fun,3)$\footnote
{
	We use $C_2(\rep,i)$ to denote the quadratic Casimir of the representation $\rep$ 		under the $i^{\textrm{th}}$ gauge group. In this paper we use
	\begin{align}
		C_2(\fun,N)
		&=C_2(\afun,N)
		=\frac{N^2-1}{2\times N},&
		C_2(\Ad,N)
		&=N
		\label{eq:Quadratic_casimir_SUN}
	\end{align}
	for $\SU{N}$ gauge groups and
	\begin{equation}
		C_2(q,1)
		=q^2
		\label{eq:Quadratic_casimir_U1}
	\end{equation}
	for $\U{1}$ gauge groups, where $q$ is the charge under that gauge group.
	\label{foot:Casmirs}
}
is the quadratic Casimir for the fundamental representation under the gauge group $\SUC$, and	
\begin{align}
	m_{\phi3}^2
	&=m_3^2+4\,m_{3D}^2+B_3, &
	m_{\si3}^2
	&=m_3^2-B_3,
	\label{eq:mass_CP_even_odd_sgluon}
\end{align}
are the soft masses squared for the \gls{CP}-even $\phi_3$ and \gls{CP}-odd $\si_3$\footnote
{
	We decompose $A_3$ as
	\begin{equation}
		A_3
		=\frac1{\sqrt2}(\phi_3+i\,\si_3).
		\label{eq:A3_decomposition}
	\end{equation}
}
that we will refer to  as the sgluon and pseudosgluon (collectively as sgluons for simplicity). Here, $m_3^2$ is the sgluon soft mass squared and $B_3$ is the sgluon bilinear term (see sec. \ref{sec:UVboundary}). The theory is then \gls{RG} evolved to the \gls{SUSY} scale $m_\textrm{SUSY}$ which we take to be the geometric stop mass
$m_\textrm{SUSY}
=\sqrt{\m{\pstn1}\m{\pstn2}}$
where the renormalisation scale dependence for the calculation of the spectrum is minimised \cite{Gauge1990,Nath1992,DeCarlos1993}.

If the majority of the squark mass is generated through integrating out the gluino and its corresponding scalar degrees of freedom, the sensitivity of electroweak parameters to the parameters defined at $M$ is reduced as the most sensitive period of running is now effectively from $m_{D3}$ rather than $M$ to $m_\textrm{SUSY}$. It is straightforward to give Dirac masses to all of the gauginos in the \gls{MSSM} in this way, each accompanied by analogous threshold corrections to the scalar spectrum, though this can introduces further complications such as tachyons and electroweak precision measurements.

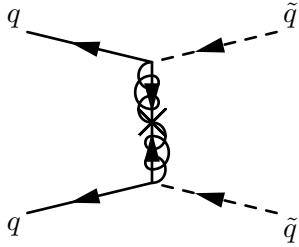
\begin{figure}[t]
\begin{center}
\condcomment{\boolean{includefeyn}}{
\parbox{41mm}{\begin{fmfgraph*}(41,24)
\fmfleft{i1,i2} \fmfright{o1,o2} 
\fmf{fermion}{c1,i1}
\fmf{fermion}{c2,i2}
\fmf{dashes_arrow}{o1,c1}
\fmf{dashes_arrow}{o2,c2}
\fmfv{label=$q$,label.dist=3}{i1}
\fmfv{label=$q$,label.dist=3}{i2}
\fmfv{label=$\qt$,label.dist=3}{o1}
\fmfv{label=$\qt$,label.dist=3}{o2}
\fmfv{decor.shape=cross}{c}
\fmf{gluino_arrow}{c1,c}
\fmf{gluino_arrow}{c2,c}
\end{fmfgraph*}}}
\end{center}
\caption{The dominant tree-level contribution to squark-squark production at the \gls{LHC} requires a chirality flipping Marjoana gluino propagator. This is absent in models with Dirac gluinos, greatly the suppressing squark-squark production cross-section.}
\label{fig:t-channelglu}
\end{figure}

The dominant diagram for squark production at the \gls{LHC} in the \gls{MSSM} is t-channel gluino exchange which requires the presence of the chirality flipping Majorana gluino propagator (see fig.~\ref{fig:t-channelglu}). With a purely Dirac gluino this is essentially negligible. For a characteristic Majorana gluino spectrum, where squarks should be at least as heavy as gluinos from a \gls{UV} perspective \cite{Jaeckel2011}, the relevant bounds from ATLAS and CMS are $\m\psq=\m\pglui\gtrsim1700 - 1800$ GeV for $\m\LSP\lesssim 700$ GeV. In a supersoft model, gluinos are naturally absent from the spectrum, the ATLAS (CMS) bounds on first two generation squarks are moderately reduced to $\m{\psq_{1,2}}\gtrsim 850$ $(780)$ GeV with $\m\LSP=0$ and no bound with $\m\LSP\gtrsim 300$ GeV, rendering them \emph{supersafe} \cite{Kribs2012,Kribs2013,Kribs2013a}. The number of events involving electroweak sparticles at the \gls{LHC} will also be reduced as the dominant processes with these sparticles as final states are those of the decay chains of the gluinos and squarks whose production is suppressed.

%Models with Dirac gauginos can also possess a $\U1_R$ symmetry if an extended Higgs sector is present. This model, dubbed the \gls{MRSSM}, significantly suppresses contributions to flavour physics beyond the \gls{SM} \cite{Kribs2008,Fok2012,Kalinowski2011}, although this improvement is only seen as moderate in models with only Dirac gluinos and in some places fares worse than the \gls{MSSM} \cite{Dudas2013}.

We will first construct two simple models that that have the following properties:
\begin{itemize}
\item Natural from the point of view of \gls{EWSB} --- electroweak sparticles all at electroweak scale.
\item A minimal set of free parameters in the \gls{UV}.
\item \emph{Supersoftess} to reduce fine tuning.
\item \emph{Supersafeness} to aleviate collider bounds.
\end{itemize}
We will then implement these models and the supersoft mechanism into a spectrum generator and perform a study, discussing the consequences for hadron collider phenomenology and fine tuning.

\section{Constrained Dirac gluino mediation}

\subsection{Overview}

\begin{figure}[t]
\begin{center}
\includegraphics[scale=1]{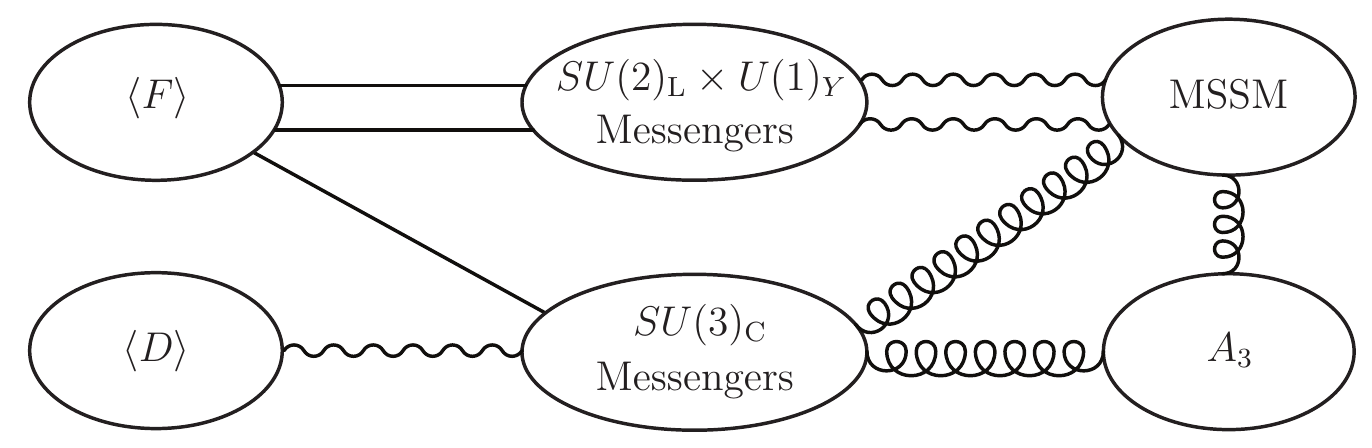}
\end{center}
\caption{The different sectors used in our setup.}
\label{fig:GaugeSchematic}
\end{figure}

As the \gls{LHC} is a proton-proton collider, the non-observation of \gls{SUSY}, and particularly of gluinos, indicates that the strongly interacting \gls{SUSY} particles should be moderately heavy to evade exclusion. To achieve this, we supplement the \gls{CMSSM} and \gls{CGGM} with a Dirac gluino. We will refer to these scenarios as \emph{Constrained Dirac gluino mediation}. Due to the one loop supersoft nature of the Dirac gauginos, the higher scale of the strong sector is not transferred to the electroweak sector through \gls{RG} running, and so electroweak sparticles can remain light (depending on the region of parameter space). Specifically, we couple $\SUC\times\SUL\times\U1_Y$ to either the \gls{CMSSM} or the \gls{CGGM}, and couple only $\SUC$ to a sector of $D$ term breaking to the mechanism of \cite{Fox2002} (see fig.~\ref{fig:GaugeSchematic} for the \gls{CGGM} setup).
The field content is the same as the \gls{MSSM} plus the chiral superfield $A_3$ detailed in table \ref{tab:FNWfields}. The $A_3$ is often referred to as an \gls{ESP} due to its potential origin as the chiral superfield component of $\NN=2$ vector multiplets.

We will now recap the effects of integrating out a messenger sector in terms of the presence of $D$ term \gls{SUSY} breaking before moving on to discuss the full \gls{UV} boundary conditions of the model.

\subsection{Boundary conditions at the Messenger scale}

\subsubsection{$D$-term breaking effective operators}
\label{sec:effectiveoperators}
\gls{FNW} \cite{Fox2002} identified two operators generated by $D$-term breaking in the presence of \glspl{ESP}
\begin{align}
	\LL^{(1)}_\textrm{Supersoft}
	=\sqrt2\,\int d^2\tta\;\frac{\WW^\prime\cdot\WW^a_3\,A^{a}_3}M
	&=\frac{D^\prime}M\,\left(i\,\pglui^a\cdot\At^a_3+\sqrt2\,A^a_3\,D^a_3\right)+\cdots
	\nonumber\\
&=m_{D3}\,\left(i\,\pglui^a\cdot\At^a_3+\sqrt2\,A^a_3\,D^a_3\right)+\cdots,\label{eq:FNWoperator1}\\
\LL^{(2)}_\textrm{Supersoft}=\int d^2\tta\;\frac{\WW^\prime\cdot\WW^\prime\,A^a_3\,A^a_3}{M^2}
&=\left(\frac{D^\prime}M\right)^2\,A_3^a\,A_3^a\nonumber\\
&=B_3\,A_3^a\,A_3^a,
\label{eq:FNWoperator2}
\end{align}
where $M$ is the scale of physics integrated out to generate the operators in eqs. \ref{eq:FNWoperator1}, and \ref{eq:FNWoperator1}, and $D^\prime$ is the \gls{VEV} of a hidden sector $U(1)^\prime$: $\vev{\WW_\al^\prime}=\tta_\al\,D^\prime$. The ``\,$\cdots$\,'' in eq. \ref{eq:FNWoperator1} correspond to operators that vanish upon including their hermitian conjugates. In a messenger setup, both of these operators are generated at one loop, leading to a tachyon in the spectrum. Indeed, this is the original reason for abandoning these models \cite{Fayet1978}. There is one further operator generated at two loops by $D$-term breaking identified by Cs\'aki et al. \cite{Csaki2013}
\begin{align}
\LL^{(1)}_\textrm{Not supersoft}=\int d^4\tta\;\frac{S^\dagger e^V S+\tilde{S}^\dagger e^{-V}\tilde{S}}{M^2}\,A_3^\dagger\,A_3
&=\left(\frac{D^\prime}M\right)^2\,A_3^\dagger A_3\nonumber\\
&=m_{A_3}^2\,A_3^\dagger A_3
\label{eq:Csakioperator}
\end{align}
where $S$ and $\tilde S$ are singlets under the \gls{SM} but charged under the $U(1)^\prime$. These give rise to the non-vanishing $D^\prime\propto|S|^2-|\tilde{S}|^2$ and break the $U(1)^\prime$ gauge symmetry. Note that the operator in eq. \ref{eq:Csakioperator} is still picks out a coefficient $\sim (D^\prime/M)^2$. Upon introducing messenger mixing, \ref{eq:Csakioperator} is generated at one loop instead of two, and then the mixing freedom can be used to tune\footnote
{
	The tuning is typically $\OO\left(\frac1{16\pi^2}\right)$.
} \ref{eq:FNWoperator2} to be two loop size  \cite{Benakli2008,Benakli2010,Csaki2013}. We then find the phenomenologically acceptable boundary conditions
\begin{equation}
m_{D3}\sim\frac1{16\pi^2}\frac{D^\prime}M,\qquad m_{A_3}^2\sim\frac1{16\pi^2}\left(\frac{D^\prime}M\right)^2,\qquad B_3\sim\,\frac\ep{16\pi^2}\left(\frac{D^\prime}M\right)^2,
\end{equation}
where $\ep \sim 1/(16\pi^2)$ is a parameter  that arises due to a cancellation between different contributions to $B_3$\footnote
{
	Strictly this is a cancellation between terms linear and quadratic in $D^\prime$, 		though this is not so important for our discussion.
}
. Note that the operator \ref{eq:Csakioperator} is not supersoft at two loops, however, and will generate
\begin{equation}K_{\textrm{Sfermion}}=\int d^4\tta\;\frac{S^\dagger e^V S+\tilde{S}^\dagger e^{-V}\tilde{S}}{M^2}\,\pq^\dagger\,\pq
\end{equation}
as can be observed from the squark two loop beta function
\begin{equation}
(16\pi^2)^2\beta_{\mm\pq2}^{(2)}=32\,g_3^2\,m_{A_3}^2+\cdots.
\label{eq:non-supersoft-beta}
\end{equation}
Supersoftness is then broken at two loops, rendering a \gls{UV} sensitivity to the scale at which the Dirac gluino mass is generated \cite{Arvanitaki2013}.

\begin{table}[t]
\begin{center}
\begin{tabular}{c|ccc}    & $\SUC$ & $\SUL$ & ${\U1}_\textrm{Y}$ \\
 \hline
$A_3$ & \tAd & \ttriv & 0
\end{tabular}
\caption{Additional field content required to give a Dirac mass to the gluino.}
\label{tab:FNWfields}
\end{center}
\end{table}

\subsubsection{Combined $D$ and $F$ term}
\label{sec:UVboundary}
Upon integrating out the messenger sector, we still have the \gls{MSSM} superpotential
\begin{equation}
W_\textrm{MSSM}=y_{\pu}\,\pau\,\pq\cdot \pHu-y_{\pd}\,\bar \pd\,\pq\cdot \pHd-y_{\pe}\,\pae\,\pl\cdot \pHd+\mu\,\pHu\cdot\pHd
\end{equation}
and a soft lagrangian conveniently decomposed into
\begin{equation}
\LL_\textrm{Soft}=\LL_\textrm{Soft}^{F}+\LL_\textrm{Soft}^{D}.
\end{equation}
$\LL_\textrm{Soft}^{F}$ is the standard soft lagrangian of the \gls{MSSM} supplemented with $A_3$
\begin{align}
-\LL_\textrm{Soft}^{F}={}&
\frac12\left(M_3\,\pglui\cdot\pglui+M_2\,\pwi\cdot\pwi+M_1\,\pbi\cdot\pbi+\hc\right)\nonumber\\
&+\left(a_{\pu}\,\pau\,\pq\cdot\pHu-a_{\pd}\,\pad\,\pq\cdot\pHd-a_{\pe}\,\pae \,\pl\cdot\pHd+\hc\right)\nonumber\\
&+\mm\pq2|\pq|^2+\mm\pu2|\pau|^2+\mm\pd2|\pad|^2+\mm\pl2|\pl|^2+\mm\pe2|\pae|^2+\mm{A^F_3}2|A_3|^2\nonumber\\
&+\left(B_3\,A_3\,A_3+\hc\right)\nonumber\\
&+\mm\pHu2|\pHu|^2+\mm\pHd2|\pHd|^2+(B_\mu\,\pHu\cdot\pHd+\hc),
\end{align}
where the $M_i$ are the Majorana gaugino massses, $a_i$ are the scalar trilinears, the $m_i^2$ are the soft masses squared and the $B_i$ are the scalar bilinears. The boundary conditions for these terms at $m_\textrm{GUT}$ in the \gls{CMSSM} are \cite{Martin1997}
\begin{align}
	m_{\tilde f}^2
	&=m_0^2, & \tilde f
	&=\pq,\pau,\pad,\pl,\pe,\pHu,\pHd,A^F_3,\\
	M_i
	&=M_{1/2}, & i
	&=1,2,3,\\
	a_i\,y_i^{-1}
	&=A_0, & i
	&=\pu,\pd,\pe,
\end{align}
with $B_\mu$ and $\mu$ determined from \gls{EWSB} at the low scale
\begin{align}
\mm\pZ2 &=\frac{\mm\pHd2-\mm\pHu2}{\sqrt{1-s^2_{2\be}}}-\mm\pHu2-\mm\pHd2-2\,|\mu|^2, \label{eq:mz2-tree}\\
s_{2\be}&=\frac{2\,B_\mu}{\mm\pHu2+\mm\pHd2+2\,|\mu|^2}
\label{eq:s2b-tree}
\end{align}
and $B_3=0$ for simplicity. The boundary conditions at $m_\textrm{Mess}$ for \gls{GGM} are
 \cite{Martin1997a}
\begin{align}
	M_i
	&=\frac{g_i^2}{16\,\pi^2}\,\Lambda_{G_i}, & i
	&=1,2,3,\\
	a_i
	&=0, & i
	&=\pu,\pd,\pe,\\
	m^2_{\tilde f}
	&=2\sum_{i=1}^3\,C_2(\rep^i_{\tilde f},i)\,k_i\,\frac{g_i^4}{(16\,\pi^2)^2}\,\Lambda_{S_i}^2, & \tilde f
	&=\pq,\pau,\pad,\pl,\pe,\pHu,\pHd,A^F_3,
	\label{eq:scalar-mass-cggm}
\end{align}
with $B_\mu$ and $\mu$ again determined from \gls{EWSB} at the low scale as in eqs. \ref{eq:mz2-tree} and \ref{eq:s2b-tree}. $C_2(\rep^i_{\tilde f},i)$ is the quadratic Casimir of the representation $\rep^i_{\tilde f}$ under the $i^{\textrm{th}}$ gauge group (see eqs.  \ref{eq:Quadratic_casimir_SUN} and \ref{eq:Quadratic_casimir_U1}) and $k_i=(3/5,1,1)$ is the standard \gls{GUT} normalisation. To compare like with like, we will take the \gls{CGGM} parameter space
\begin{align}
	\Lda_{G_i}&=\Lda_G, & \Lda_{S_i}&=\Lda_S, & i&=1,2,3,
\end{align}
and looking along the line $\Lda_S=\Lda_G$ gives the boundary conditions of the \gls{mGMSB} \cite{Dine1993,Dine1995,Dine1996} subspace of models originally developed in \cite{Dimopoulos1981a,Dine1981,Dine1982,Dine1982a,Nappi1982,Alvarez-Gaume1982}.  We concede that we have not solved the $B_\mu$ problem of \gls{GMSB}. With a future study one could take supplement \gls{GMSB} with a Dirac gluino. Then as was studied in \cite{Abel2009,Abel2010,Dolan2011,Grellscheid2011} $t_\be$ would be taken as an output rather than input, and a small value of $B_\mu$ would be specified at the high scale. $\LL_\textrm{Soft}^{D}$ contains the operators, including the non-standard soft  terms \cite{Jack1999a} generated by the $D$-term SUSY breaking discussed in \ref{sec:effectiveoperators}
\begin{align}
-\LL_\textrm{Soft}^D={}&
\left(i\,m_{D3}\,\pglui^a\cdot\At_3^a+\hc\right)
+\frac{m_{\phi 3}^2}2\,\phi_3^2
+\frac{m_{\si 3}^2}2\,\si_3^2\nonumber\\
&+2\,g_3\,m_{D3}\,\phi_3^a\,\left(\pq^\dagger\,T^a\,\pq+\pau^\dagger\,T^a\,\pau+\pad^\dagger\,T^a\,
\pad\right),
\label{eq:LD}
\end{align}
where $\phi_3$, $\si_3$, $m_{\phi_3}^2$ and $m_{\si_3}^2$ are as defined in eqs. \ref{eq:A3_decomposition} and \ref{eq:mass_CP_even_odd_sgluon} and the $T^a$ are the generators of $\SUC$ in the fundamental representation. The second line in \ref{eq:LD} is the origin of the supersoftness of these models, and provides the additional interaction required for the diagram on the second line of \ref{eq:Squark_mass_1-loop}, cutting off the sensitivity to the \gls{UV} scale where $m_{D3}$ is generated.  Finally, for both the \gls{CMSSM} and \gls{CGGM} we take
\begin{equation}
m_{D3}=\frac{1}{16\pi^2}\Lda_D,\qquad
m_{A_3^D}^2=\frac{c_1^2}{16\pi^2}\,\Lda_D^2,\qquad
B_3=0,
\end{equation}
where $c_1$ represents $\OO(1)$ mixings in the messenger sector that have been tuned to make $B_3$ phenomenologically negligible as already discussed.

\subsection{One loop threshold corrections at the Dirac gluino mass}

\subsubsection{Significance}

The Dirac gluinos and the sgluons play the role of messengers $D$-term \gls{SUSY} for the strongly interacting sparticles. As our calculation will be performed in \DRbar, a mass-independent scheme, in order to treat the large hierarchy between the gluino mass and the rest of the \gls{SUSY} spectrum correctly, we integrate out the gluino and the sgluons, resulting in shifts of the parameters at the gluino mass. This leads to a different behaviour of the \gls{RG} compared to the \gls{MSSM}. The most important contributions to take into account are the corrections to squark masses and to the strong gauge coupling $g_3$. We will see that this alters where \gls{EWSB} occurs and can increase the naturalness of these models.

\subsubsection{Threshold corrections}

\label{sec:thresholds}

\paragraph{Squark masses:} 

The gluino in these models is not pure Dirac, although in some regions of parameter space this may be approximately true. Consequently, instead of using the analytic formulae in eq. \ref{eq:Squark_mass_1-loop}, we will numerically compute the full 1-loop threshold correction to squark masses\footnote
{
	There is no contribution from $\Pi^{\si_3}_{\psq}(\m\psq)$ as the $\si_3$ coupling to squarks is zero.
}
\begin{equation}
	\mm\psq2	\rightarrow\mm\psq2
	-\Pi^{\pglui}_{\psq}(\m\psq)
	-\Pi^{\phi_3}_{\psq}(\m\psq)
\end{equation}
where
\begin{align}
	\Pi_{\psq}^{\pglui}(p)
	&=\frac{g_3^2}{6\,\pi^2}\,\left|(Z_g)_{i,1}\right|^2\,G_0(p,\m{\pglui_i},0),&
	\Pi_{\psq}^{\phi_3}(p)
	&=\frac{g_3^2}{3\,\pi^2}\,m_{D3}^2\,B_0(p,\m\psq,m_{\phi_3})
\end{align}
and $Z_g$ is the matrix that diagonalises the gluino mass matrix $\m\pglui$
\begin{align}
	\m\pglui
	&=\begin{pmatrix} M_3 & m_{D3} \\ m_{D3} & 0 \end{pmatrix},&
	Z_g\,\m\pglui\,Z_g^\dagger&=\textrm{diag}(\m{\pglui_1},\m{\pglui_2})
\end{align}
where $\m\pglui$ is in the $(\pglui,\tilde{A}_3)$ basis. $B_0$ and $G_0$ are scalar integrals \cite{Pierce1996,Ellis2008}. 

\bigskip

\paragraph{Strong gauge coupling:}
The 1-loop threshold corrections to $g_3$ at $m_{D3}$ are \cite{Hall1981}
\begin{equation}
	g_3
	\rightarrow g_3
	\left\{
	1\pm\frac{g_3^2}{16\,\pi^2}
	%\frac1{16\pi^2}\times\frac16\times\frac32 comes from one half of the dynkin index plus a factor of 1/6 for each real scalar
		\left[
		\sum_i\log\left(\frac{\mm{\pglui_i}2}{m_{D3}^2}\right)
		+\frac14\log\left(\frac{m_{\phi_3}^2}{m_{D3}^2}\right)
		%+\frac14\log\left(\frac{m_{\si_3}^2}{m_{D3}^2}\right)
		\right]
	\right\}
\end{equation}
where the positive (negative) contribution occurs when running from the \gls{UV} (\gls{IR}) to the \gls{IR} (\gls{UV}) and all parameters are evaluated at the renomalisation scale $\mu(m_{D3})=m_{D3}$.

\paragraph{Quark masses:} We do not implement the quark mass threshold corrections from the gluinos and sgluons. To correctly do this would be quite technical and we anticipate that the overall impact on the areas we are interested in (such as the \gls{SUSY} spectrum, \gls{EWSB} and tuning) should be minimal; correction of this kind must be proportional to chiral symmetry breaking and since the quarks are essentially massless at $m_{D3}$ the remaining correction is proportional to the Majorana gluino mass. For the top quark \cite{Pierce1996}
\begin{equation}
\de\m\pt
=-\frac{g_3^2}{12\,\pi^2}\sin(\tta_{\pst})\,M_3
\left[
B_0(0,M_3,\m{\pstn1})
-B_0(0,M_3,\m{\pstn2})
\right],
\end{equation}
where $\tta_{\pst}$ is the stop mixing angle. This will alter the yukawa couplings in the \gls{UV} and hence only affect the running of 
\gls{UV} parameters that depend on the yukawa couplings. We expect the low energy physics to be largely unaffected however, and instead we include the loop contributions to the quark masses from gluinos and sgluons at $\m\pZ$ and $m_\textrm{SUSY}$. By doing this we making a systematic error proportional to $(16\pi^2)^{-2}\times\log(m_{D3}/m_\textrm{SUSY})\times\log(m_{GUT}/m_\textrm{SUSY})\lesssim 0.1\%$.
\section{Numerical setup}

We use the standard top-down approach where we fix a set of \gls{UV} boundary conditions at either $m_\textrm{GUT}$ in the \gls{CMSSM} or $m_\textrm{Mess}$ in \gls{CGGM}. The low energy spectrum is found through \gls{RG} evolution, and then the corresponding flavour observables and fine tuning are calculated. 

To achieve this, we have used the Mathematica package \SARAH 4.3.0 \cite{Staub2008a,Staub2010,Staub2011c,Staub2012,Staub2013,Porod2014}
to generate source code for the spectrum generator \SPheno 3.3.2 \cite{Porod2011,Porod2012}. \SPheno solves the \gls{RG} equations taking into account the presence of the Dirac gluino at one and two loops. This program then calculates the one loop masses for all particles in the model, the branching ratios for all kinematically allowed two body decays and the branching ratios for three body decays involving intermediate $\PW$ and $\PZ$ bosons.

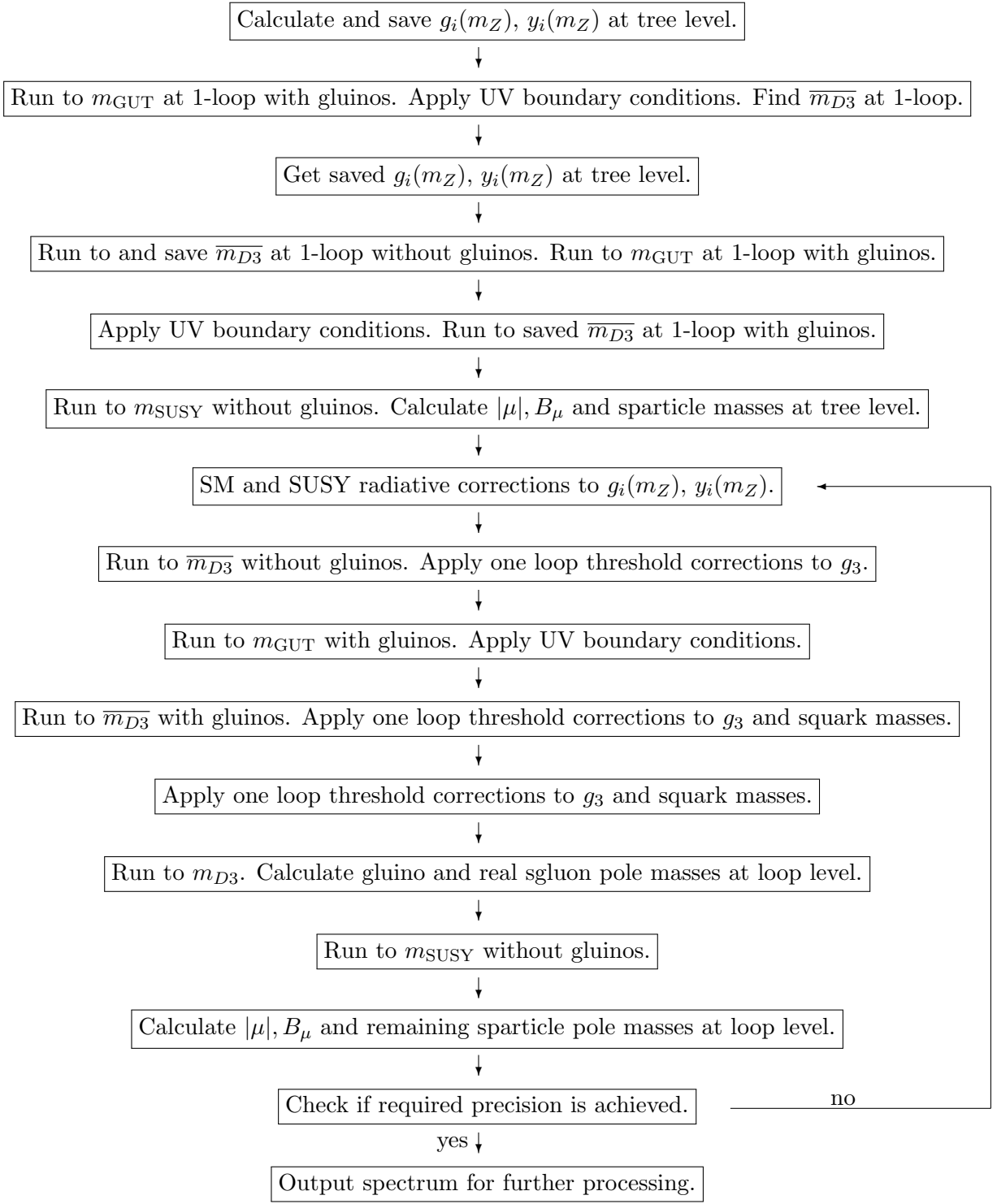
\begin{figure}[h]
\begin{center}
\setlength{\unitlength}{0.8pt}
\begin{picture}(350,700)
\put(40,675){\makebox(280,10)[c]{\fbox{Calculate and save $g_i(m_Z)$, $y_i(m_Z)$ at tree level.}}}
\put(175,665){\vector(0,-1){12}}
\put(40,630){\makebox(280,10)[c]{\fbox{Run to $m_\textrm{GUT}$ at 1-loop with gluinos. Apply UV boundary conditions. Find $\overline{m_{D3}}$ at 1-loop.}}}
\put(175,620){\vector(0,-1){12}}
\put(40,585){\makebox(280,10)[c]{\fbox{Get saved $g_i(m_Z)$, $y_i(m_Z)$ at tree level.}}}
\put(175,575){\vector(0,-1){12}}
\put(40,540){\makebox(280,10)[c]{\fbox{Run to and save $\overline{m_{D3}}$ at 1-loop without gluinos. Run to $m_\textrm{GUT}$ at 1-loop with gluinos.}}}
\put(175,530){\vector(0,-1){12}}
\put(40,495){\makebox(280,10)[c]{\fbox{Apply UV boundary conditions. Run to saved $\overline{m_{D3}}$ at 1-loop with gluinos.}}}
\put(175,485){\vector(0,-1){12}}
\put(40,450){\makebox(280,10)[c]{\fbox{Run to $m_\textrm{SUSY}$ without gluinos. Calculate $|\mu|,B_\mu$ and sparticle masses at tree level.}}}
\put(175,440){\vector(0,-1){12}}
\put(40,405){\makebox(280,10)[c]{\fbox{SM and SUSY radiative corrections to $g_i(m_Z)$, $y_i(m_Z)$.}}}
\put(175,395){\vector(0,-1){12}}
\put(40,360){\makebox(280,10)[c]{\fbox{Run to $\overline{m_{D3}}$ without gluinos. Apply one loop threshold corrections to $g_3$.}}}
\put(175,350){\vector(0,-1){12}}
\put(40,315){\makebox(280,10)[c]{\fbox{Run to $m_\textrm{GUT}$ with gluinos. Apply UV boundary conditions.}}}
\put(175,305){\vector(0,-1){12}}
\put(40,270){\makebox(280,10)[c]{\fbox{Run to $\overline{m_{D3}}$ with gluinos. Apply one loop threshold corrections to $g_3$ and squark masses.}}}
\put(175,260){\vector(0,-1){12}}
\put(40,225){\makebox(280,10)[c]{\fbox{Apply one loop threshold corrections to $g_3$ and squark masses.}}}
\put(175,215){\vector(0,-1){12}}
\put(40,180){\makebox(280,10)[c]{\fbox{Run to $m_{D3}$. Calculate gluino and real sgluon pole masses at loop level.}}}
\put(175,170){\vector(0,-1){12}}
\put(40,135){\makebox(280,10)[c]{\fbox{Run to $m_\textrm{SUSY}$ without gluinos.}}}
\put(175,125){\vector(0,-1){12}}
\put(40,70){\makebox(280,50)[c]{\fbox{Calculate $|\mu|,B_\mu$ and remaining sparticle pole masses at loop level.}}}
\put(175,80){\vector(0,-1){12}}
\put(40,45){\makebox(280,10)[c]{\fbox{Check if required precision is achieved.}}}
\put(20,23){\makebox(280,10)[c]{yes}}
\put(225,49){\makebox(320,10)[c]{no}}
\put(470,410){\vector(-1,0){100}}
\put(470,49){\line(-1,0){150}}
\put(470,49){\line(0,1){361}}
\put(175,35){\vector(0,-1){12}}
\put(40,0){\makebox(280,10)[c]{\fbox{Output spectrum for further processing.}}}
\end{picture}
\setlength{\unitlength}{1.0pt}
\end{center}
\caption{Algorithm used to calculate the spectrum. Adapted from fig. 1 in \cite{Porod2011}. Note that apart from where it explicitly states running to a saved value of $\overline{m_{D3}}$, the scale is found by requiring a solution to $\mu(m_{D3})=m_{D3}$. This typically updates with each iteration since it depends on the behaviour of $g_3$ whose running is determined by the location of $\overline{m_{D3}}$.}
\label{fig:algorithm}
\end{figure}

The UV boundary conditions discussed in section \ref{sec:UVboundary} are implemented we can solve the \gls{EWSB} minimisation conditions for $\mu$ and $B_\mu$. We only study the $\mu>0$ case in order to maximise the effect from stop mixing upon the Higgs sector. The \SPheno code has been modified to include an intermediate step in \gls{RG} running where the gluino and its corresponding scalar degrees of freedom are integrated out at the gluino mass. This implements the \gls{EWSB} mechanism of supersoft models outlined in \cite{Fox2002}. The masses for the gluino and the real sgluon are calculated at the this intermediate scale instead of $m_\textrm{SUSY}$. A schematic of this algorithm is shown in fig. \ref{fig:algorithm}.

\begin{table}[t]
\begin{center}
	\begin{tabular}{c|c|c}
	 Sparticle  & Lower Mass Limit at 95 \% CL (GeV) & Reference \\ \hline
	 Neutralino (stable) & $45.5$ & \cite{Beringer2012} \\
	 Neutralino (unstable) & $96.8$ & \cite{TheDELPHICollaboration2005,TheOPALCollaboration2006} \\	
	 Sneutrino  & $41$ & \cite{Beringer2012} \\ 
	 Chargino   & $103.5 $                & \cite{LEP22001,LEP22002}\\
	 Sleptons   & $100.2$& \cite{LEPSUSYWG}\\
	\end{tabular}
\end{center}
\caption{The strongest most model independent non-hadron collider limits on \gls{LOSP} and \gls{NLSP} masses. The lightest neutralino $\pneuz$ is assumed to be bino-like, and allowed to decay to the gravitino $\tilde{G}$ in \gls{GMSB}, emitting a photon.}
\label{tab:sparticlelimits}
\end{table}

\section{Spectra}
\label{sec:spectra}

On each of the parameter space plots we include the relevant limits on \gls{SUSY} particle masses. As production cross section is suppressed for all \gls{SUSY} particles in models with Dirac gluinos (shown in section \ref{sec:XSecs}), we take only the strongest most model independent limits available set by lepton colliders, outlined in table~\ref{tab:sparticlelimits}. For the \gls{CMSSM}, the stable neutralino limit is applied, whereas for \gls{CGGM} the unstable limit is used instead. The red, purple and green solid lines indicate the limit on the slepton, neutralino and sneutrino masses, and the blue dashed line indicates the limit on the chargino masses. 
\clearpage
\noindent
Points below and to the left of these lines are excluded at the 95\% \gls{CL}.
We will present three types of graphs in the $(m_0,M_{1/2})$ and $(\Lda_G,\Lda_S)$ planes to illustrate the similarities and differences between spectra:
\begin{itemize}
\item Gradients of Higgs boson masses with contours of the parameters entering into the one loop Higgs mass approximation in eq.
\ref{eq:1loopmh}.
\item \gls{LOSP} species with mass contours of the typical candidates.
\item \gls{NLOSP} species.
\end{itemize}

In the \gls{MSSM} the two loop contribution from gluinos gives quite a significant contribution. Because the two loop Higgs mass has not yet been computed in the presence of a Dirac gluino, we will not impose achieving the correct value as a strict requirement, as we would be incorrectly ruling out viable regions of parameter space. Although the full calculation will be completed in the future \cite{Goodsell2014}, the effective field theory framework used here requires a different approach. At the gluino scale, one would need to match the theory onto a theory with broken \gls{SUSY} with \glspl{RGE}. This requires removing the approximation that e.g. the stop-Higgs quartic coupling and the Higgs-top yukawa terms remain equal along the \gls{RG} flow
\begin{equation*}
y_{\Ptop}\,\APtop\,\Pq\cdot \pHu\,\longleftrightarrow |y_{\Ptop}|^2|\PStop|^2|\pHu|^2.
\end{equation*}
Instead, the coefficients of the operators $\APtop\,\Pq\cdot \pHu$ and $|\PStop|^2|\pHu|^2$ should have different \glspl{RGE} below the Dirac gluino mass. After applying threshold corrections to each coupling, flowing down from the gluino mass to the \gls{SUSY} scale would then correctly include the two loop contributions to the Higgs mass with gluino integrated out. With the new non-\gls{SUSY} \gls{RGE} calculators becoming available \cite{Lyonnet2013,Staub2013}, the possibility to correctly incorporate these kinds of particle threshold effects into spectrum generators in the future is a very interesting possibility

Only a subset of the scans are presented in the body of the text. The remaining parameter configurations can be found in appendix \ref{app:additional_plots}. The generic dependence of the spectrum and low energy parameters on the UV boundary conditions can be inferred by analysing the cases we present.
\label{sec:CMSSM-scans}

We first present the comparison of the \gls{CMSSM} with and without a Dirac gluino. We scan
\begin{align}
0 \; \textrm{TeV} \; \leq m_0 & \leq 6 \; \textrm{TeV} &
0 \; \textrm{TeV} \; \leq M_{1/2} & \leq 4 \; \textrm{TeV}
\end{align}
and take a moderate and large $t_\be = 10, 25$. In the presence of a Dirac gluino, we set $m_{D3}(m_\textrm{GUT}) = 5, 7.5, 10$ TeV which, due to \gls{RG} running, lead to a significant spread of physical Dirac gluino masses that can be estimated using
\begin{equation}
	\overline{m_{D3}}|_\textrm{approx}	=\left[m_{D3}(\Lda)\,\Lda^{\frac{3\,g_3^2(\Lda)}{8\,\pi^2}}\right]^{\frac{1}{1+\frac{3\,g_3^2(\Lda)}{8\,\pi^2}}}
\end{equation}
where $\Lda$ can be any scale, but is most conveniently taken as the \gls{UV} scale.

The first thing to note is that there is a new region of parameter space in the $(m_0,M_{1/2})$ plane opening up for very low $M_{1/2}$ but non-zero $m_0$ in the presence of a Dirac gluino. This region isn't populated in the \gls{MSSM} due to an absence of \gls{EWSB} when $\mm\pHu2$ isn't pushed negative enough for a positive $|\mu|^2$ solution; at this point in parameter space in the \gls{CMSSM} one needs extra logs from $M_3$ to push the squark mass up along the \gls{RG} trajectory. In the case of a Dirac gluino, one can essentially ignore the need for a Majorana gluino mass, as the threshold correction on its own is enough to lift the squark mass in the \gls{IR}, triggering \gls{EWSB} for even zero $M_{1/2}$. Here however, the \gls{LEP} bound on the chargino mass becomes important, putting an experimental lower limit on $M_{1/2}$ of $\OO(100)\,\textrm{GeV}$.

\begin{figure}[t]
\begin{center}
\includegraphics[scale=0.65]{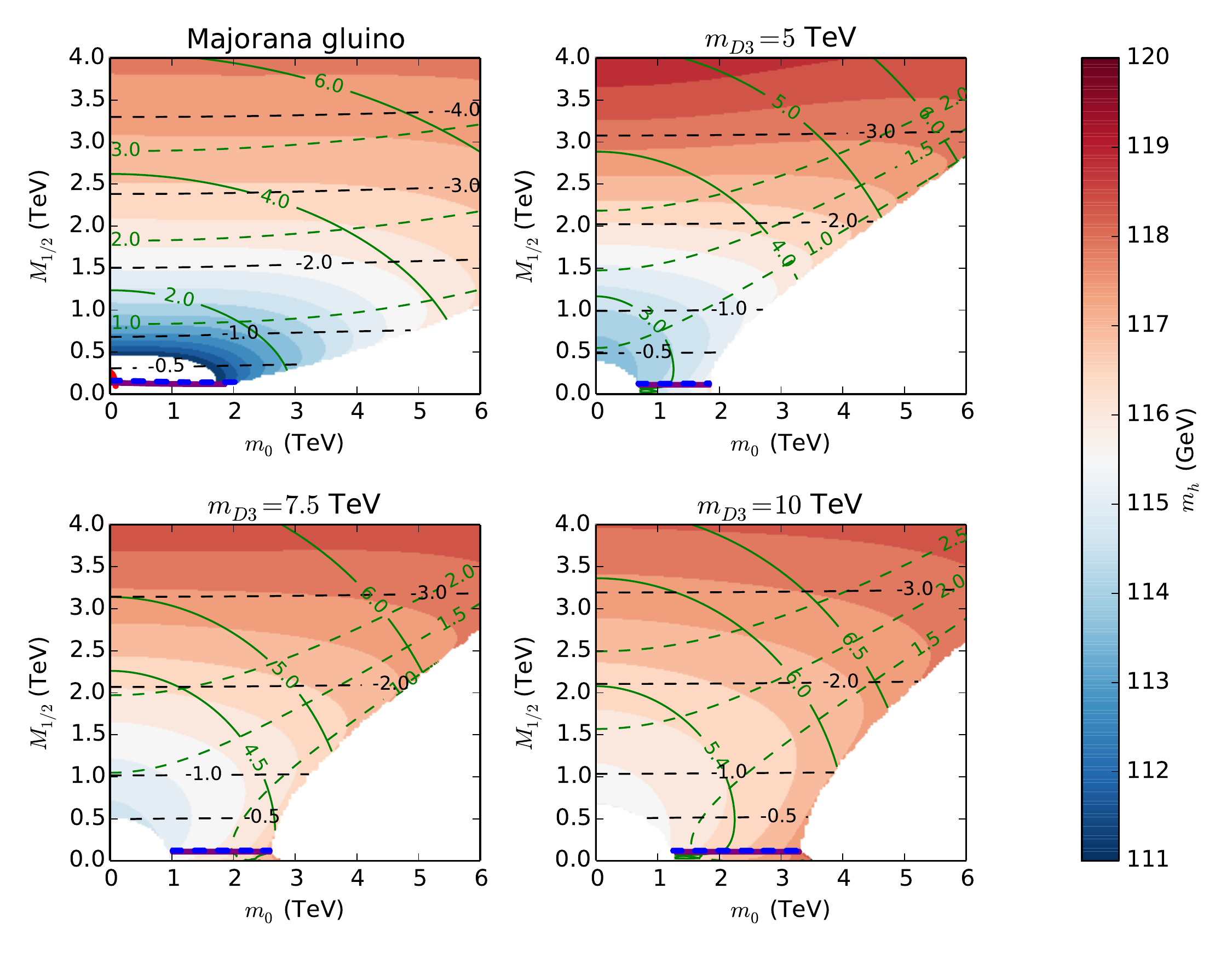}
\end{center}
\caption{Higgs sector parameters in the \gls{CMSSM} with $t_\be = 10$ and $m_{D3}$ fixed as indicated. The gradient indicates the Higgs mass. The black dashed, green dashed and green solid lines are contours of $a_{\pt}(m_\textrm{SUSY})$, $\mu(m_\textrm{SUSY})$, and $m_\textrm{SUSY}$ respectively. All contours unless otherwise specified are in TeV.}
\label{fig:HiggsCMSSMtb10}
\end{figure}

\begin{figure}[t]
\begin{center}
\includegraphics[scale=0.55]{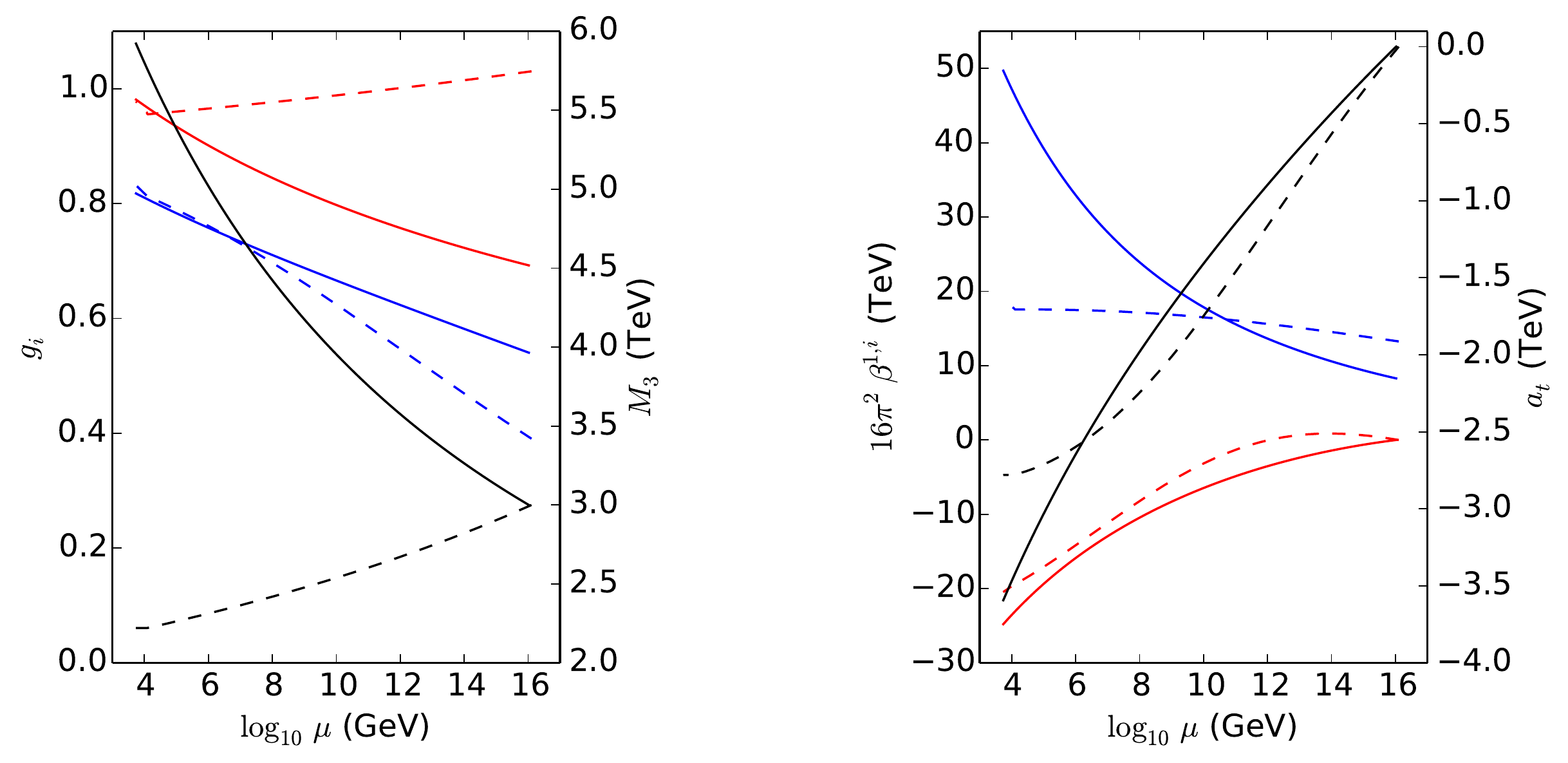}
\end{center}
\caption{\gls{RG} evolution of dominant parameters contributing to the running of $a_{\pt}$ in the \gls{CMSSM} with $m_0 = 4.5$ TeV, $M_{1/2} = 4$ TeV, $m_{D3} = 5$ TeV and $t_\be=25$. Solid lines correspond to the \gls{CMSSM} and dashed lines correspond to the \gls{CMSSM} supplemented with a Dirac gluino. \textbf{Left}: The blue, red and black lines show the evolution of $y_{\pt}$, $g_3$ and $M_3$ respectively. \textbf{Right}: The blue, red and black lines show the evolution of $\frac{32}3\,y_{\pt}\,g_3^2\,M_3$, $a_{\pt}\left(18\,|y_{\pt}|^2-\frac{16}3\,g_3^2\right)$ and $a_{\pt}$ respectively.}
\label{fig:RGEsAt}
\end{figure}

\paragraph{Higgs:} In figures \ref{fig:HiggsCMSSMtb10} and \ref{fig:HiggsCMSSMtb25} we show the Higgs mass and the parameters entering the one loop Higgs mass formula in eq. \ref{eq:1loopmh}. Even though we are taking $A_0(m_\textrm{GUT})=0$, a non-zero value is generated by running. In the large $y_t$ limit (see eq. \ref{eq:beta-au-1loop} for the complete expression)
\begin{equation}
	16\pi^2\be_{a_{\pt}}^{(1)}
	\supset a_{\pt}\left[
	18\,|y_{\pt}|^2+\frac{16}3\,\left(\ttag-2\right)\,g_3^2
	\right]
	+\frac{32}3\,y_{\pt}\,g_3^2\,M_3\,\ttag,
	\label{eq:beta-at}
\end{equation}
where
\begin{align}
	\ttag&=1\quad\textrm{if}\quad\mu\geq\overline{m_{D3}}, & 
	\ttag&=0\quad\textrm{if}\quad\mu<\overline{m_{D3}}.
\end{align}
with the precise definitions given in appendix \ref{sec:RGEs}. In the \gls{CMSSM} without a gluino, $\ttag = 1$ always in eq. \ref{eq:beta-at}. Note that we do not observe the more negative values of $a_{\pt}$ in the presence of the gluino that were found in \cite{Bhattacharyya2013}. This can be understood by considering the running of the Majorana gluino mass in the presence of a Dirac gluino
\begin{align}
16\pi^2\be^{(1)}_{M_3}
&= -6\,g_3^2\,M_3& & \textrm{MSSM}\\
16\pi^2\be^{(1)}_{M_3}
&=0& & \textrm{MSSM with Dirac Gluino}.
\end{align}
Because we are taking $a_{\pt}(m_\textrm{GUT})=0$ then the gluino term dominates for most of the flow, and in the \gls{CMSSM}, this term becomes larger that in the \gls{CMSSM} with a Dirac gluino as demonstrated in fig. \ref{fig:RGEsAt}.
\begin{figure}[t]
\begin{center}
\includegraphics[scale=0.65]{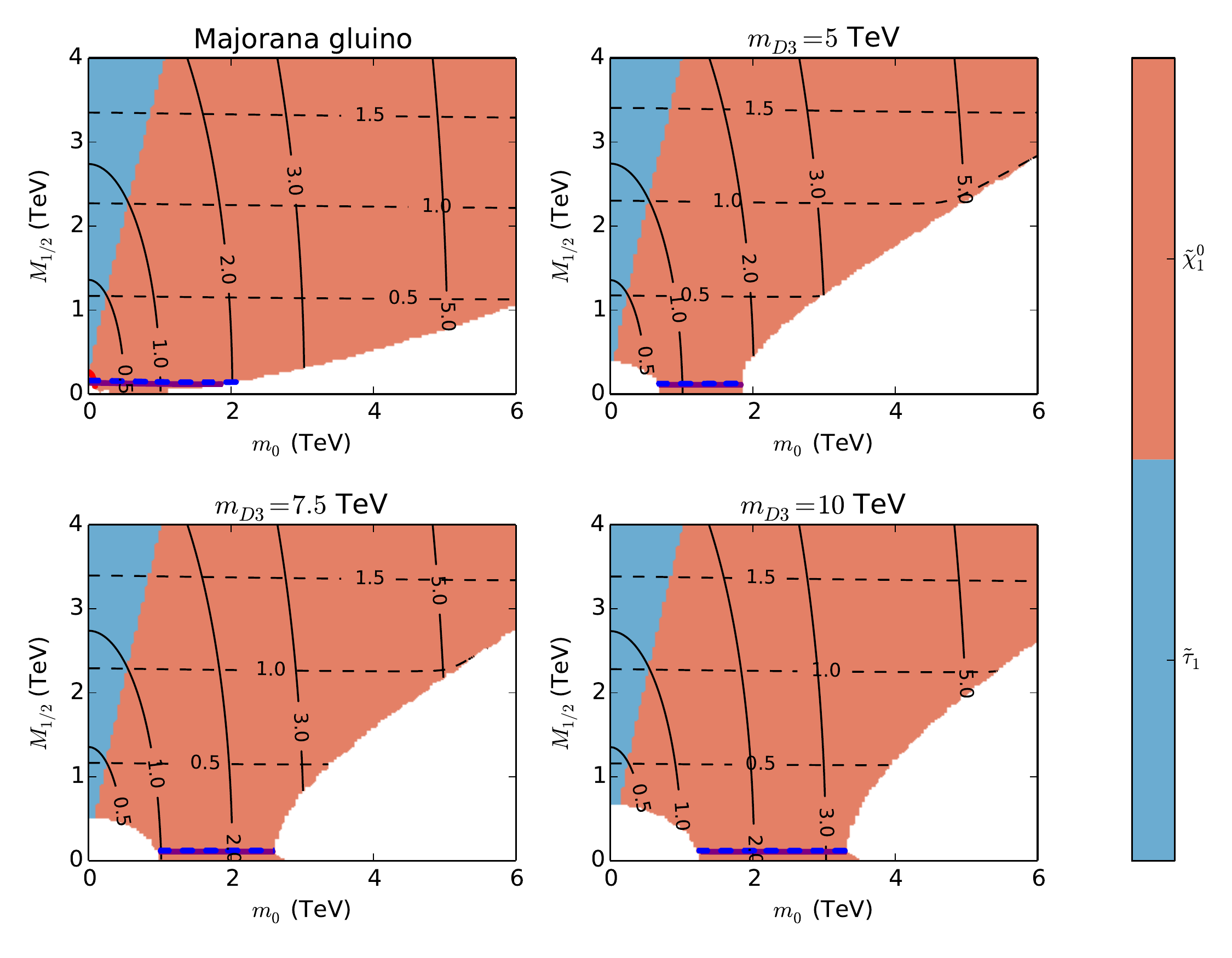}
\end{center}
\caption{\gls{LOSP} species in the \gls{CMSSM} with $t_\be = 10$ and $m_{D3}$ fixed as indicated. The black dashed and black solid lines are contours of lightest neutralino mass $\m\pneuz$ and stau mass $\m\pstau$ in TeV.}
\label{fig:CMSSM-LSP-tb10}
\end{figure}
The contours of $m_\textrm{SUSY}$ in the presence of a Dirac gluino are increased to the minimum squark mass possible in the model (i.e. determined by eq.	\ref{eq:Squark_mass_1-loop}). For large values of $m_0$ and $M_{1/2}$ contours of $m_{\textrm{SUSY}}$ across the different models approach each other. 

The $\mu$ parameter is seen to increase with increasing Dirac gluino mass. This can be understood by considering the \gls{EWSB} conditions in the large $t_\be$ limit
\begin{equation}
|\mu|^2
=-\mm\pHu2-\frac{\mm\pZ2}2+\OO(t_\be^{-2}).
\end{equation}
$\mm\pHu2$ is driven negative by the squark soft scalar masses
\begin{align}
	16\pi^2\be^{(1)}_{\mm\pHu2}
	&\supset6\,|y_{\pt}|^2\left(\mm\pq2+\mm\pt2\right)
	\label{eq:betamHu2}
\end{align}
which are in turn determined by the Dirac gluino mass through eq. \ref{eq:Squark_mass_1-loop}. The values of $\mu$ in the \gls{MSSM} for moderate $(m_0,M_{1/2})$ are actually lower with a Dirac gluino than without. Considering the \gls{RG} equation for $y_{\pt}$
\begin{equation}
16\pi^2\be^{(1)}_{y_{\pt}}
\supset\frac{8}3(\ttag-3)\,g_3^2.
\end{equation}
This term causes $y_{\pt}$ to decrease in the flow from the \gls{IR} to the \gls{UV}. In the \gls{MSSM}, the strong interactions retain asymptotic freedom, whereas with a Dirac gluino present, $g_3$ remains roughly constant along the entire flow. In the Dirac gluino case, this causes $y_{\pt}$ to decrease much more rapidly, and so the the integrated term of eq. \ref{eq:betamHu2} with a Dirac gluino than without.

The lower limit on squark masses translates into a lower limit on the Higgs mass. Apart from at low $(m_0,M_{1/2})$ where we get a separation between the strong and electroweak sectors it is difficult to distinguish the \gls{CMSSM} with and without a gluino. The presence of a Dirac gluino allows us, for a given Higgs mass, to realise a lighter electroweak scalar spectrum for low $(m_0,M_{1/2})$.

\begin{figure}[t]
\begin{center}
\includegraphics[scale=0.65]{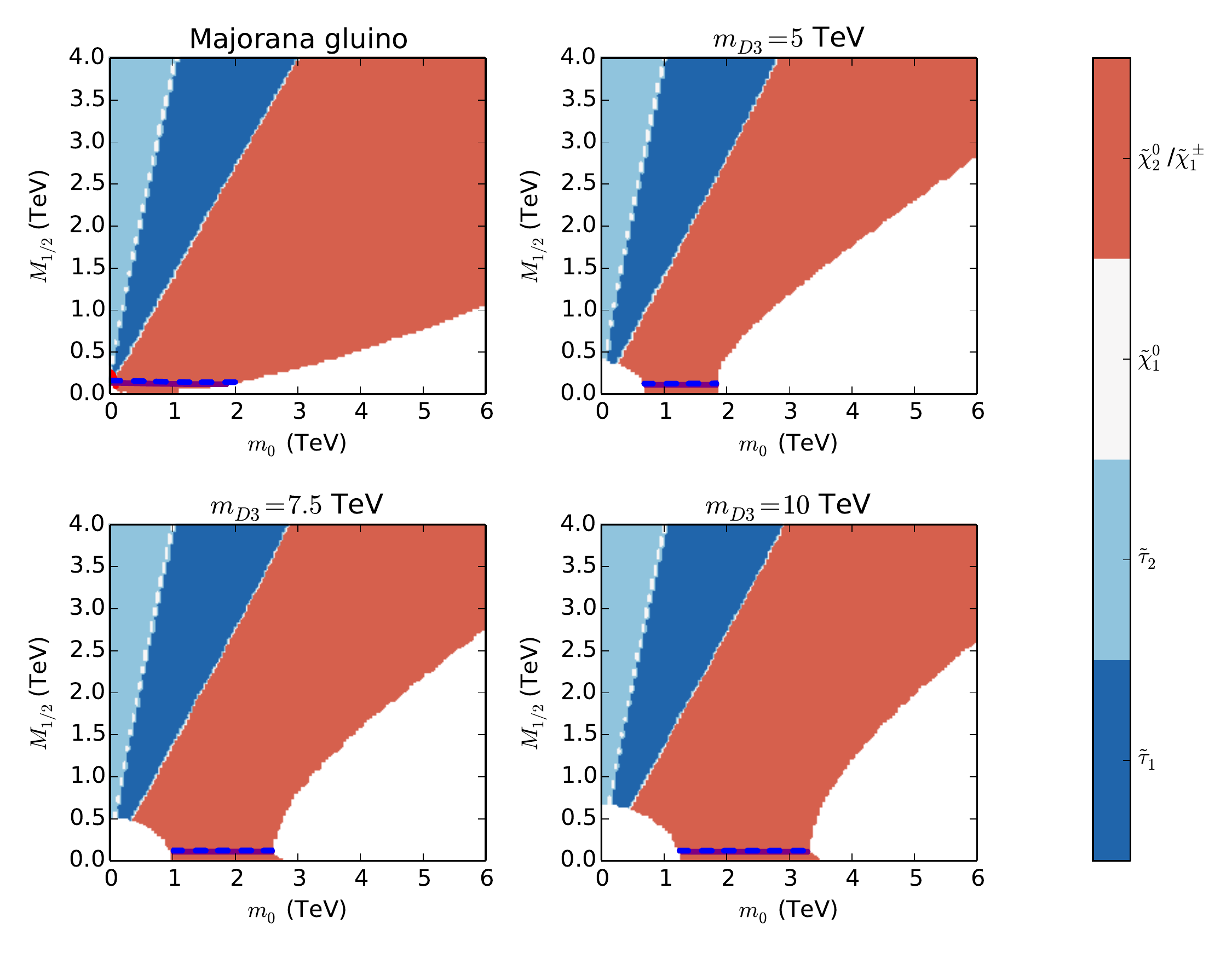}
\end{center}
\caption{\gls{NLOSP} species in the \gls{CMSSM} with $t_\be = 10$ and $m_{D3}$ fixed as indicated.}
\label{fig:CMSSM-NLSP-tb10}
\end{figure}

\begin{figure}[t]
\begin{center}
\includegraphics[scale=0.65]{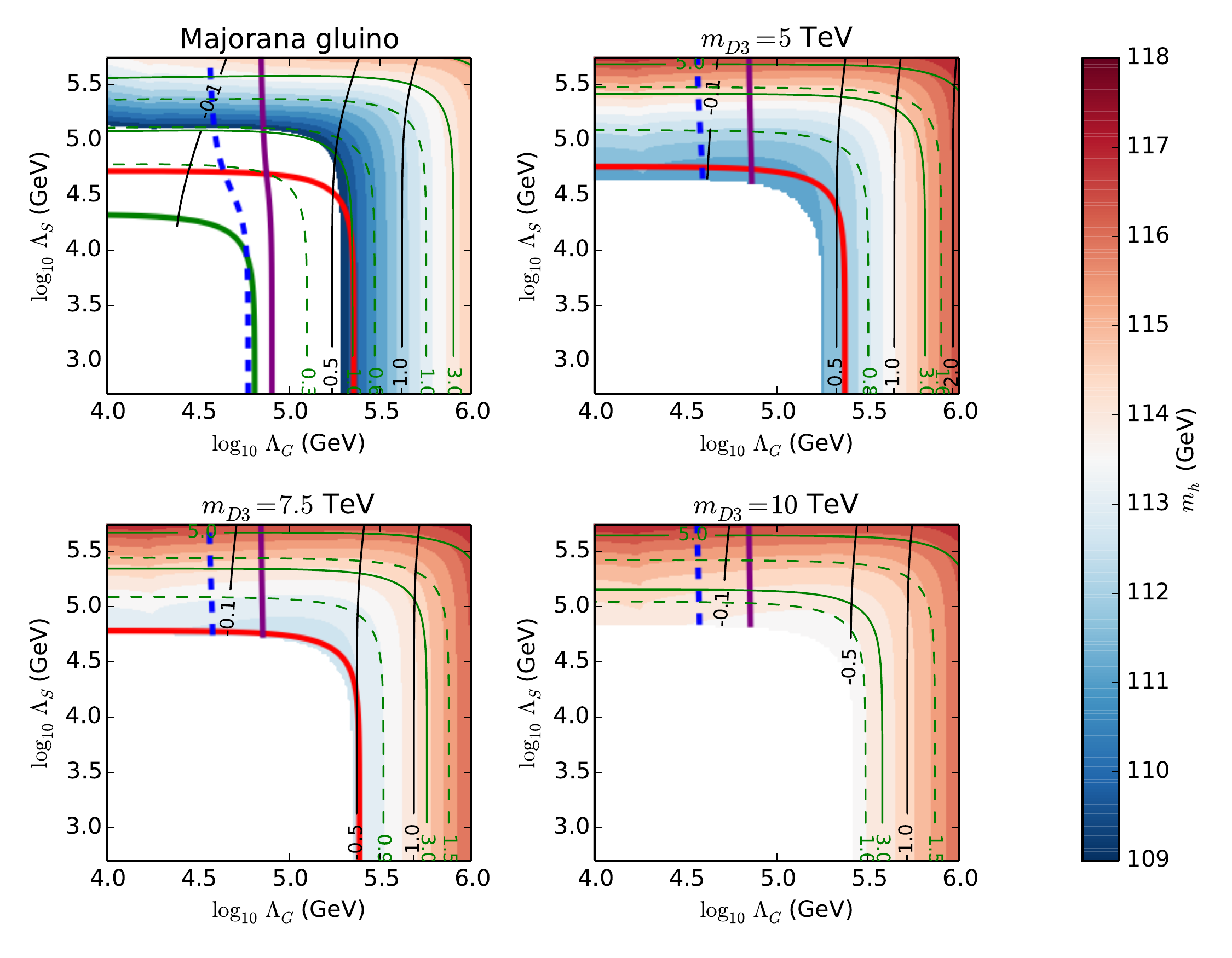}
\end{center}
\caption{Higgs sector parameters in \gls{CGGM} with $t_\be = 10$, $m_\textrm{Mess}=10^7$ GeV and $m_{D3}$ fixed as indicated. The gradient indicates the Higgs mass. The black dashed, green dashed and green solid lines are contours of $a_{\pt}(m_\textrm{SUSY})$, $\mu(m_\textrm{SUSY})$, and $m_\textrm{SUSY}$ respectively. All contours unless otherwise specified are in TeV.}
\label{fig:CGGM-Higgs-tb10-M7}
\end{figure}

\paragraph{LOSP:} The \gls{LOSP} candidate in the presence of a Dirac gluino is essentially unchanged in the \gls{CMSSM}. The blue regions in figs. \ref{fig:CMSSM-LSP-tb10} and \ref{fig:CMSSM-LSP-tb25} have a charged stau $\pstau_1$ as the \gls{LOSP} and so are excluded. The remainder of the parameter space is entirely bino-like neutralino $\pneuz$ \gls{LOSP}, a good dark matter candidate.

\paragraph{NLOSP:} The \gls{NLOSP} candidate in the presence of a Dirac gluino is similarly relatively unchanged essentially when compared to the Majorana case. The light blue regions in figs. \ref{fig:CMSSM-NLSP-tb10} and \ref{fig:CMSSM-NLSP-tb25} have the second lightest stau $\pstau_2$ as the \gls{NLOSP} but are excluded as the corresponding region has a lightest stau $\pstau_1$ \gls{LOSP}. The dark blue region has lightest stau $\pstau_1$ \gls{LOSP} and leads to one lepton and \gls{MET} or jets and \gls{MET} in the final state, as does the red region with wino-like chargino $\PSino^{\pm}$ \gls{NLOSP}. This chargino $\PSino^{\pm}$ is also coincident with the wino-like neutralino  $\PSino^0_2$ which instead leads to either entirely \gls{MET} in the final state or \gls{MET} with either two leptons of opposite sign or a jet.

It is clear that nature of the light spectrum is largely unaffected by the presence of a Dirac gluino, except that it is now possible to raise the strongly interacting sector almost\footnote{There will always arise terms proportional to $(16\,\pi^2)^{-1}\log(m_{D3}/m_\textrm{SUSY})$, and there two loop sensitivity to the sgluon soft mass in eq. \ref{eq:non-supersoft-beta} present.} independently of the electroweak sector, giving some freedom to aleviate the tension with results at hadron collider experiments to date. 
\subsection{Constrained General Gauge Mediation}

\begin{figure}[t]
\begin{center}
\includegraphics[scale=0.65]{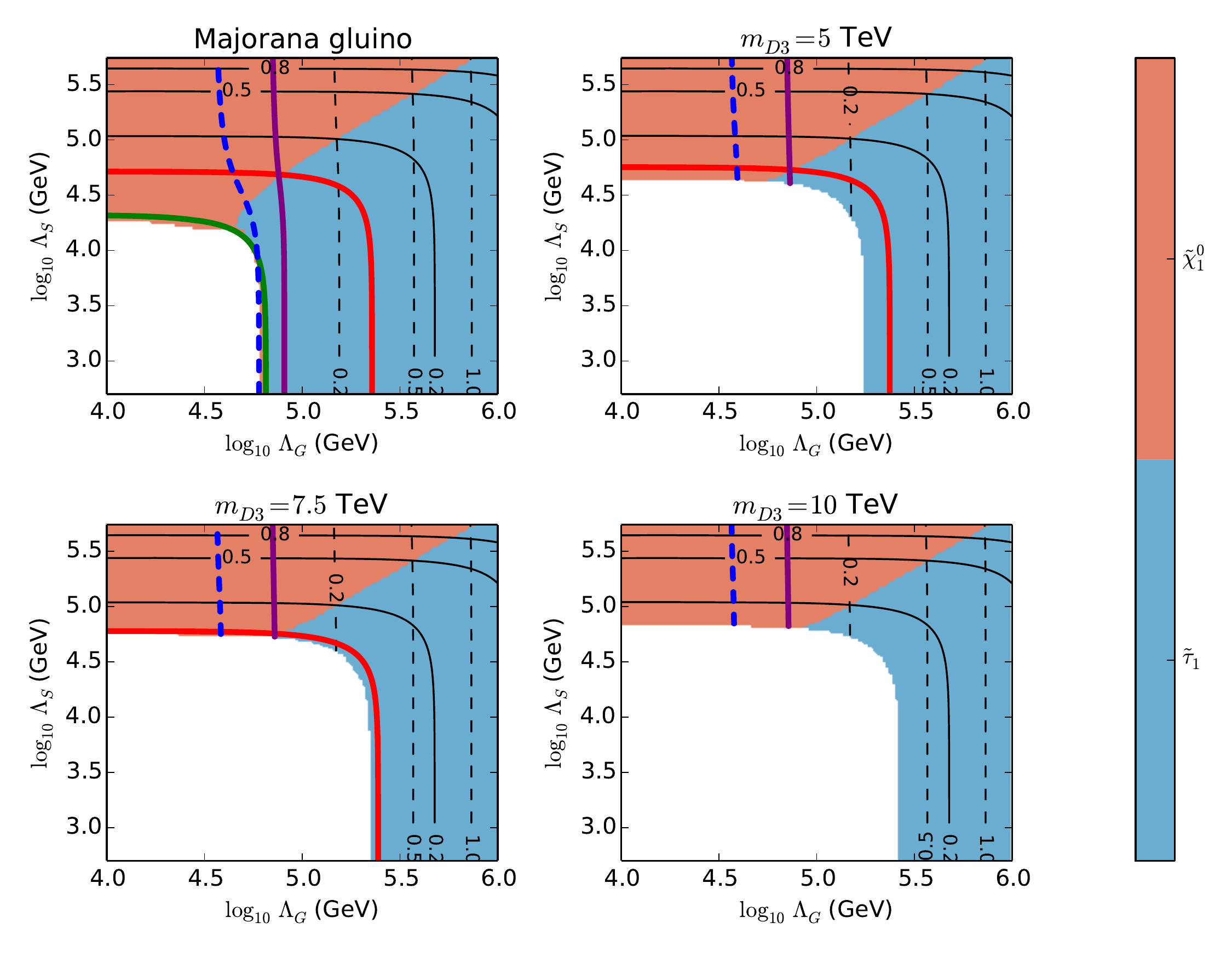}
\end{center}
\caption{\gls{LOSP} species in \gls{CGGM} with $t_\be = 10$, $m_\textrm{Mess} = 10^7$ GeV and $m_{D3}$ fixed as indicated. The black dashed and black solid lines are contours of lightest neutralino mass $\m\pneuz$ and stau mass $\m\pstau$ in TeV.}
\label{fig:CGGM-LSP-tb10-M7}
\end{figure}

We now present the comparison of \gls{CGGM} with and without Dirac gluino. A recent comprehensive study of the parameter space of \gls{CGGM} was done in \cite{Grajek2013a}. We scan
\begin{align}
10^3 \; \textrm{GeV} \; \leq \Lda_G & \leq 10^7 \; \textrm{GeV} &
10^3 \; \textrm{GeV} \; \leq \Lda_S & \leq 10^7 \; \textrm{GeV}
\end{align}
whilst taking $t_\be = 10, 25$ and again we again take $m_{D3}(m_\textrm{GUT}) = 5, 7.5, 10$ TeV in the presence of a Dirac gluino. We take two messenger scales $m_\textrm{Mess}=10^7 \; \textrm{GeV}$ and $ 10^{12} \; \textrm{GeV}$ to represent  short and long periods of running.

The theoretically allowed parameter space is reduced by the presence of a Dirac gluino as is seen in fig. \ref{fig:CGGM-LSP-tb10-M7}. Although viable \gls{EWSB} is  occurring, the the lightest stau $\pstau_1$ is being driven tachyonic for a larger portion of the parameter space. This is induced by the Dirac gluino much for much higher \gls{UV} stau mass set by eq. \ref{eq:scalar-mass-cggm}. This is caused by  larger values of $|\mu|^2$ for a given $(\Lda_G,\Lda_S)$ by the threshold corrections at the Dirac gluino scale, driving the smallest eigenvalue of the stau mass matrix
\begin{equation}
m^2_{\Ptau,\textrm{mat}}=
\begin{pmatrix}
\mm{\pl_{3,3}}2 + D\,\textrm{terms} & v\,(a_{\Ptau}^*\,c_\be - \mu\, y_{\Ptau}\,s_\be) \\ 
v\,(a_{\Ptau}c_\be - \mu^*\, y_{\Ptau}\,s_\be) & \mm{\pe_{3,3}}2 + D\,\textrm{terms}
\end{pmatrix}
\end{equation}
negative.

\paragraph{Higgs:} In figures \ref{fig:CGGM-Higgs-tb10-M7}, \ref{fig:CGGM-Higgs-tb25-M7}, \ref{fig:CGGM-Higgs-tb10-M12} and \ref{fig:CGGM-Higgs-tb25-M12} we show the Higgs mass and the parameters entering the one loop Higgs mass formula in eq. \ref{eq:1loopmh}. The characteristic properties here are essentially unchanged from the \gls{CMSSM} counterpart as we have only considered the \gls{CMSSM} case $A_0=0$.

\begin{figure}[t]
\begin{center}
\includegraphics[scale=0.65]{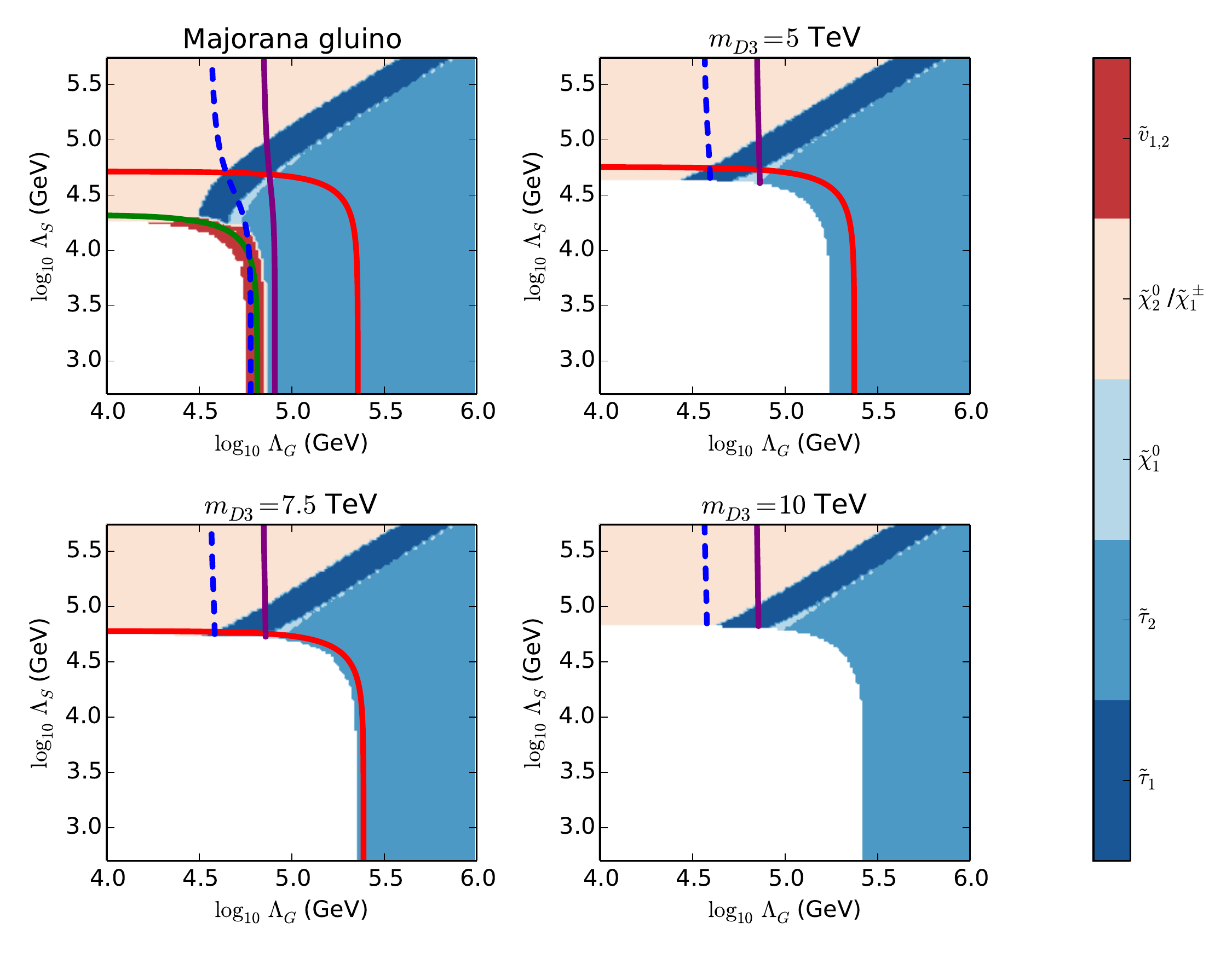}
\end{center}
\caption{\gls{NLOSP} species in \gls{CGGM} with $t_\be = 10$, $m_\textrm{Mess} = 10^7$ GeV and $m_{D3}$ fixed as indicated. The black dashed and black solid lines are contours of lightest neutralino mass $\m\pneuz$ and stau mass $\m\pstau$ in TeV.}
\label{fig:CGGM-NLSP-tb10-M7}
\end{figure}

\paragraph{LOSP:} The \gls{LOSP} candidates  in \gls{CGGM} with and without a Dirac gluino are similar to those of the \gls{CMSSM} as can be seen in figs. \ref{fig:CGGM-LSP-tb10-M7}, \ref{fig:CGGM-LSP-tb25-M7}, \ref{fig:CGGM-LSP-tb10-M12} and \ref{fig:CGGM-LSP-tb25-M12}. The difference here is that the blue regions that correspond to stau $\pstau_1$ \gls{LOSP} are now viable as the \gls{LSP} in these models is the gravitino $\tilde{G}$. The stau can either be long lives produce a missing energy signature or it can undergo the decay $\pstau \rightarrow \tilde G \; \Ptau$ inside the detector depending on its mass. If it does decay it will lead to one lepton and \gls{MET} or jets and \gls{MET}.  The remainder of the parameter space is has entirely bino-like neutralino $\pneuz$ \gls{LOSP}. This could appear a dark matter candidate on collider time-scales, but it may also undergo the decay $\pneuz \rightarrow \tilde G \; \Pphoton$. This decay is responsible for the stronger lower bounds on the neutralino mass $\m\pneuz$ in \gls{CGGM}.

\paragraph{NLOSP:} In \gls{CGGM} we have a sneutrino \gls{NLOSP} candiate in addition to those found in the \gls{CMSSM}. These are shown in figs. \ref{fig:CGGM-NLSP-tb10-M7}, \ref{fig:CGGM-NLSP-tb25-M7}, \ref{fig:CGGM-NLSP-tb10-M12} and \ref{fig:CGGM-NLSP-tb25-M12}. This only happens without a Dirac gluino however, as in the region where a sneutrino $\PSnu$ \gls{NLOSP} would be achieved, the lightest stau $\PStau_1$ has already been pushed tachyonic. The region with sneutrino $\PSnu$ \gls{NLOSP} is ruled out by collider searches. The remaining \gls{NLOSP} candidates have the same decays as seen in the \gls{CMSSM} except that they may be accompanied by an additional photon in the final state.

\subsection{Overview}
\label{sec:overview}

\begin{figure}[t]
\begin{center}
\includegraphics[scale=0.65]{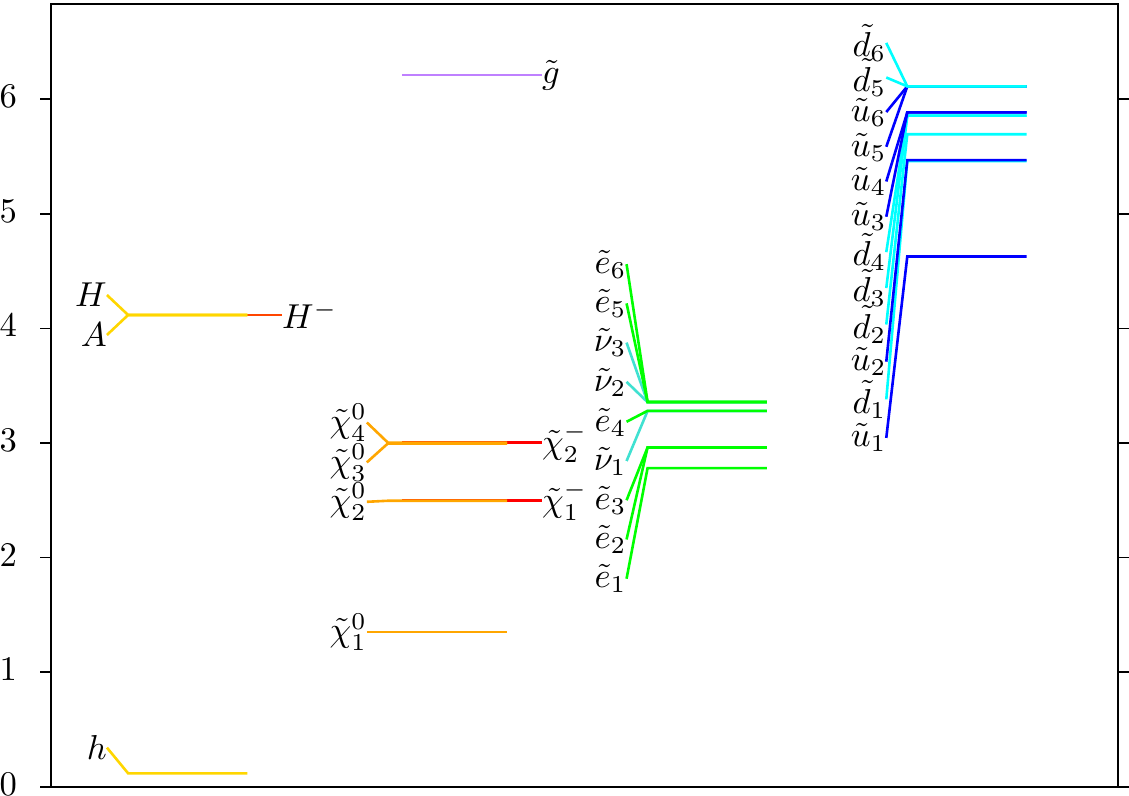}\;
\includegraphics[scale=0.65]{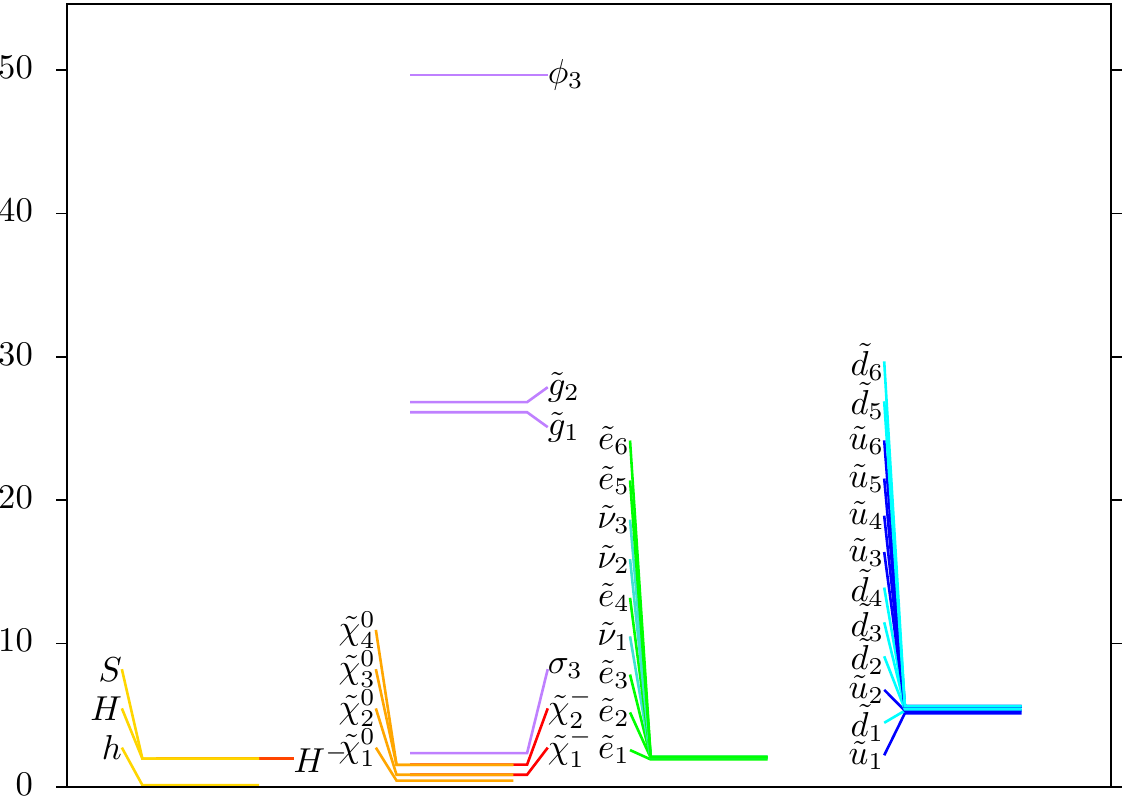}
\end{center}
\caption{Sparticle spectra the \gls{CMSSM} (left) and the \gls{CMSSM} with a Dirac gluino (right) for the benchmark points in table \ref{tab:RGE_benchmark}}
\label{fig:spectra}
\end{figure}

Overall, one sees that when each the \gls{CMSSM} and \gls{CGGM} are supplemented with a Dirac gluino, very little changes in the electroweak spectrum. This is of course by construction since the effective theory is essentially the \gls{MSSM} without a gluino. The Higgs mass however, is raised across the whole parameter space and can be made largely independent of $(m_0,M_{1/2})$ or $(\Lda_G,\Lda_S)$ at sufficiently low values of these parameters. Note that this is different to having non-universal scalar masses and gaugino masses, since giving a large mass to squarks and or gluinos in the \gls{UV} will lead to a very large value for $\mu$, giving very heavy Higgsinos and non-\gls{SM}-like Higgses as well as being accompanied by considerable fine tuning. The Wino mass will also be lifted along the \gls{RG} flow since 
\begin{equation}
(16\,\pi^2)^2\beta_{M_2}\supset 48\,(g_2\,g_3)^3\,M_3
\end{equation}
causes $M_2$ to increase by $\sim$ 500 GeV for a 10 TeV Majorana gluino. A characteristic plot of the spectra in the \gls{CMSSM} with and without a Dirac gluino is shown in fig. \ref{fig:spectra}. Since the overall result is a light set of electroweak particles with the neutralino as the \gls{LOSP}, the detailed phenomenology is expected to be very similar to that of the \emph{well-tempered neutralino} \cite{Arkani-Hamed2006,Brock2014}. One could also take all of the orderings of our electroweak states and map them on to the analysis in \cite{Berggren2013}.

\section{Cross sections}
\label{sec:XSecs}

\begin{figure}[t]
\begin{center}
\includegraphics[scale=0.55]{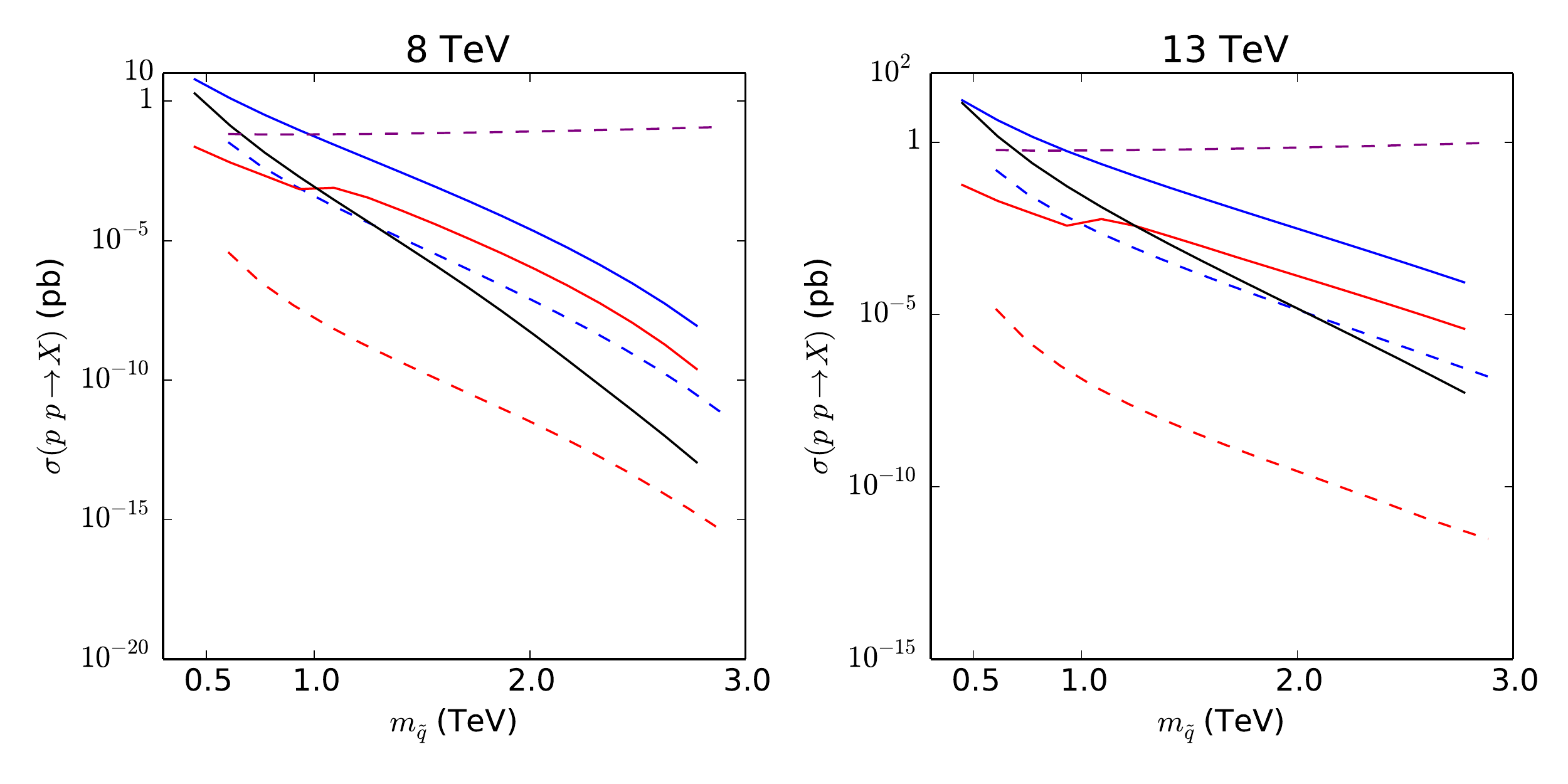}
\end{center}
\caption{\gls{LO} cross sections for various processes at 8 TeV (left) and 13 TeV (right) \gls{LHC}. The solid and dashed lines indicate cross sections in the \gls{CMSSM} with and without a Dirac gluino respectively. The blue, black and purple lines indicate total disquark $(\PSq_i\,\PSq_j)$, digluino $(\PSgluino_i\,\PSgluino_j)$ dipseudosgluon production $(\sigma_3\,\sigma_3)$. The red lines indicate cross section $\times$ branching ratio for processes beginning with disquark production and ending up with two jets, two same sign leptons and \gls{MET} in the final state. The cross sections were calculated using \MG using the MSTW2008lo68cl \gls{PDF} set. All two and three body branching ratios were calculated using \SPheno 3.3.2.}
\label{fig:XSecs}
\end{figure}

Here we present the \gls{LO} cross sections at 8 and 13 TeV \gls{LHC} with and without a Dirac gluino in the \gls{CMSSM}. We fixed $t_\be=10$, $m_0=200\,\textrm{GeV}$ and scanned over
\begin{align*}
M_{1/2}\in[200,1600]\,\textrm{GeV} & & \textrm{CMSSM}\\
M_{1/2}=400,\;m_{D3}\in[500,5000]\,\textrm{GeV} & & \textrm{CMSSM with Dirac gluino}
\end{align*}
leading to the spread of squark masses shown in fig. \ref{fig:XSecs}. For disquark production, we can see that there is suppression in the Dirac gluino case of approximately two orders of magnitude. Note that this is only true for disquark production, but is not true for squark-antisquark production as the dominant diagrams required for these processes do not involve the Majorana nature of gluinos as was discussed in \cite{Kribs2012,Kribs2013}.

Digluino production is only displayed for the \gls{CMSSM} without a Dirac gluino, since for the parameter space displayed, di-gluino production is kinematically forbidden in the Dirac gluino case. Similarly, disgluon production is kinematically forbidden. The dipseudosgluon production rate, however, is relatively high due to its light mass and its large $\SUC$ charge. Since this particle is the lightest strongly interacting sparticle and is \gls{CP}-odd, it has no particles to decay into \footnote{This is the $\phi_O=0$ case of \cite{Goodsell2014a} since we have a pure pseudoscalar in our setup and therefore no $\sigma\,\PSq\,\PSq^\dagger$ couplings that allow the loop decay to gluons.} in the \gls{CMSSM} this will be stable particle. The pseudosgluon could possibly form colour singlet bound states in an analogous way to the gluon-gluino bound states $R_0$ in \cite{Raby1997,Raby1999a}. It was also shown in \cite{Baer1999} that until the non-perturbative aspects of $\PSg\,\PSg$ in the early universe production are understood, is is not possible to constrain gluino \gls{LOSP} as a dark matter candidate. We leave this curious investigation for future study. In any case, the \gls{LOSP} will be the dominant dark matter component as many more of them are expected to be produced in the early universe. Direct searches for dark matter should not be an issue as $\si_3$ only interacts strongy, and the $\SUC$ charge of the nucleus is zero. It is also simple to allow both the sgluon and the pseudosgluon undergo the loop decay via squarks studied in \cite{Goodsell2014a} by taking $m_{D3}$ complex.

Finally we display the product of branching ratios approximation for the cross section for two jets, two same sign leptons and missing energy
\begin{align*}
2\times\Big[
\sum_{i\leq j}\sigma(\Pp\,\Pp\rightarrow \PSup_i\,\PSup_j)&\times 
\textrm{Br}(\PSup_i\rightarrow \textrm{jet}+\Pl^++\slashed{\textrm{E}})\times
\textrm{Br}(\PSup_j\rightarrow \textrm{jet}+\Pl^++\slashed{\textrm{E}})+\\
\sum_{i\leq j}\sigma(\Pp\,\Pp\rightarrow \PSdown_i\,\PSdown_j)&\times 
\textrm{Br}(\PSdown_i\rightarrow \textrm{jet}+\Pl^-+\slashed{\textrm{E}})\times
\textrm{Br}(\PSdown_j\rightarrow \textrm{jet}+\Pl^-+\slashed{\textrm{E}})
\Big],
\end{align*}
where the squark branching ratios are given by all possible combinations of kinematically allowed decays leading to one jet, one lepton and missing energy
\begin{equation}
\textrm{Br}(\PSup_i\rightarrow \textrm{jet}+\Pl^++\slashed{\textrm{E}})
\sim\textrm{Br}(\PSup_i\rightarrow \Pdown \; \PSino^+_1)
\times\textrm{Br}(\PSino^+_1 \rightarrow \Pl^+\,\nu\,\PSino^0_1)+\cdots.
\end{equation}
Although this approximation misses effects coming from off-shell intermediate sparticles in the decay chain that increase the cross section $\times$ branching ratio, it can still serve as an indicator of what to expect if one simulated the high multiplicity final states fully. All branching ratios are calculated as a function of the parameter space scanned by \SPheno. All other branching ratios are \gls{SM} branching ratios which can are given in table \ref{tab:branching_ratios}. All decay products in the chain considered are displayed in table \ref{tab:decay_products}. Whilst the Majorana case still allows a number of events visible at the \gls{LHC} given an integrated luminosity of $23.26\,\textrm{fb}^{-1}$ such that the same sign lepton analyses \cite{Collaboration2013a} are sensitive in the direct squark (via sleptons) models, the case with a Dirac gluino is far beyond producing any same sign dileptons plus two jet events at the \gls{LHC} with the current integrated luminosity. In addition, the Majorana digluino production is the dominant process leading to two same-sign dipletons with $\PSg\rightarrow 2 \,\textrm{jets}+l^\pm+\slashed{E}$. This decay is simply absent with a heavy Dirac gluino.

\begin{table}[t]
\begin{center}
	\begin{tabular}{c|c|c}
	 Decaying particle   & Decay products                                & Branching fraction \\ \hline
	 $\PZ$      & invisible & $0.2000 \pm 0.0006$ \\ \hline
	 $\PWp$     & $\APelectron\,\Pnue$ & $0.1075 \pm 0.0013$ \\
	            & $\APmuon\,\Pnum$ & $0.1057 \pm 0.0015$ \\
	            & $\APtauon\,\Pnut$ & $0.1125 \pm 0.0020$ \\ \hline
	 $\APtauon$ & $\APnut\,\APelectron\,\Pnue$ & $0.1783 \pm 0.0004$ \\
	            & $\APnut\,\APmuon\,\Pnum$ & $0.1741 \pm 0.0004$ \\ \hline
	 $\Ptop^-$    & $\PWp\,\Pbottom$ & $0.91 \pm 0.04$   
	\end{tabular}
\end{center}
\caption{\gls{SM} branching ratios used in calculation of branching ratios $\times$ cross sections. All are the world averages taken from \cite{Beringer2012}.}
\label{tab:branching_ratios}
\end{table}

\begin{table}[t]
\begin{center}
	\begin{tabular}{c|c}
	 Particle   & Relevant Decay Products \\ \hline
	 $\PSup_{1,\ldots,6}$ & $\Pdown_{1,2,3}\,\PSino^+_{1,2}$ \\
	 $\PSdown_{1,\ldots,6}$ & $\Pup_{1,2}\,\PSino^-_{1,2}$ \\
	 $\PSino^+_{2}$ & $\Pe^+_{1,\ldots,3}\,\PSnu_{1,2,3}$; $\Pnu_{1,2,3}\,\PSe^+_{1,\ldots,6}$; $\PWp\,\PSino^0_{1,2}$; $\PZ\,\PSino^+_{1}$ \\
	 $\PSino^+_{1}$ & $\Pe^+_{1,\ldots,3}\,\PSnu_{1,2,3}$; $\Pnu_{1,2,3}\,\PSe^+_{1,\ldots,6}$; $\PWp\,\PSino^0_{1,2}$ \\
	 $\PSino^0_{2}$ & $\PZ\,\PSino^0_{1}$ \\
$\PSe^-_{1,\ldots,6}$ & $\Pe^-_{1,\ldots,3}\,\PSino^0_{1,2}$; $\Pnu_{1,\ldots,3}\,\PSino^+_{1}$\\
$\PSnu_{1,2,3}$ & $\PSino^+_{1,2}\,\Pe^-_{1,2,3}$; $\PSino^0_{1,2}\,\Pnu_{1,2,3}$%\\
%$\PZ$ & $\Pnu_{1,2,3}\,\APnu_{1,2,3}$ \\
%$\PWp$ & $\Pe^+_{1,\ldots,3}\,\Pnu_{1,3}$ \\
%$\APtauon$ & $\APnut\,\Pe^+_{1,2}\,\Pnu_{1,2}$ \\
%$\Ptop^-$ & \PWp\,\Pbottom
	\end{tabular}
	\end{center}
\caption{Decays considered for the squark to one jet, one lepton and \gls{MET}.}
\label{tab:decay_products}
\end{table}

One feature to note is that in the \gls{MSSM}, there is a rise in the branching ratio $\times$ cross section for 1 TeV squarks in both the 8 and 13 TeV cases. This doesn't occur in with a Dirac gluino in the parameter space studied. In the \gls{MSSM}, we are raising $M_{1/2}$ in order to raise the squark masses. As this happens, a gap between the lightest chargino and the sneutrino masses opens up. The chains that involving $\PSino_1^+\rightarrow \PSnu\,\Pl^+$ account for 10 \% of the overall branching ratio of a squark into one lepton, one jet and \gls{MET} and only turn on once $M_{1/2}$ becomes large enough. In the Dirac gluino case this channel never opens up as we raise $m_{D3}$ to raise the squark masses instead of $M_{1/2}$.

\section{Electroweak symmetry breaking and fine tuning}

\begin{figure}[t]
\begin{center}
\includegraphics[scale=0.55]{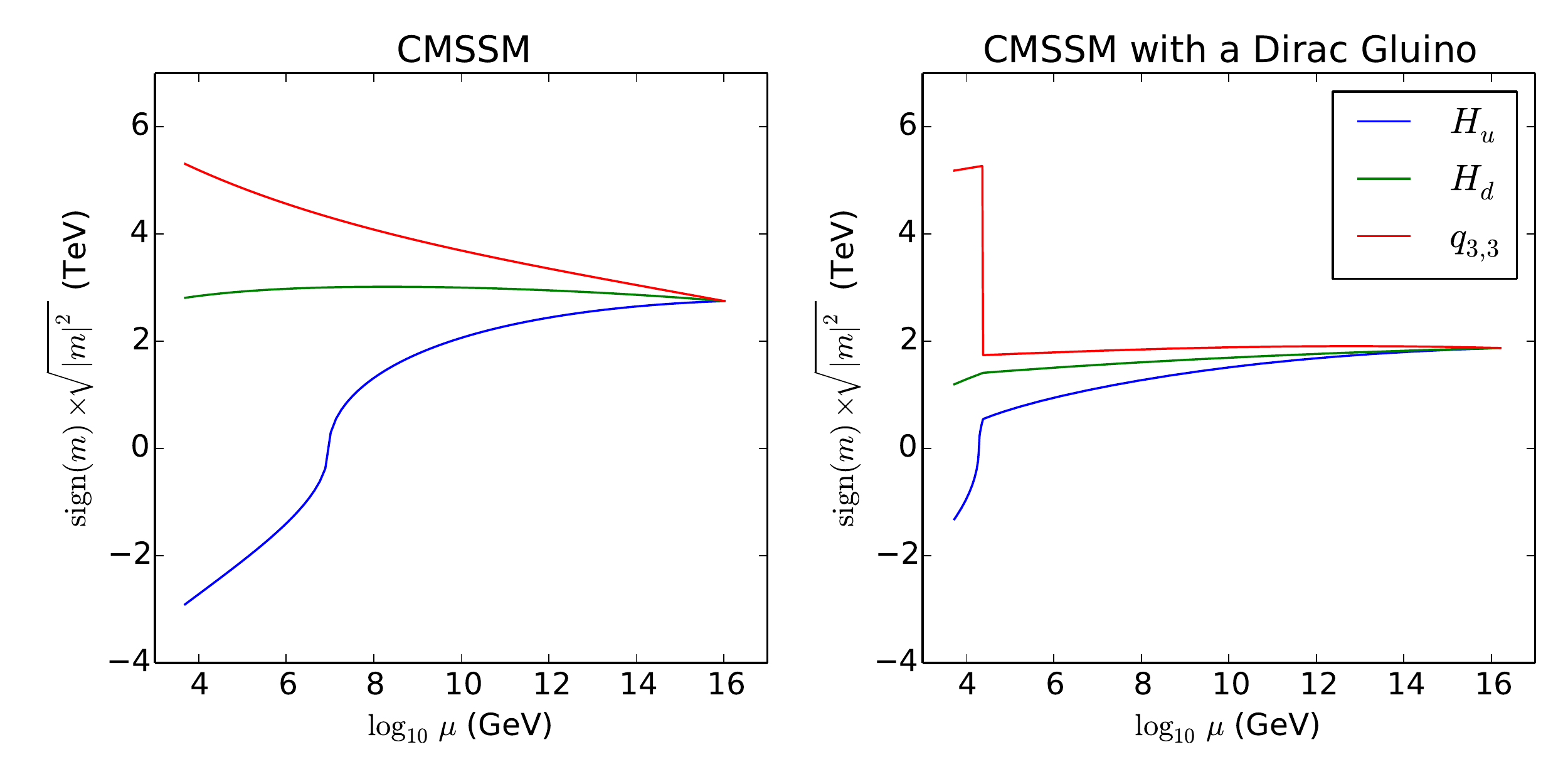}
\end{center}
\caption{\gls{RGE} of $\mm\pHu2$ (blue), $\mm\pHd2$ (green) and $\mm{\psq_{3,3}}2$ (red) from the \gls{GUT} scale to the \gls{SUSY} scale in the \gls{CMSSM} (left) and the \gls{CMSSM} with a Dirac gluino (right) for the benchmark points given in table \ref{tab:RGE_benchmark}
}
\label{fig:RGEs}
\end{figure}
\begin{savenotes}
\begin{table}[t]
\begin{center}
	\begin{tabular}{c|ccccc}
	 Model	         & $m_0$ (TeV) & $M_{1/2}$ (TeV) & $m_{D3}$ (TeV) & $m_{\ph}^{(1)}$ (GeV) & $m_{\ph}^{(2)}$ (GeV) \\ \hline
	 \gls{CMSSM}     & 2.750       & 3.000               & N/A            & 118.1                 & 127.4                 \\
	 \gls{CMSSM} + DG & 1.875       & 1.000               & 10.00          & 117.3                 & unknown
	\end{tabular}
\end{center}
\caption{Benchmark points for the \gls{RGE} of parameters in the CMSSM with and without a Dirac gluino shown in figure \ref{fig:RGEs}.}
\label{tab:RGE_benchmark}
\end{table}
\end{savenotes}

As has already been indicated, \gls{EWSB} in a model with a Dirac gluino is triggered much closer to the electroweak scale. As is well understood in most \gls{SUSY} models, it is the stop mass (and at two loops a Majorana gluino mass) that causes this to happen. The same is true with a Dirac gluino. The difference here is that the stop mass can be negligible along the whole \gls{RG} flow until the Dirac gluino mass is reached. The supersoft contribution from integrating out the gluino is applied to the squark masses, and they drive $\mm\pHu2$ negative for the remainder of the flow through its \gls{RG} equation given in eq. \ref{eq:betamHu2}. This effect is demonstrated in fig. \ref{fig:RGEs}. The upshot is that for a particularly large final squark mass, there is some control over how large $\mm\pHu2$ (and consequently $|\mu|^2$) is. In the \gls{LL} approximation at one loop we find
\begin{equation}
\mm\pHu2(m_\textrm{SUSY})
=
\mm\pHu2(m_\textrm{GUT})-\be^{(1)}_{\mm\pHu2}\times\log\left(\frac{m_{\textrm{GUT}}}{m_\textrm{SUSY}}\right)
\approx
m_0^2\left[1-\frac{3\,|y_{\Ptop}|^2}{4\,\pi^2}\right]\times\log\left(\frac{m_{\textrm{GUT}}}{m_0}\right)
\end{equation}
in the \gls{CMSSM} and 
\begin{equation}
\mm\pHu2(m_\textrm{SUSY})\approx m_0^2-(m_0^2+\mm\pq2)\times\frac{3\,|y_{\Ptop}|^2}{4\,\pi^2} \times\log\left(\frac{m_{D3}}{m_0+\m\pq}\right)
\label{eq:approxmHu2SUSY}
\end{equation}
in the \gls{CMSSM} with a Dirac gluino where $\mm\pq2$ is given by eq. \ref{eq:Squark_mass_1-loop}. Since $\mm\pHd2$ is so linked to the electroweak \gls{UV} sensitivity, it is reasonable to expect that Dirac gluinos have the ability to reduce the amount of fine tuning in the presence of larger squark masses.

\begin{figure}[t]
\begin{center}
\includegraphics[scale=0.55]{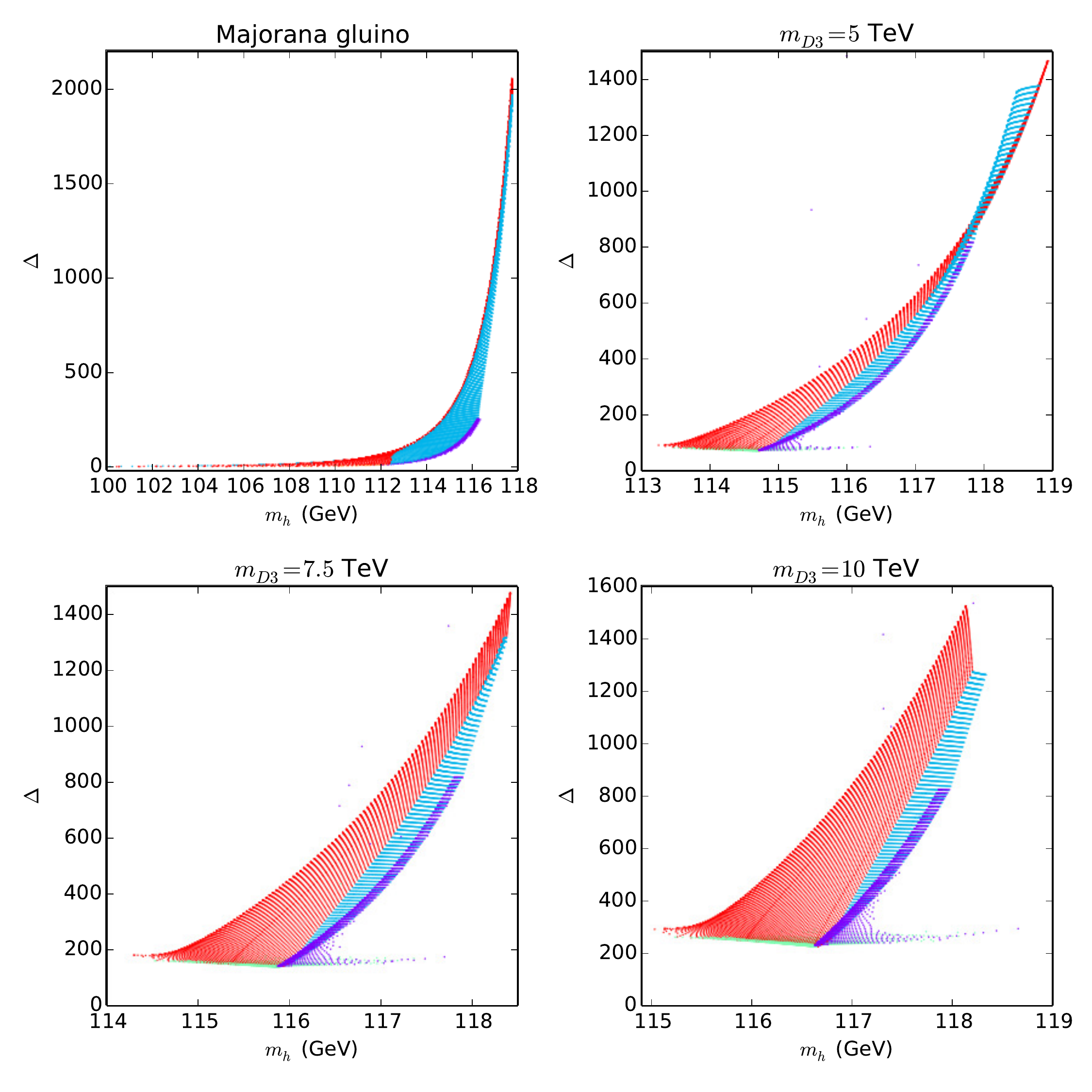}
\end{center}
\caption{Fine tuning in the \gls{CMSSM} with $t_\be = 10$ and $m_{D3}$ fixed as indicated. The red, purple, blue, and green regions correspond to $\mu$, $m_0$, $M_{1/2}$ and $m_{D3}$ as the dominant source of tuning.}
\label{fig:CMSSM-FT-tb10}
\end{figure}

To quantify the impact this difference in triggering \gls{EWSB} has on fine tuning, we take the measure $\Delta$ from \cite{Barbieri1988}
\begin{align}
\Delta&\equiv\,\textrm{max}\left[\textrm{Abs}(\Delta_\OO)\right],&
\Delta_\OO&\equiv \frac{\partial\,\log\,v^2}{\partial\,\log\,\OO}
\end{align}
such that $\Delta^{-1}$ gives a measure of how tuned the parameters $\OO$ need to be tuned to achieve the observed \gls{EWSB} scale $v$. This was calculated at the \gls{SUSY} using the routines generated by \SARAH modified to include the thresholds discussed in section \ref{sec:thresholds} where appropriate. Since we are interested in \gls{UV} sensitivity, we take the $\OO$s as the set of parameters that would be fixed by the \gls{UV} model at either the \gls{GUT} scale in \gls{CMSSM} or the messenger scale in \gls{CGGM}. These are
\begin{equation}
\OO|_\textrm{CMSSM}\in\{m_0,M_{1/2},\mu,B_\mu,m_{D3}\},\quad
\OO|_\textrm{CGGM}\in\{\Lda_G,\Lda_S,m_\textrm{Mess},\mu,B_\mu,m_{D3}\}.
\end{equation}

\begin{figure}[t]
\begin{center}
\includegraphics[scale=0.55]{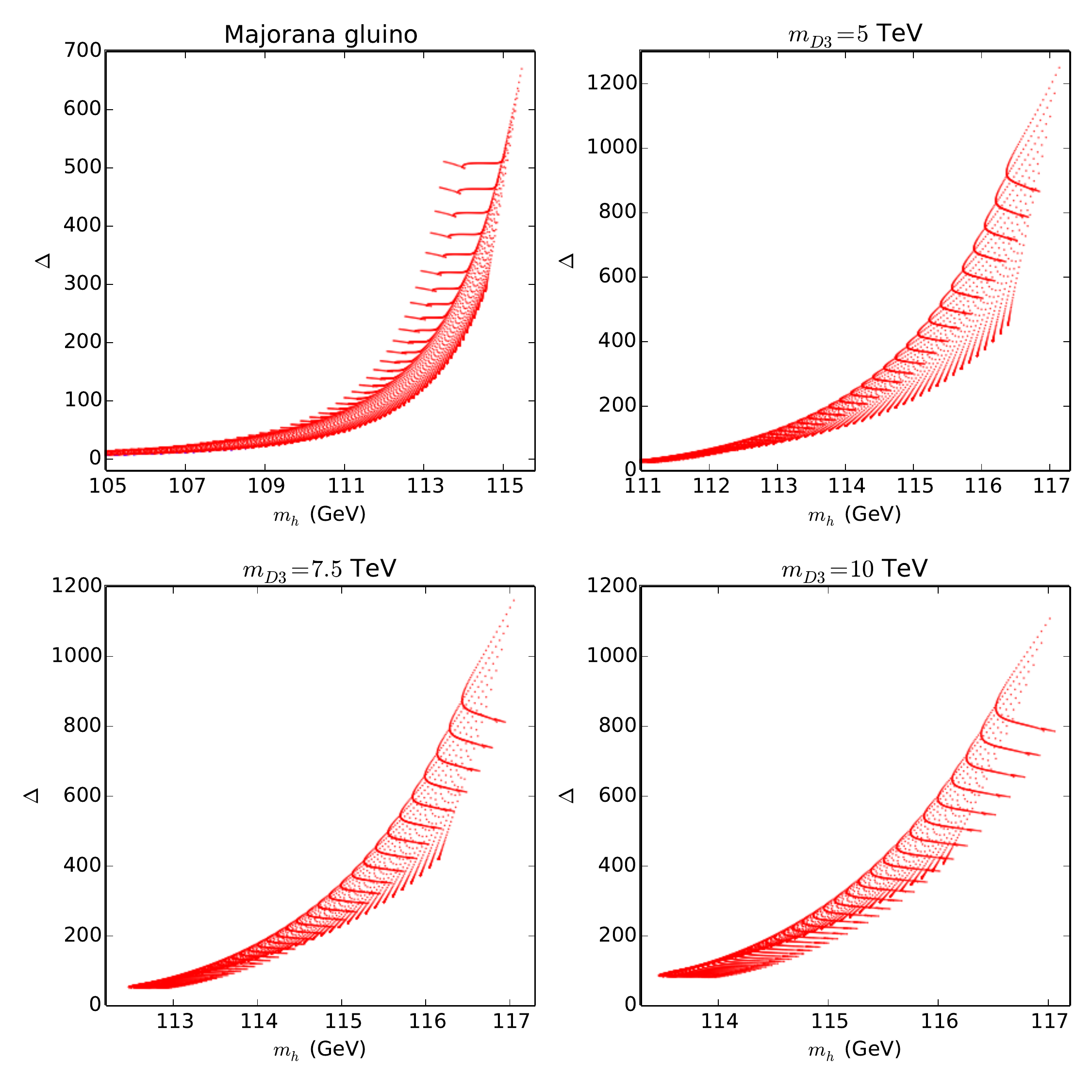}
\end{center}
\caption{Fine tuning in the \gls{CMSSM} with $t_\be = 10$, $m_\textrm{Mess} = 10^{7}$ GeV and $m_{D3}$ fixed as indicated. The dominant source of tuning is entirely from the $\mu$ parameter.}
\label{fig:CGGM-FT-tb10-M7}
\end{figure}

The tuning in the \gls{CMSSM} for the parameter space investigated in section \ref{sec:spectra} is shown in figs. \ref{fig:CMSSM-FT-tb10} and \ref{fig:CMSSM-FT-tb25}, and the tuning in \gls{CGGM} is shown in figs. \ref{fig:CGGM-FT-tb10-M7}, \ref{fig:CGGM-FT-tb25-M7}, \ref{fig:CGGM-FT-tb10-M12} and \ref{fig:CGGM-FT-tb25-M12}.

In the \gls{CMSSM} and in \gls{CGGM} it is observed that, for a given Higgs mass, new points exist with a reduction in fine tuning of typically up to a factor of two or three. In the \gls{CMSSM} also a line of points opening up with moderately large Higgs mass mass but low $(\Delta \sim 200)$ fine tuning. These points occur where the two terms in eq. \ref{eq:approxmHu2SUSY} approximately cancel, giving low --- $\OO (0.5 - 1\,\textrm{TeV})$ --- values of $\m\pHu$ and $\mu$. The strip is very thin, since an increase in either $m_0$ or $M_{1/2}$ makes the right hand side become more positive in eq. \ref{eq:approxmHu2SUSY}, leaving no \gls{EWSB} and decreasing $m_0$ or $M_{1/2}$ leads to a reduction in the Higgs mass. Unfortunately since these points are at very low values of $M_{1/2}$ that give rise to neutralino and chargino masses that are excluded by \gls{LEP}. 

The reduction in tuning in \gls{CGGM} is less drastic than that seen in the \gls{CMSSM}. This is because the mechanism reduces tuning through making logarithms smaller. Where in the \gls{CMSSM} we have the log reduced $\log(m_{\textrm{GUT}}/m_{\textrm{SUSY}}) \rightarrow \log(m_{D3}/m_{\textrm{SUSY}})$, whereas in \gls{CGGM} this is only the factor $\log(m_{\textrm{Mess}}/m_{\textrm{SUSY}}) \rightarrow \log(m_{D3}/m_{\textrm{SUSY}})$. Similarly, the reduction in fine tuning in \gls{CGGM} is less drastic in the case of the lower messenger scale than the higher messenger scale. In the \gls{CMSSM} one can see the full range of \gls{UV} parameters becoming the dominant source of tuning whereas in \gls{CGGM} it is mainly the $\mu$ parameter across the entire space. However, in both the \gls{CMSSM} and \gls{CGGM}, all the underlying \gls{UV} parameters considered do have associated tunings across the respective parameter spaces.

\section{Conclusions}

We have constructed a set of simple \gls{UV} models with the supersoft mechanism introduced in \cite{Fox2002} by extending the \gls{MSSM} field content by only what was required to give the gluino a Dirac mass. We then performed the first implementation of the supersoft mechanism into a state of the art spectrum generator and carried out an analysis of the spectra, the production rates at \gls{LHC}8 and \gls{LHC}13, and fine tuning.

In the presence of a Dirac gluino, we find that it is possible to essentially decouple the strong sparticles without affecting the electroweak spectrum except that one finds that the pseudosgluon usually remains light and may even be a novel dark matter candidate by forming neutral bound states with other strongly interacting particles. 

The decoupling of the strongly interacting sparticles from the electroweak sparticles has been shown to give a handle on the production cross sections at the \gls{LHC}. Using a product of branching ratios approximation, we have shown that the Dirac gluino completely removes the same sign dilepton sign dilepton as a visible signature in current \gls{LHC} data. A full simulation of the decay chain needs to be done to confirm this and it should also include the usually subdominant purely electroweak contributions to these events as these may now be important. It would also be interesting to investigate how many charginos and neutralinos are still produced in these cases with t-channel squarks. 

Taking account the spectra and cross section suppression, we find that the final states of these models at the \gls{LHC} are therefore altered in the following way:
\begin{itemize}
\item The number of events involving the Majorana gluino propagator are suppressed by roughly two orders of magnitude. This includes the same sign dilepton events.
\item Events involving the pair production of gluinos are absent.
\item The mass hierarchy between the strong and electroweak sectors causes hard jets in a \gls{SUSY} cascade to be harder than usual.
\item \gls{LOSP} candidates are typical, yielding a number of leptons and missing energy in the final stages of a cascade. In the case of \gls{CGGM} this may also include the emission of a photon.
\item The number of events with jets and missing energy will increase in the case of a stable pseudosgluon.
\end{itemize}
Unfortunately there are no smoking gun signatures for these models. Their main distinguishing characteristic is that there are different numbers of each type of visible event compared to models without a Dirac gluino --- generally fewer. Note that for models of this type, a new lepton collider such as the \gls{ILC} or \gls{CLIC} would be able to simply bypass the strong sparticle sector and directly probe the much lighter accessible electroweak states.

Finally, the allowed tuning in these models is found to reduced. In allowed regions of parameter space, the reduction for a given Higgs mass is generally by a factor of two or three, although one has to keep in consideration that a reduction in fine tuning is being achieved whilst the gluino mass is being taken up to ten times of greater that which is usually considered for precisely reasons of tuning.

There are two obvious extensions of this study:
\begin{itemize}
\item The accuracy of the Higgs mass calculation needs improving in order to say something more concrete and more tightly constrain the model. In order to achieve this, the full set of general broken \gls{SUSY} two loop \glspl{RGE} should be used below the Dirac gluino mass, allowing a two loop accurate Higgs mass prediction. This should be possible with the general two loop \gls{RGE} calculators on the market \cite{Lyonnet2013,Staub2013}. Since these calculations are in \MSbar scheme, one would need to take care to convert to the \DRbar scheme before implementing them into a \gls{SUSY} spectrum generator \cite{Martin1993a}.
\item In our study of the \gls{CMSSM}, we kept the $A$ terms zero for simplicity. As was noted in \cite{Bhattacharyya2013}, the presence of additional scalar octets allows $g_3$ to remain much larger over the \gls{RG} flow, and can consequently generate large negative $A$ terms in the \gls{IR} providing one starts with a negative $A$ term. This model has the potential to reduce tuning much further by allowing a reduction in the squark masses and at the same time the length of flowing between the Dirac gluino mass and the \gls{SUSY} scale.
\end{itemize}
\section{Acknowledgements}
I would like to thank Alberto Mariotti for useful discussions, Florian Lyonnet for guidance with \pyrate and Florian Staub for help with \SARAH. I would also like to thank Valya Khoze for many fruitful discussions, helpful comments and support during this project.
DB is supported by a U.K. Science and Technology Facilities Council (STFC) studentship.

\newpage 

\appendix

%\section{Scalar integrals for one loop threshold corrections}
%\label{App:scalar_integrals}

%Here we present the scalar integrals involved in the evalutation of one loop %threshold corrections due to the gluino and sgluons. They are calculated the in %$\overline{\text{DR}}$ scheme and regularised in $4-2\,\ep$ dimensions with %renormalisation scale $\mu$ \cite{Pierce1996,Ellis2008}.

%\begin{align}
%A_0(m)&=
%m^2\left[\frac1\ephat+1-\log\left(\frac{m^2}{\mu^2}\right)\right]\label{eq:basis1}\\
%B_0(p,m_1,m_2)&=
%\frac1\ephat-\log\left(\frac{p^2}{\mu^2}\right)-f_B(x_+)-f_B(x_-)
%\label{eq:basis2}
%\end{align}
%where
%\begin{equation}
%x_\pm\equiv\frac{s\pm\sqrt{s^2-4\,p^2(m_1^2-i\,\epsilon)}}{2p^2},\qquad f_B(x)\equiv \log(1-x)-x\,\log(1-x^{-1})-1
%\end{equation}
%with $\frac1\ephat\equiv\frac1\ep-\gamma_E+\log(4\pi)$ and $s=p^2+m_1^2-m_2^2$ and the remaining scalar integral required expressible as combinations of eqs. \ref{eq:basis1} and \ref{eq:basis2}
%\begin{equation}
%	G_0(p,m_1,m_2)
%	=(p^2-m_1^2-m_2^2)\,B_0(p,m_1,m_2)-A_0(m_1)-A_0(m_2).
%\end{equation}

\section{Additional plots}
\label{app:additional_plots}

\subsection{Constrained Minimal Supersymmetric Standard Model}

\vspace{2.2cm}
\begin{figure}[h]
\begin{center}
\includegraphics[scale=0.65]{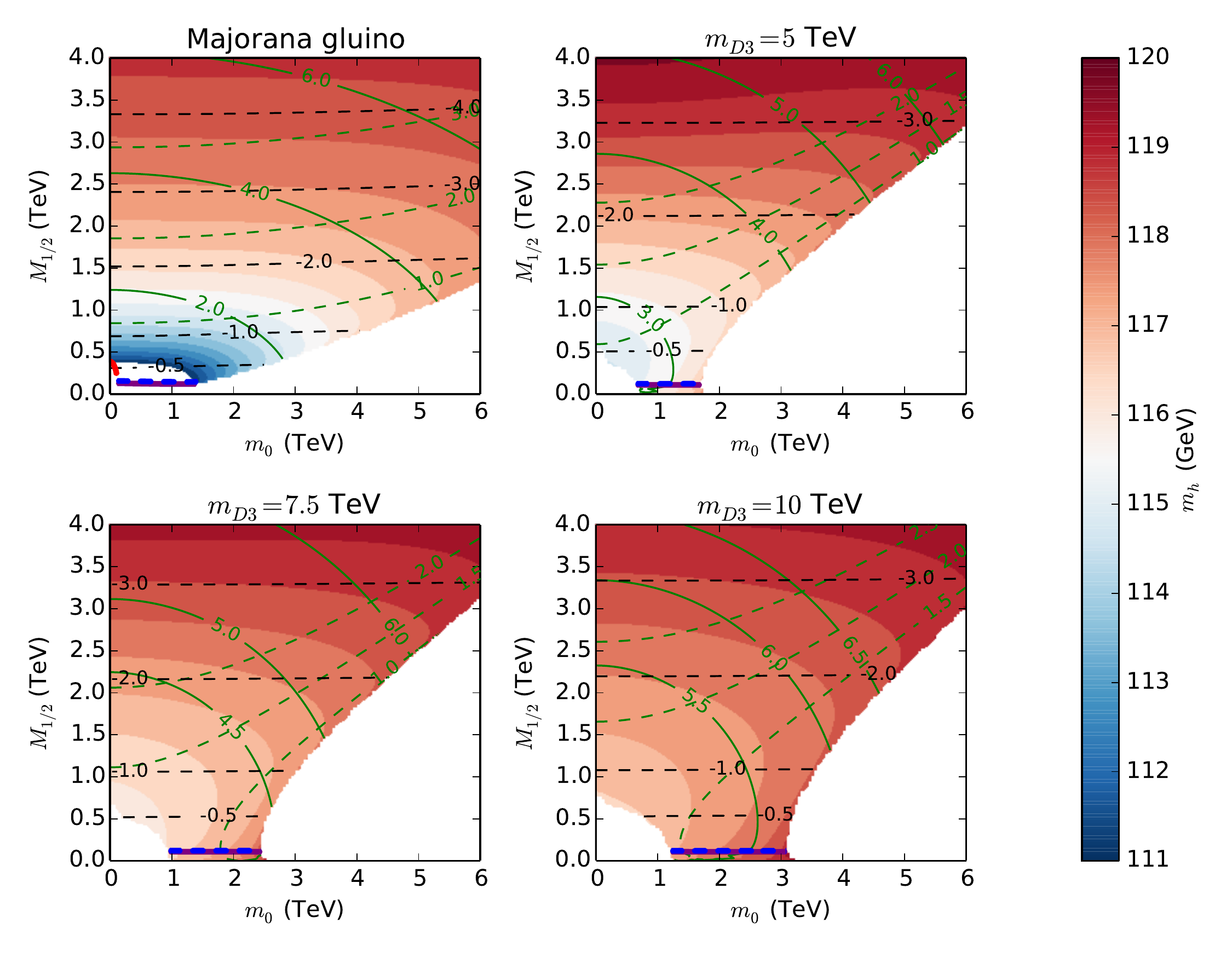}
\end{center}
\caption{Higgs sector parameters in the \gls{CMSSM} with $t_\be = 25$ and $m_{D3}$ fixed as indicated. The gradient indicates the Higgs mass. The black dashed, green dashed and green solid lines are contours of $a_{\pt}(m_\textrm{SUSY})$, $\mu(m_\textrm{SUSY})$, and $m_\textrm{SUSY}$ respectively. All contours unless otherwise specified are in TeV.}
\label{fig:HiggsCMSSMtb25}
\end{figure}

\begin{figure}[t]
\begin{center}
\includegraphics[scale=0.65]{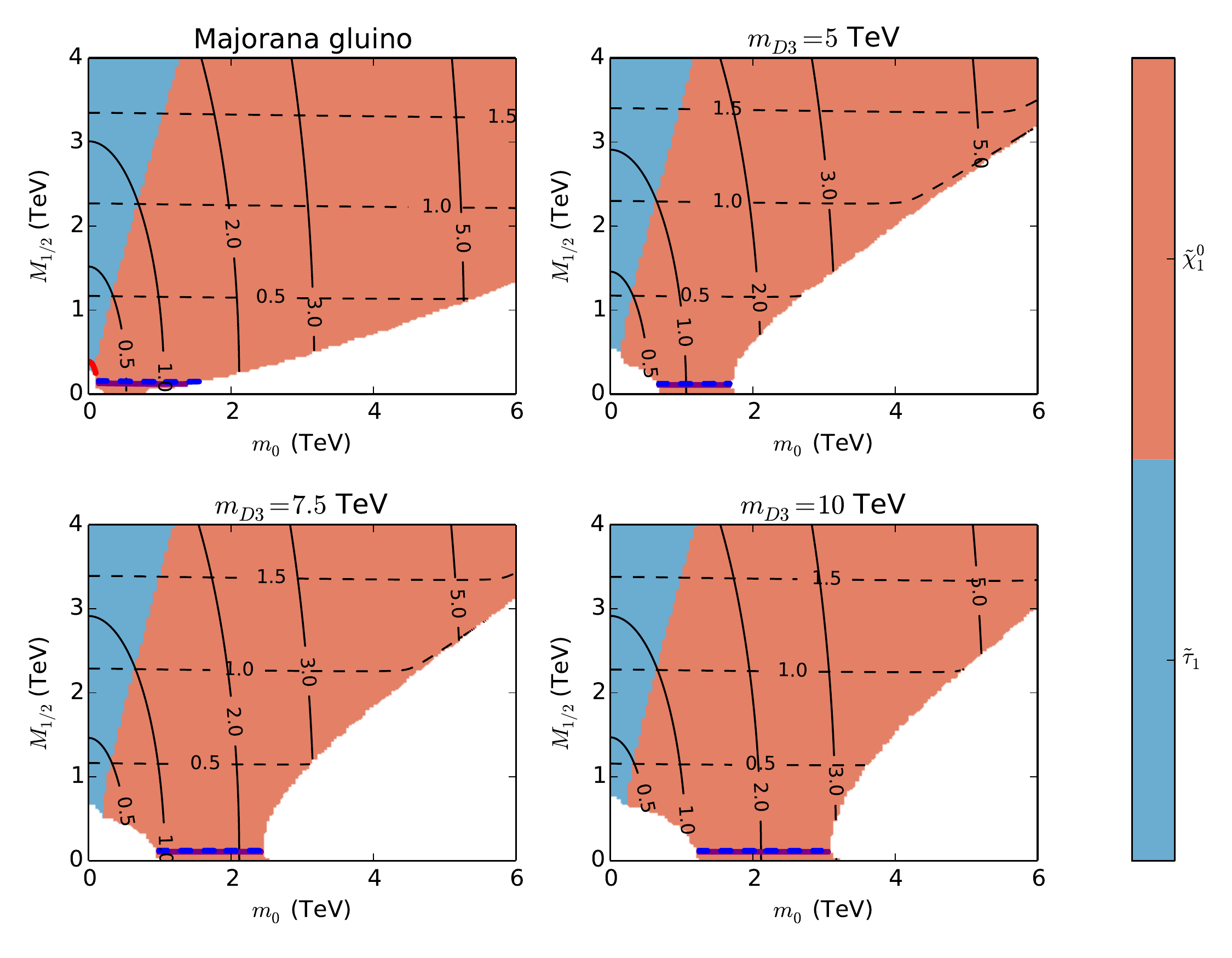}
\end{center}
\caption{\gls{LOSP} species in the \gls{CMSSM} with $t_\be = 25$ and $m_{D3}$ fixed as indicated. The black dashed and black solid lines are contours of lightest neutralino mass $\m\pneuz$ and stau mass $\m\pstau$ in TeV.}
\label{fig:CMSSM-LSP-tb25}
\end{figure}

\begin{figure}[t]
\begin{center}
\includegraphics[scale=0.65]{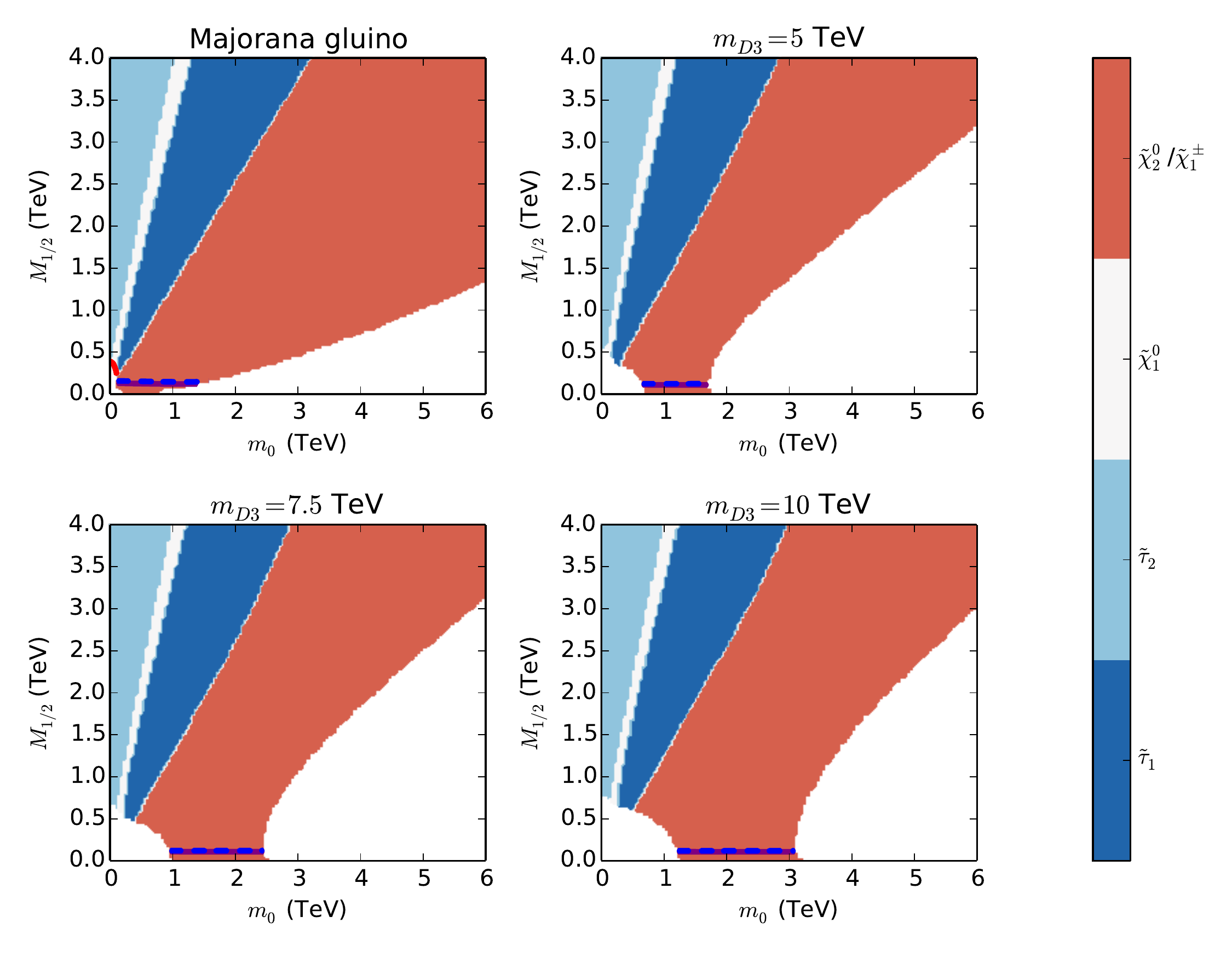}
\end{center}
\caption{\gls{NLOSP} species in the \gls{CMSSM} with $t_\be = 25$ and $m_{D3}$ fixed as indicated}
\label{fig:CMSSM-NLSP-tb25}
\end{figure}

\begin{figure}[t]
\begin{center}
\includegraphics[scale=0.55]{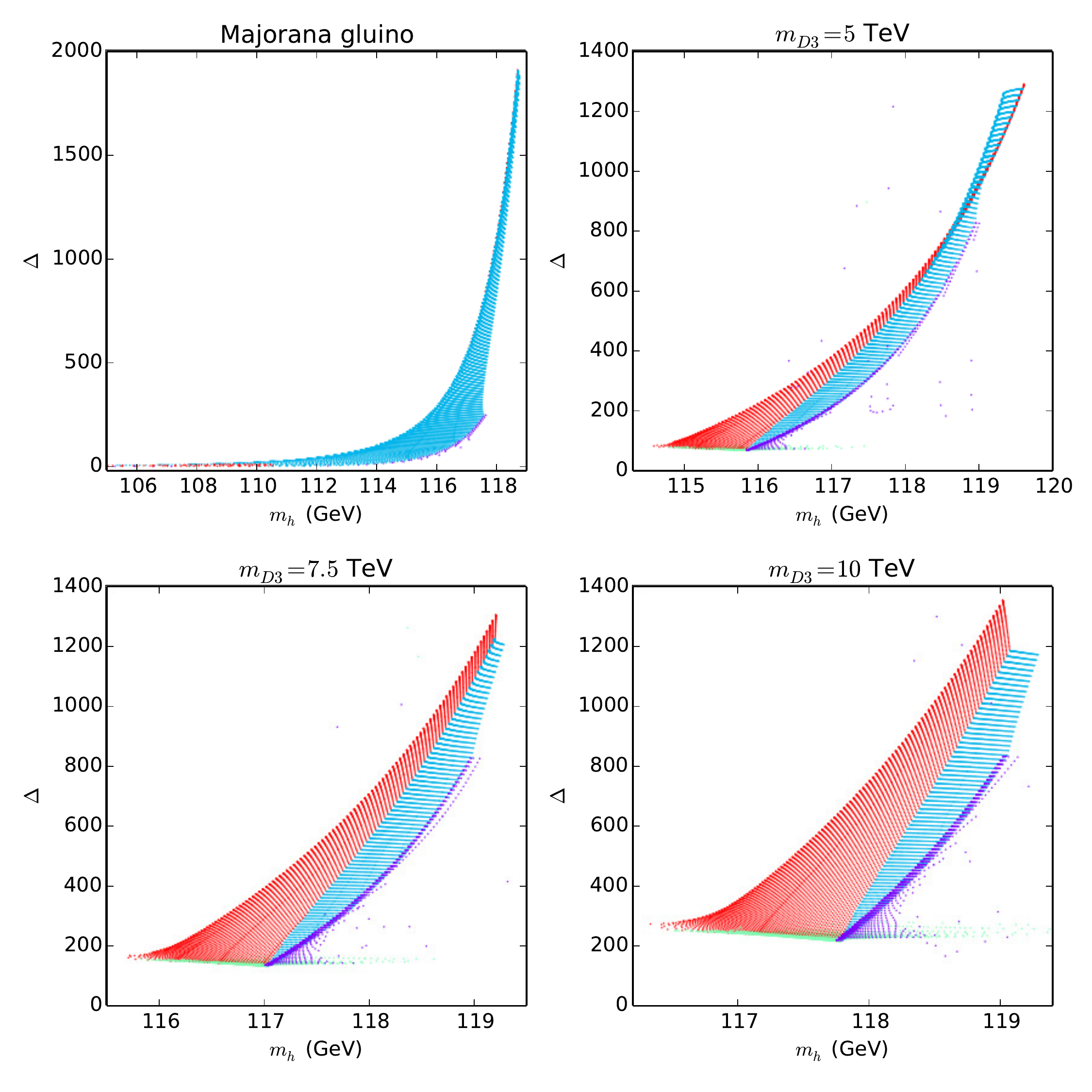}
\end{center}
\caption{Fine tuning in the \gls{CMSSM} with $t_\be = 25$ and $m_{D3}$ fixed as indicated. The red, purple, blue, and green regions correspond to $\mu$, $m_0$, $M_{1/2}$ and $m_{D3}$ as the dominant source of tuning.}
\label{fig:CMSSM-FT-tb25}
\end{figure}

\clearpage

\subsection{Constrained General Gauge Mediation}
\vspace{3.1cm}
\begin{figure}[h]
\begin{center}
\includegraphics[scale=0.65]{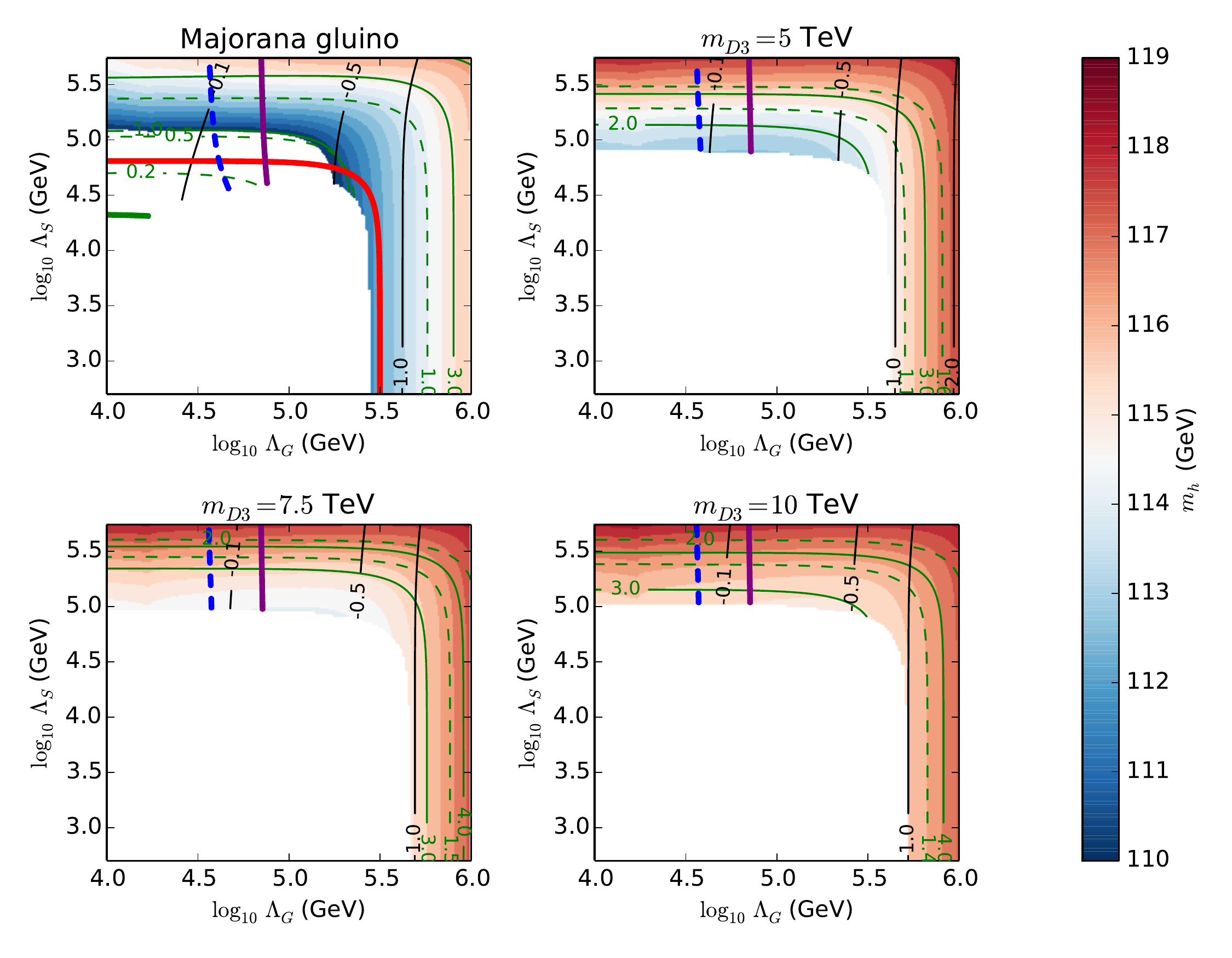}
\end{center}
\caption{Higgs sector parameters in \gls{CGGM} with $t_\be = 25$, $m_\textrm{Mess}=10^7$ GeV and $m_{D3}$ fixed as indicated. The gradient indicates the Higgs mass. The black dashed, green dashed and green solid lines are contours of $a_{\pt}(m_\textrm{SUSY})$, $\mu(m_\textrm{SUSY})$, and $m_\textrm{SUSY}$ respectively. All contours unless otherwise specified are in TeV.}
\label{fig:CGGM-Higgs-tb25-M7}
\end{figure}

\begin{figure}[t]
\begin{center}
\includegraphics[scale=0.65]{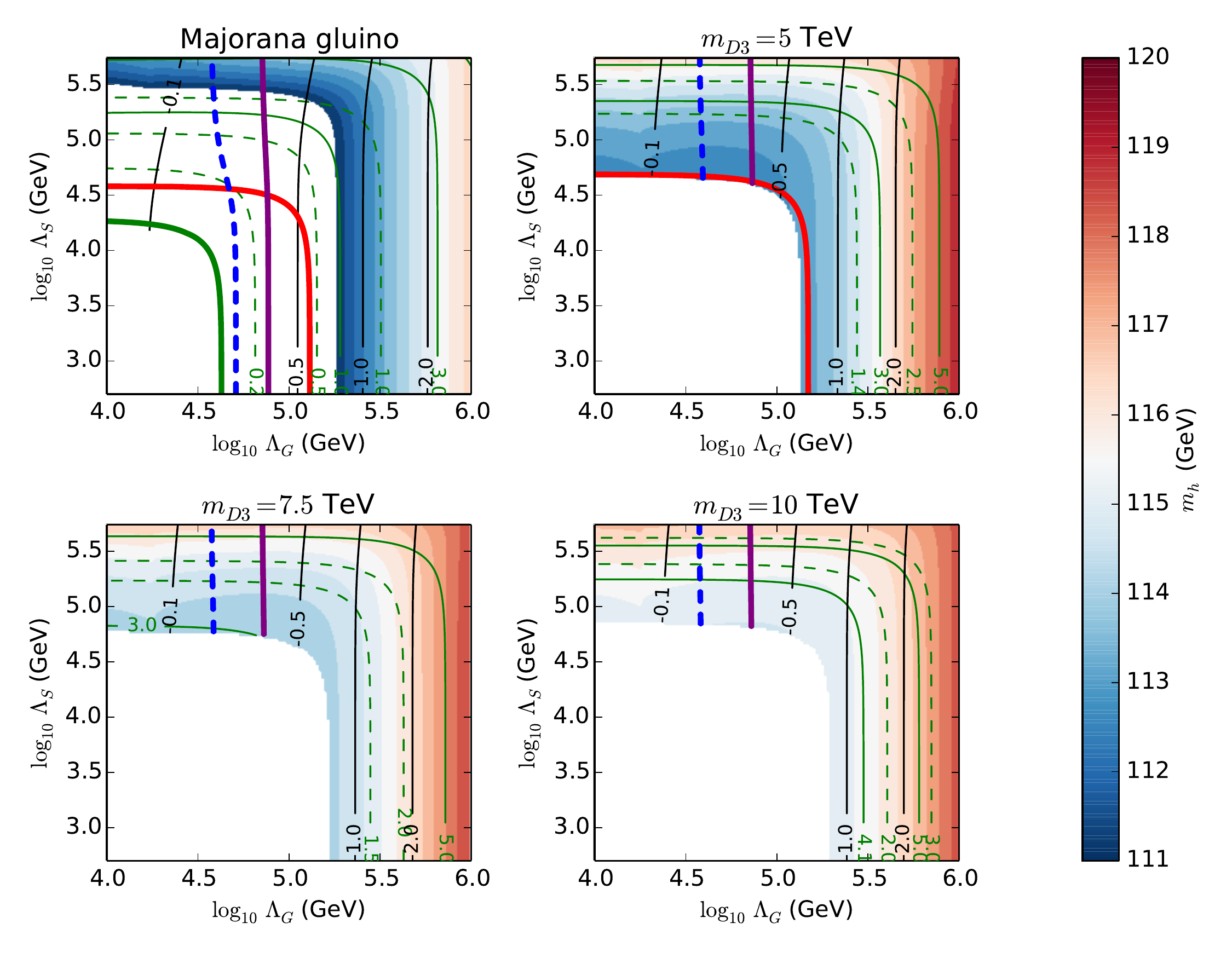}
\end{center}
\caption{Higgs sector parameters in \gls{CGGM} with $t_\be = 10$, $m_\textrm{Mess}=10^{12}$ GeV and $m_{D3}$ fixed as indicated. The gradient indicates the Higgs mass. The black dashed, green dashed and green solid lines are contours of $a_{\pt}(m_\textrm{SUSY})$, $\mu(m_\textrm{SUSY})$, and $m_\textrm{SUSY}$ respectively. All contours unless otherwise specified are in TeV.}
\label{fig:CGGM-Higgs-tb10-M12}
\end{figure}

\begin{figure}[t]
\begin{center}
\includegraphics[scale=0.65]{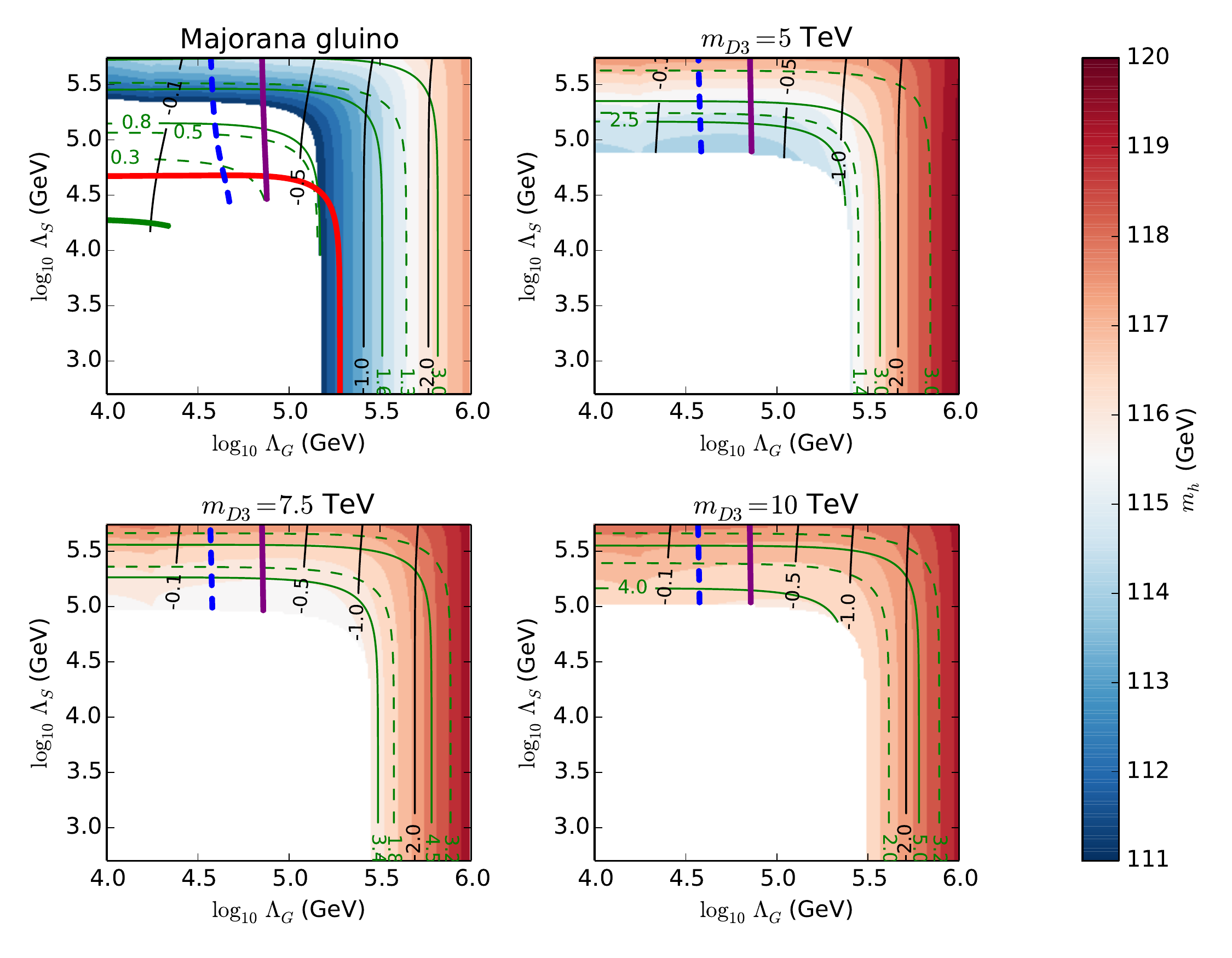}
\end{center}
\caption{Higgs sector parameters in \gls{CGGM} with $t_\be = 25$, $m_\textrm{Mess}=10^{12}$ GeV and $m_{D3}$ fixed as indicated. The gradient indicates the Higgs mass. The black dashed, green dashed and green solid lines are contours of $a_{\pt}(m_\textrm{SUSY})$, $\mu(m_\textrm{SUSY})$, and $m_\textrm{SUSY}$ respectively. All contours unless otherwise specified are in TeV.}
\label{fig:CGGM-Higgs-tb25-M12}
\end{figure}

\begin{figure}[t]
\begin{center}
\includegraphics[scale=0.65]{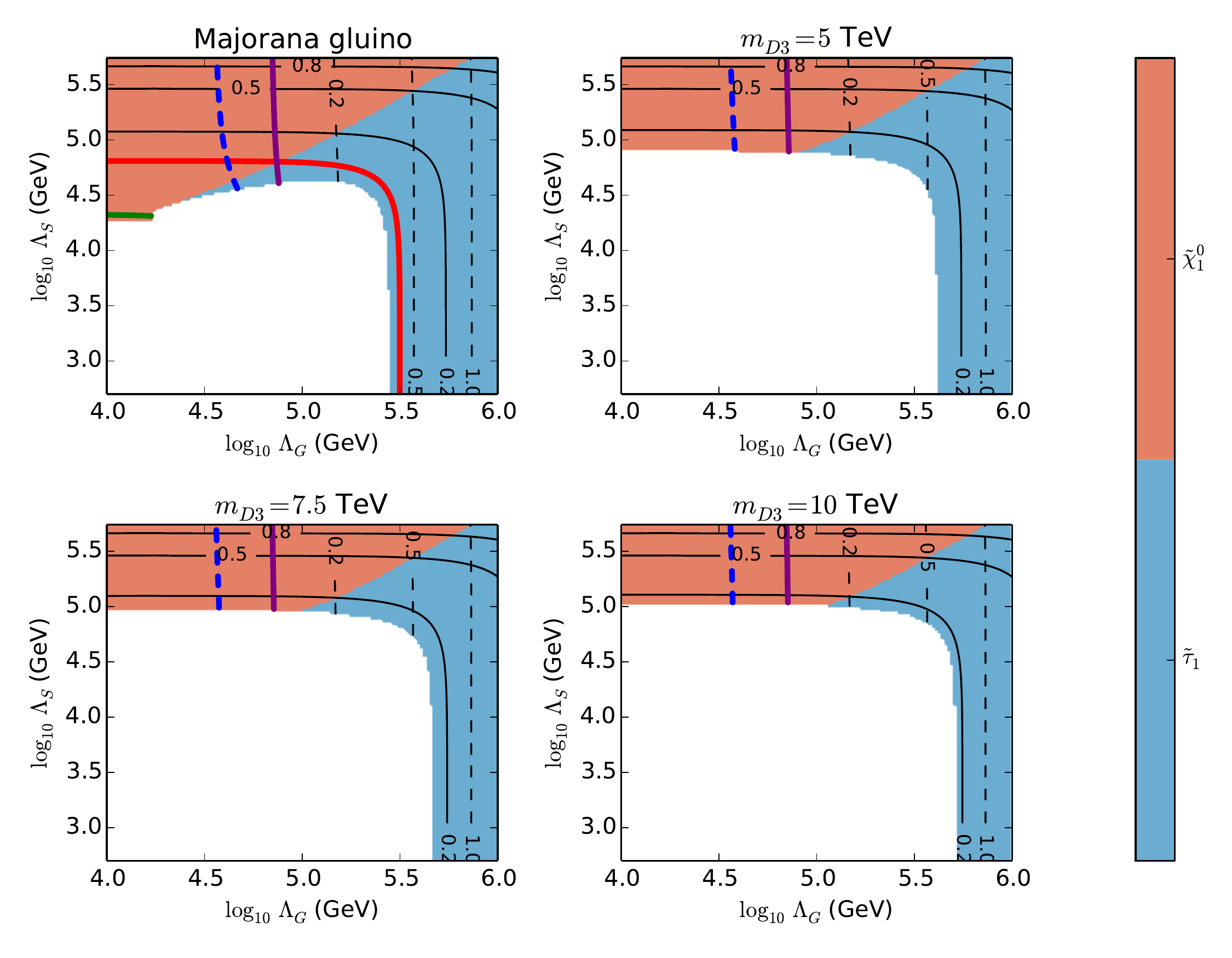}
\end{center}
\caption{\gls{LOSP} species in \gls{CGGM} with $t_\be = 25$, $m_\textrm{Mess} = 10^7$ GeV and $m_{D3}$ fixed as indicated. The black dashed and black solid lines are contours of lightest neutralino mass $\m\pneuz$ and stau mass $\m\pstau$ in TeV.}
\label{fig:CGGM-LSP-tb25-M7}
\end{figure}

\begin{figure}[t]
\begin{center}
\includegraphics[scale=0.65]{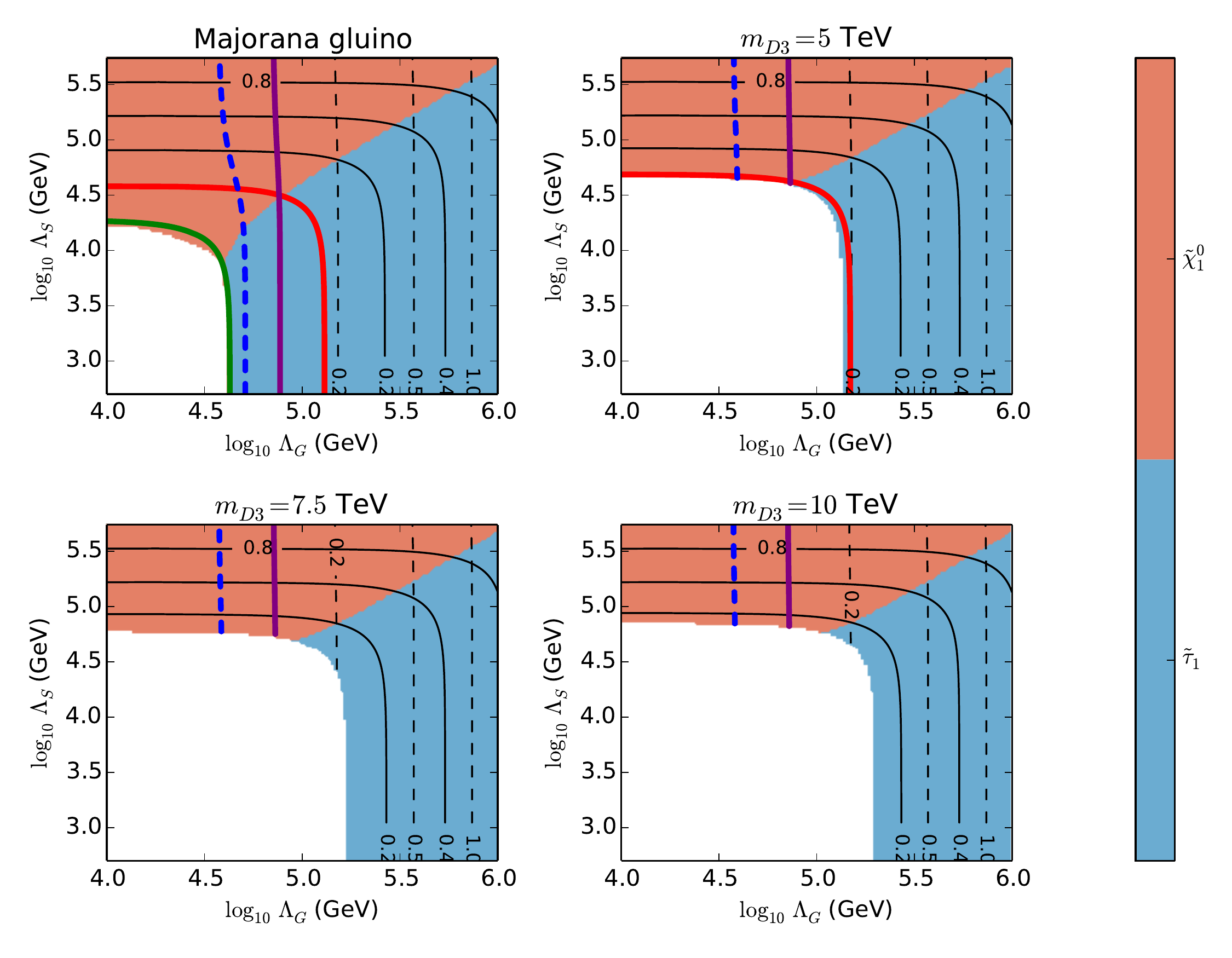}
\end{center}
\caption{\gls{LOSP} species in \gls{CGGM} with $t_\be = 10$, $m_\textrm{Mess} = 10^{12}$ GeV and $m_{D3}$ fixed as indicated. The black dashed and black solid lines are contours of lightest neutralino mass $\m\pneuz$ and stau mass $\m\pstau$ in TeV.}
\label{fig:CGGM-LSP-tb10-M12}
\end{figure}

\begin{figure}[t]
\begin{center}
\includegraphics[scale=0.65]{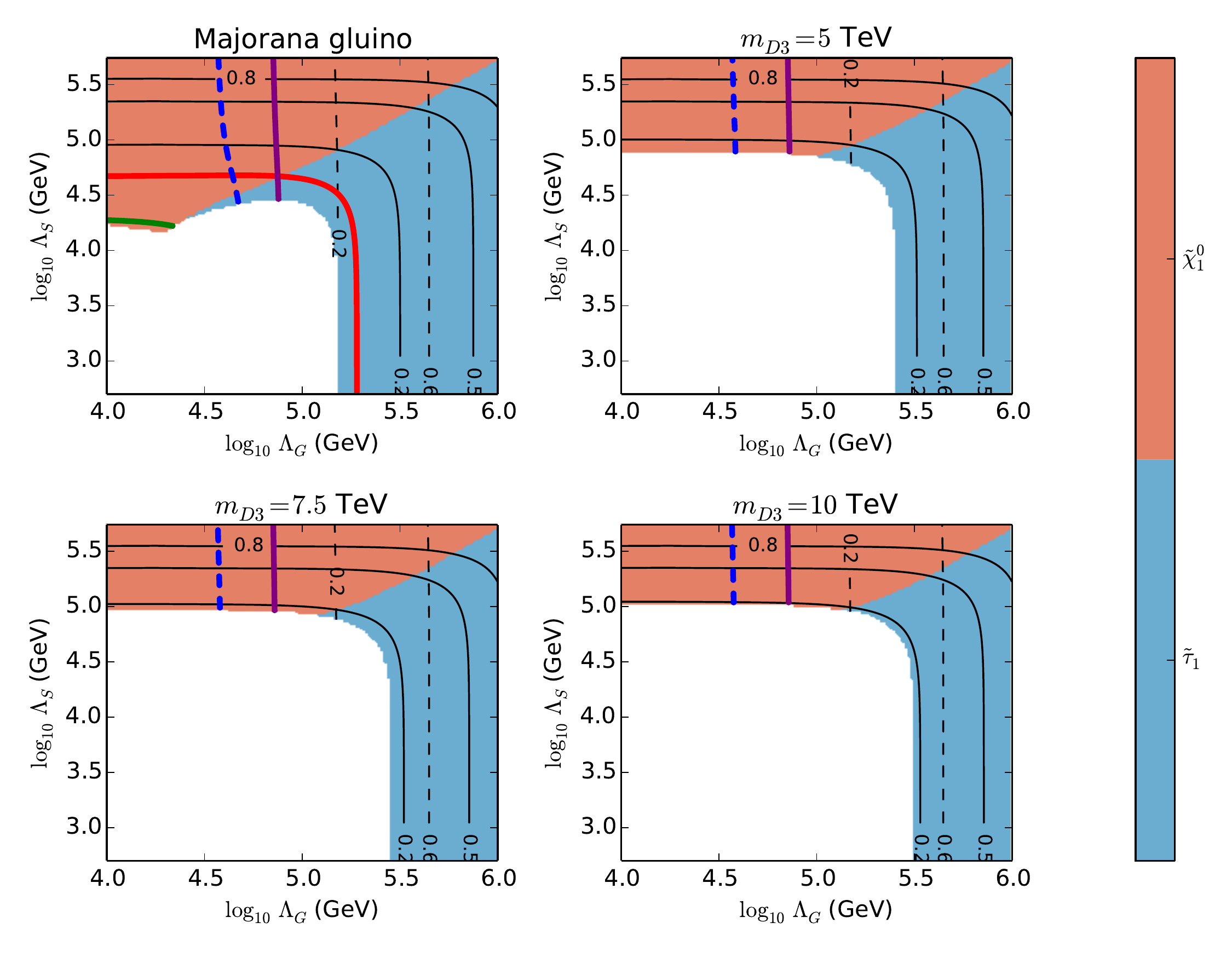}
\end{center}
\caption{\gls{LOSP} species in \gls{CGGM} with $t_\be = 25$, $m_\textrm{Mess} = 10^{12}$ GeV and $m_{D3}$ fixed as indicated. The black dashed and black solid lines are contours of lightest neutralino mass $\m\pneuz$ and stau mass $\m\pstau$ in TeV.}
\label{fig:CGGM-LSP-tb25-M12}
\end{figure}

\begin{figure}[t]
\begin{center}
\includegraphics[scale=0.65]{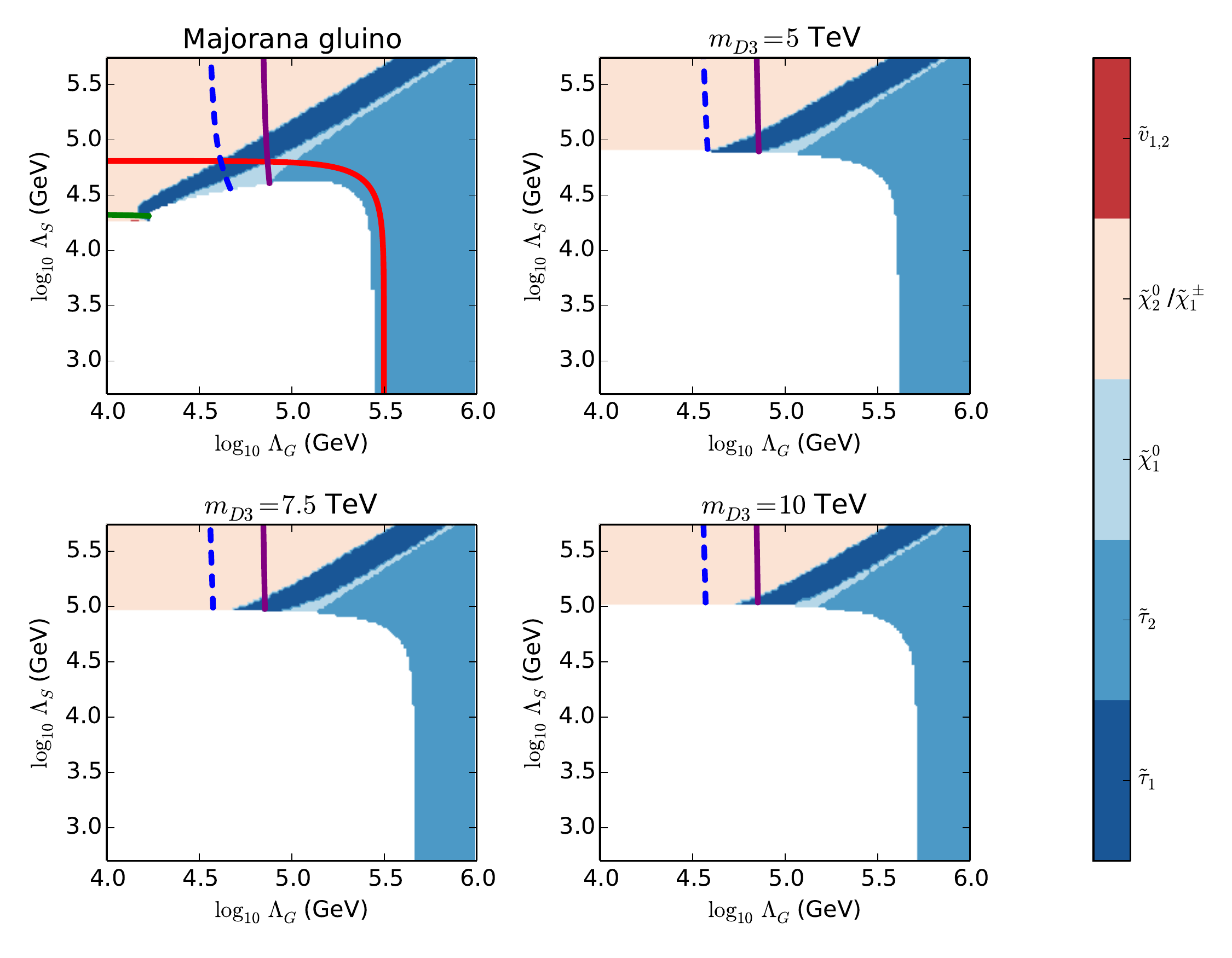}
\end{center}
\caption{\gls{NLOSP} species in \gls{CGGM} with $t_\be = 25$, $m_\textrm{Mess} = 10^7$ GeV and $m_{D3}$ fixed as indicated. The black dashed and black solid lines are contours of lightest neutralino mass $\m\pneuz$ and stau mass $\m\pstau$ in TeV.}
\label{fig:CGGM-NLSP-tb25-M7}
\end{figure}

\begin{figure}[t]
\begin{center}
\includegraphics[scale=0.65]{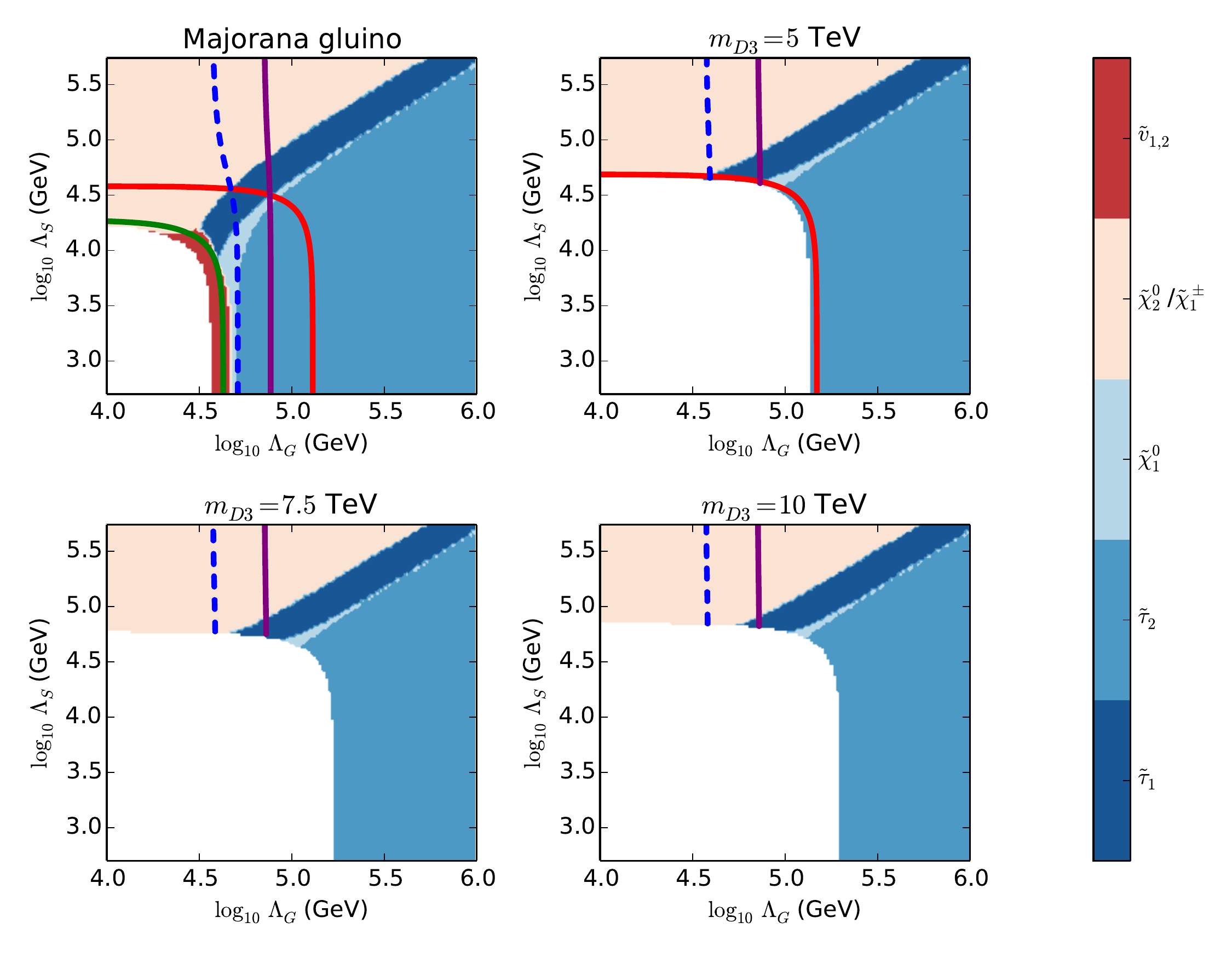}
\end{center}
\caption{\gls{NLOSP} species in \gls{CGGM} with $t_\be = 10$, $m_\textrm{Mess} = 10^{12}$ GeV and $m_{D3}$ fixed as indicated. The black dashed and black solid lines are contours of lightest neutralino mass $\m\pneuz$ and stau mass $\m\pstau$ in TeV.}
\label{fig:CGGM-NLSP-tb10-M12}
\end{figure}

\begin{figure}[t]
\begin{center}
\includegraphics[scale=0.65]{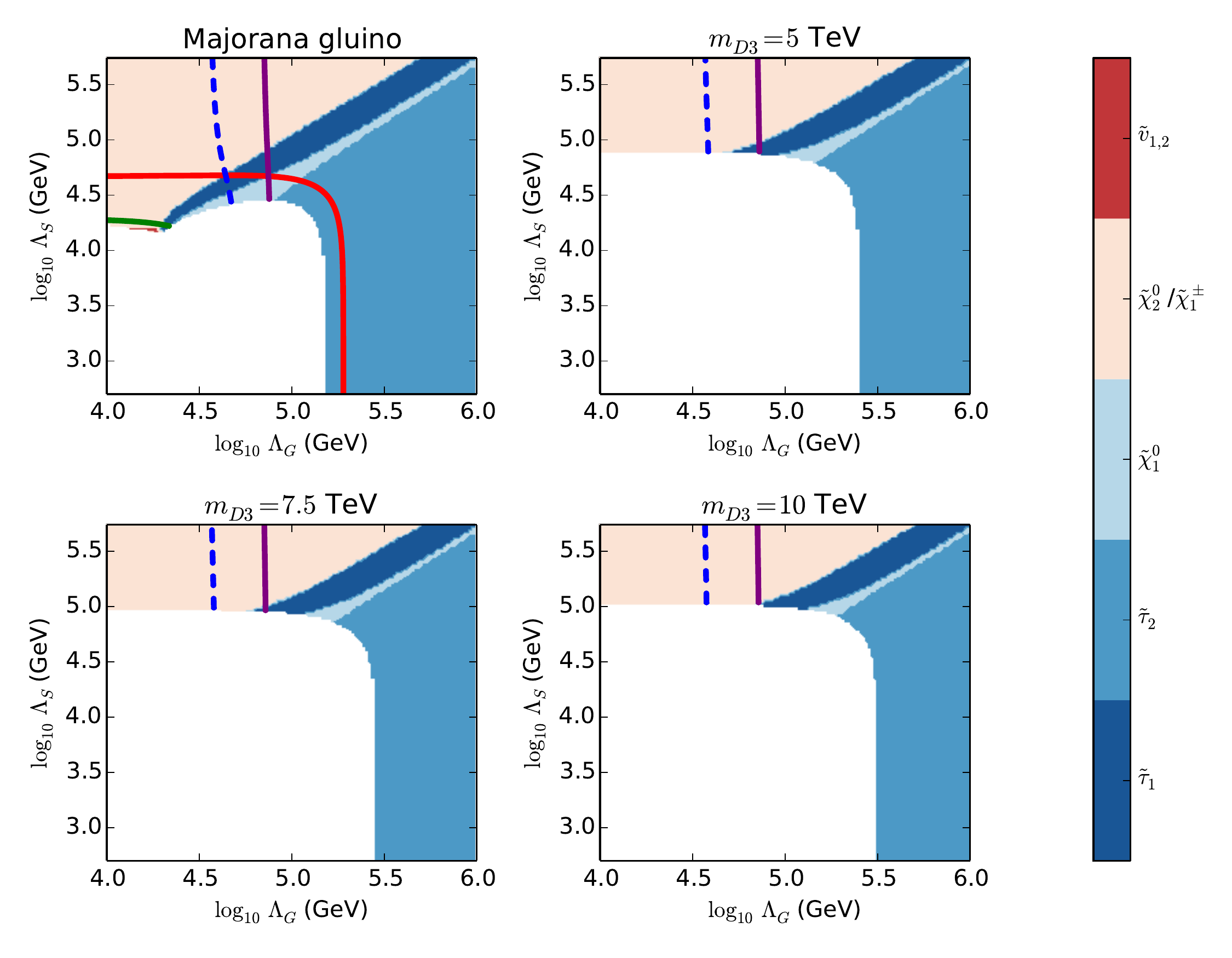}
\end{center}
\caption{\gls{NLOSP} species in \gls{CGGM} with $t_\be = 25$, $m_\textrm{Mess} = 10^{12}$ GeV and $m_{D3}$ fixed as indicated. The black dashed and black solid lines are contours of lightest neutralino mass $\m\pneuz$ and stau mass $\m\pstau$ in TeV.}
\label{fig:CGGM-NLSP-tb25-M12}
\end{figure}

\begin{figure}[t]
\begin{center}
\includegraphics[scale=0.55]{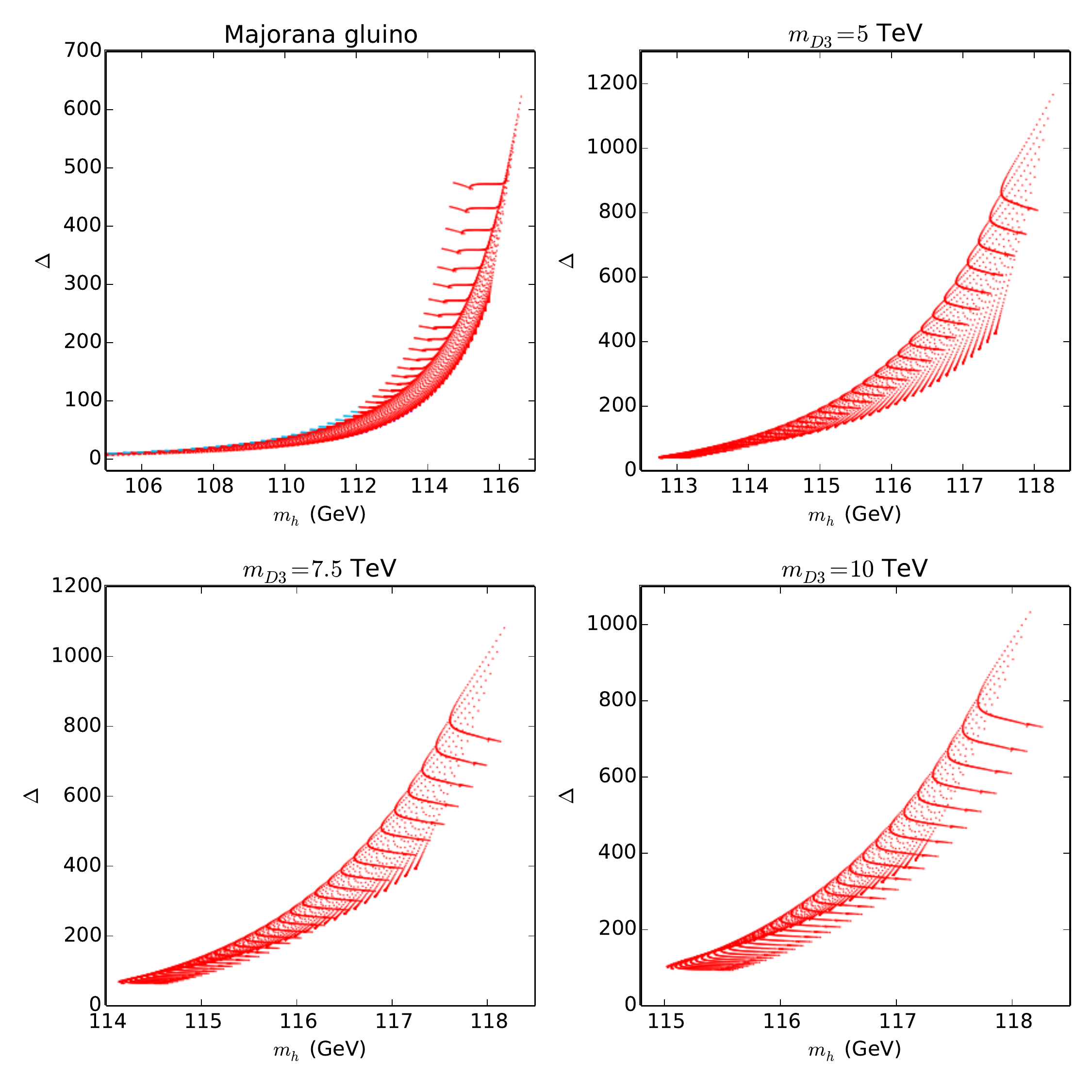}
\end{center}
\caption{Fine tuning in \gls{CGGM} with $t_\be = 25$, $m_\textrm{Mess} = 10^{7}$ GeV and $m_{D3}$ fixed as indicated. The red and blue regions correspond to $\mu$ and $\Lda_S$ as the dominant source of tuning.}
\label{fig:CGGM-FT-tb25-M7}
\end{figure}

\begin{figure}[t]
\begin{center}
\includegraphics[scale=0.55]{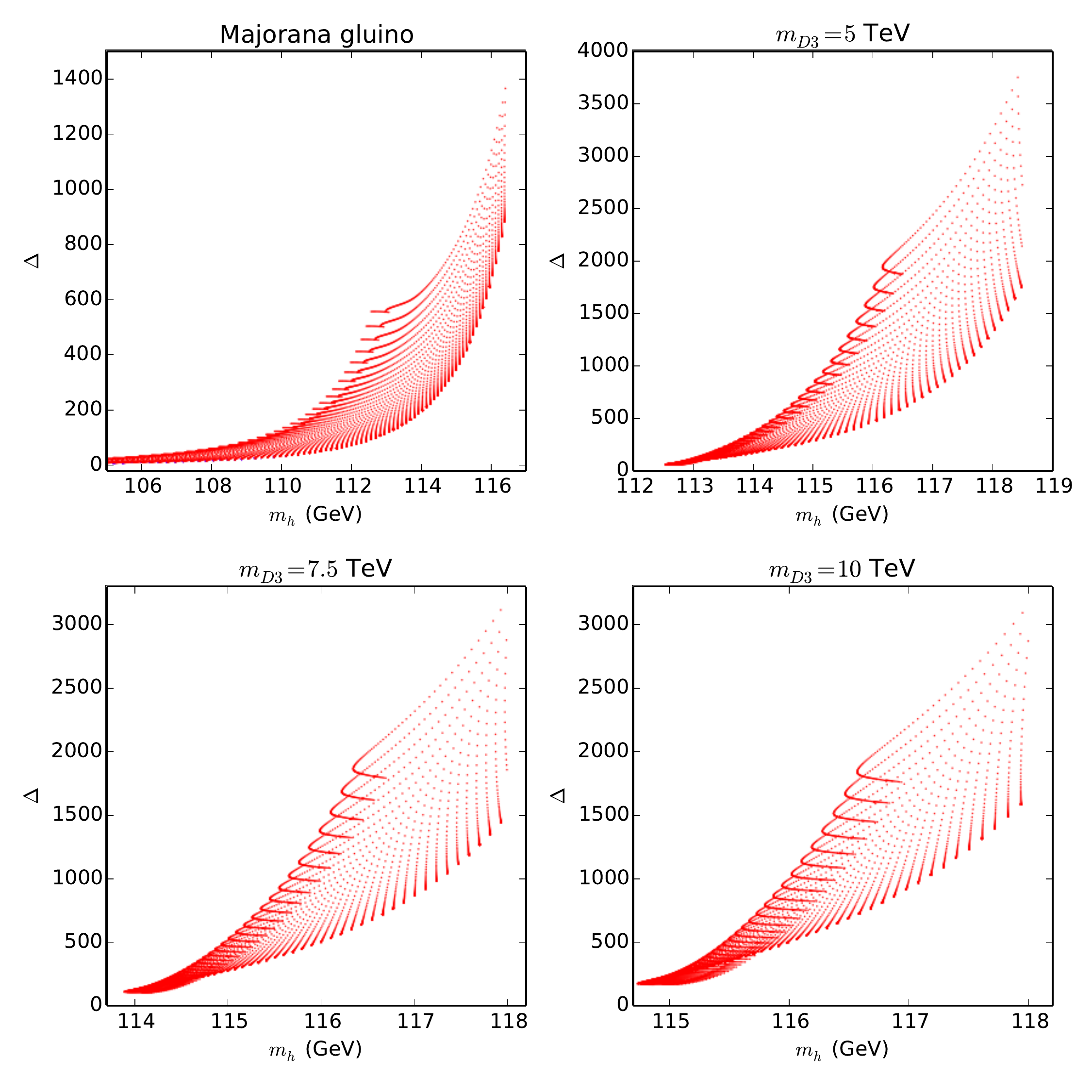}
\end{center}
\caption{Fine tuning in \gls{CGGM} with $t_\be = 10$, $m_\textrm{Mess} = 10^{12}$ GeV and $m_{D3}$ fixed as indicated. The dominant source of tuning is entirely from the $\mu$ parameter.}
\label{fig:CGGM-FT-tb10-M12}
\end{figure}

\begin{figure}[t]
\begin{center}
\includegraphics[scale=0.55]{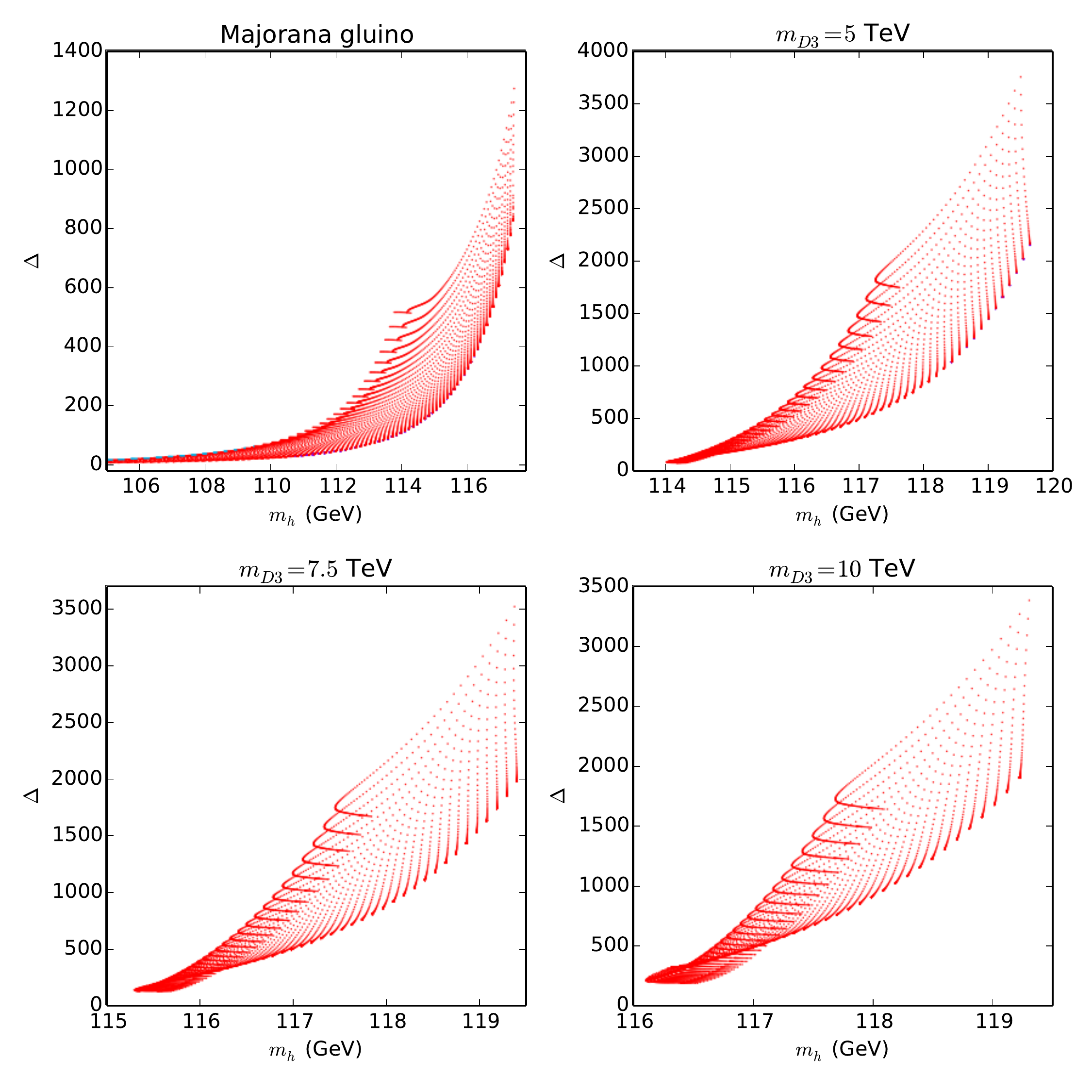}
\end{center}
\caption{Fine tuning in \gls{CGGM} with $t_\be = 25$, $m_\textrm{Mess} = 10^{12}$ GeV and $m_{D3}$ fixed as indicated. The red and blue regions correspond to $\mu$ and $\Lda_S$ as the dominant source of tuning.}
\label{fig:CGGM-FT-tb25-M12}
\end{figure}

\clearpage

\section{Renormalisation Group equations with Dirac gluino decoupling}
\label{sec:RGEs}

These \glspl{RGE} were calculated using a combination of \SARAH, \pyrate \cite{Lyonnet2013} and results from \cite{Box2010a,Box2010}. We decouple the gluino and the sgluons at renormalisation scales $\mu$ below $\mu(m_{D3})=m_{D3}\equiv \overline{m_{D3}}.$ We therefore define
\begin{align}
	\ttag&=1\quad\textrm{if}\quad\mu\geq\overline{m_{D3}}, & 
	\ttag&=0\quad\textrm{if}\quad\mu<\overline{m_{D3}}.
\end{align}
Decoupling is achieved at two loop accuracy for the gauge coupling for all particles, whereas the decoupling for the remaining terms is correct to one loop for all particles and correct to two loop for the sgluons and right handed gluino.

\subsection{SUSY parameters}

\paragraph{Gauge couplings}
\begin{align} 
	\beta_{g_1}^{(1)}
	&=\frac{33}{5} g_{1}^{3},\\ 
	\beta_{g_1}^{(2)}
	&=\frac{1}{25}g_{1}^{3}
	\Big[
		-130\,\tr\,\Big({y_{\pu}  y_{\pu}^{\dagger}}\Big)  + 135 g_{2}^{2}  + 199 g_{1}^{2}  + 220(3-\ttag) g_{3}^{2}  -70 \,\tr\,\Big({y_{\pd}  y_{\pd}^{\dagger}}\Big)  \nonumber \\&-90 \,\tr\,\Big({y_{\pe}  y_{\pe}^{\dagger}}\Big) 
	\Big],\\ 
	\beta_{g_2}^{(1)}
	&=g_{2}^{3},\\ 
	\beta_{g_2}^{(2)} & =  
	\frac{1}{5} g_{2}^{3} \Big[-10 \,\tr\,\Big({y_{\pe}  y_{\pe}^{\dagger}}\Big)  + 60(3-\ttag) g_{3}^{2}  + 125 g_{2}^{2}  -30 \,\tr\,\Big({y_{\pd}  y_{\pd}^{\dagger}}\Big)  -30 \,\tr\,\Big({y_{\pu}  y_{\pu}^{\dagger}}\Big)  + 9 g_{1}^{2} \Big],\\ 
	\beta_{g_3}^{(1)} & =  
	-\frac92(1-\ttag)g_{3}^{3},\\ 
	\beta_{g_3}^{(2)}
	&=\frac{1}{5} g_{3}^{3} 
	\Big[
		11 g_{1}^{2}  -20 \,\tr\,\Big({y_{\pd}  y_{\pd}^{\dagger}}\Big)  -20 \,\tr\,\Big({y_{\pu}  y_{\pu}^{\dagger}}\Big)  + 5(39+29\,\ttag) g_{3}^{2}  + 45 g_{2}^{2}
	\Big].
\end{align} 

\paragraph{Yukawa couplings}
\begin{align} 
	\beta_{y_{\pd}}^{(1)}
	&= 3 {y_{\pd}  y_{\pd}^{\dagger}  y_{\pd}}  + y_{\pd} \Big[-3 g_{2}^{2}  + 3 \,\tr\Big({y_{\pd}  y_{\pd}^{\dagger}}\Big) +\frac83\Big(\ttag-3\Big) g_{3}^{2}  -\frac{7}{15} g_{1}^{2}  + \,\tr\Big({y_{\pe}  y_{\pe}^{\dagger}}\Big)\Big] + {y_{\pd}  y_{\pu}^{\dagger}  y_{\pu}},\\ 
	\beta_{y_{\pd}}^{(2)} 
	& = \frac{4}{5} g_{1}^{2} {y_{\pd}  y_{\pu}^{\dagger}  y_{\pu}} -4 {y_{\pd}  y_{\pd}^{\dagger}  y_{\pd}  y_{\pd}^{\dagger}  y_{\pd}} -2 {y_{\pd}  y_{\pu}^{\dagger}  y_{\pu}  y_{\pd}^{\dagger}  y_{\pd}} -2 {y_{\pd}  y_{\pu}^{\dagger}  y_{\pu}  y_{\pu}^{\dagger}  y_{\pu}} \nonumber \\ 
	 &+{y_{\pd}  y_{\pd}^{\dagger}  y_{\pd}} \Big[6 g_{2}^{2}-3 \,\tr\Big({y_{\pe}  y_{\pe}^{\dagger}}\Big)  -9 \,\tr\Big({y_{\pd}  y_{\pd}^{\dagger}}\Big)  + \frac{4}{5} g_{1}^{2} \Big]-3 {y_{\pd}  y_{\pu}^{\dagger}  y_{\pu}} \,\tr\Big({y_{\pu}  y_{\pu}^{\dagger}}\Big) +y_{\pd} \Big[\frac{287}{90} g_{1}^{4}\nonumber \\ 
	 & +g_{1}^{2} g_{2}^{2} +\frac{15}{2} g_{2}^{4} +\frac{8}{9} g_{1}^{2} g_{3}^{2} +8 g_{2}^{2} g_{3}^{2} +\frac{128}9 g_{3}^{4} -\frac{2}{5} \Big( g_{1}^{2}-40 g_{3}^{2}\Big)\,\tr\Big({y_{\pd}  y_{\pd}^{\dagger}}\Big) \nonumber \\ 
	 &+\frac{6}{5} g_{1}^{2} \,\tr\Big({y_{\pe}  y_{\pe}^{\dagger}}\Big) -9 \,\tr\Big({y_{\pd}  y_{\pd}^{\dagger}  y_{\pd}  y_{\pd}^{\dagger}}\Big) -3 \,\tr\Big({y_{\pd}  y_{\pu}^{\dagger}  y_{\pu}  y_{\pd}^{\dagger}}\Big) -3 \,\tr\Big({y_{\pe}  y_{\pe}^{\dagger}  y_{\pe}  y_{\pe}^{\dagger}}\Big) \Big],\\ 
	\beta_{y_{\pe}}^{(1)} 
	& = 3 {y_{\pe}  y_{\pe}^{\dagger}  y_{\pe}}  + y_{\pe} \Big[-3 g_{2}^{2}  + 3 \,\tr\Big({y_{\pd}  y_{\pd}^{\dagger}}\Big)  -\frac{9}{5} g_{1}^{2}  + \,\tr\Big({y_{\pe}  y_{\pe}^{\dagger}}\Big)\Big],\\ 
	\beta_{y_{\pe}}^{(2)}
	&=-4 {y_{\pe}  y_{\pe}^{\dagger}  y_{\pe}  y_{\pe}^{\dagger}  y_{\pe}} +{y_{\pe}  y_{\pe}^{\dagger}  y_{\pe}} \Big[-3 \,\tr\Big({y_{\pe}  y_{\pe}^{\dagger}}\Big)  + 6 g_{2}^{2}  -9 \,\tr\Big({y_{\pd}  y_{\pd}^{\dagger}}\Big) \Big]\nonumber \\ 
	 &+\frac{1}{10} y_{\pe} \Big\{ 3 \Big[45 g_{1}^{4} +6 g_{1}^{2} g_{2}^{2} +25 g_{2}^{4} +4 g_{1}^{2} \,\tr\Big({y_{\pe}  y_{\pe}^{\dagger}}\Big) -30 \,\tr\Big({y_{\pd}  y_{\pd}^{\dagger}  y_{\pd}  y_{\pd}^{\dagger}}\Big) \nonumber \\ 
	 &-10 \,\tr\Big({y_{\pd}  y_{\pu}^{\dagger}  y_{\pu}  y_{\pd}^{\dagger}}\Big)-10 \,\tr\Big({y_{\pe}  y_{\pe}^{\dagger}  y_{\pe}  y_{\pe}^{\dagger}}\Big) \Big]-4 \Big(-40 g_{3}^{2}  + g_{1}^{2}\Big)\,\tr\Big({y_{\pd}  y_{\pd}^{\dagger}}\Big)\Big\},\\ 
	\beta_{y_{\pu}}^{(1)}
	&= 3 {y_{\pu}  y_{\pu}^{\dagger}  y_{\pu}}  -\frac{1}{15} y_{\pu} \Big[13 g_{1}^{2}  + 45 g_{2}^{2}  -45 \,\tr\Big({y_{\pu}  y_{\pu}^{\dagger}}\Big)  + 40\Big(3-\ttag\Big) g_{3}^{2} \Big] + {y_{\pu}  y_{\pd}^{\dagger}  y_{\pd}},\\ 
	\beta_{y_{\pu}}^{(2)}
	&=  
	\frac{2}{5} g_{1}^{2} {y_{\pu}  y_{\pu}^{\dagger}  y_{\pu}} +6 g_{2}^{2} {y_{\pu}  y_{\pu}^{\dagger}  y_{\pu}} -2 {y_{\pu}  y_{\pd}^{\dagger}  y_{\pd}  y_{\pd}^{\dagger}  y_{\pd}} -2 {y_{\pu}  y_{\pd}^{\dagger}  y_{\pd}  y_{\pu}^{\dagger}  y_{\pu}} \nonumber \\ 
	 &-4 {y_{\pu}  y_{\pu}^{\dagger}  y_{\pu}  y_{\pu}^{\dagger}  y_{\pu}} +{y_{\pu}  y_{\pd}^{\dagger}  y_{\pd}} \Big[-3 \,\tr\Big({y_{\pd}  y_{\pd}^{\dagger}}\Big)  + \frac{2}{5} g_{1}^{2}  - \,\tr\Big({y_{\pe}  y_{\pe}^{\dagger}}\Big) \Big]-9 {y_{\pu}  y_{\pu}^{\dagger}  y_{\pu}} \,\tr\Big({y_{\pu}  y_{\pu}^{\dagger}}\Big) \nonumber \\ 
	 &+y_{\pu} \Big[\frac{2743}{450} g_{1}^{4} +g_{1}^{2} g_{2}^{2} +\frac{15}{2} g_{2}^{4} +\frac{136}{45} g_{1}^{2} g_{3}^{2} +8 g_{2}^{2} g_{3}^{2} +\frac{128}9 g_{3}^{4}  \nonumber \\&+\frac{4}{5} \Big(20 g_{3}^{2}  + g_{1}^{2}\Big)\,\tr\Big({y_{\pu}  y_{\pu}^{\dagger}}\Big)-3 \,\tr\Big({y_{\pd}  y_{\pu}^{\dagger}  y_{\pu}  y_{\pd}^{\dagger}}\Big) -9 \,\tr\Big({y_{\pu}  y_{\pu}^{\dagger}  y_{\pu}  y_{\pu}^{\dagger}}\Big) \Big].
\end{align}

\paragraph{\gls{SUSY} masses}
\begin{align} 
	\beta_{\mu}^{(1)} 
	& = 3 \mu \,\tr\Big({y_{\pd}  y_{\pd}^{\dagger}}\Big)  -\frac{3}{5} \mu \Big(5 g_{2}^{2}  -5 \,\tr\Big({y_{\pu}  y_{\pu}^{\dagger}}\Big)  + g_{1}^{2}\Big) + \mu \,\tr\Big({y_{\pe}  y_{\pe}^{\dagger}}\Big), \\ 
	\beta_{\mu}^{(2)} 
	& = \frac{1}{50} \mu 
	\Big[
		207 g_{1}^{4} +90 g_{1}^{2} g_{2}^{2} +375 g_{2}^{4} -20 \Big(-40 g_{3}^{2}  + g_{1}^{2}\Big)\,\tr\Big({y_{\pd}  y_{\pd}^{\dagger}}\Big) +60 g_{1}^{2} \,\tr\Big({y_{\pe}  y_{\pe}^{\dagger}}\Big) \nonumber \\ 
		 &+800 g_{3}^{2} \,\tr\Big({y_{\pu}  y_{\pu}^{\dagger}}\Big) -450 \,\tr\Big({y_{\pd}  y_{\pd}^{\dagger}  y_{\pd}  y_{\pd}^{\dagger}}\Big) -300 \,\tr\Big({y_{\pd}  y_{\pu}^{\dagger}  y_{\pu}  y_{\pd}^{\dagger}}\Big) -150 \,\tr\Big({y_{\pe}  y_{\pe}^{\dagger}  y_{\pe}  y_{\pe}^{\dagger}}\Big) \nonumber \\ 
		 &+40 g_{1}^{2} \,\tr\Big({y_{\pu}  y_{\pu}^{\dagger}}\Big)-450 \,\tr\Big({y_{\pu}  y_{\pu}^{\dagger}  y_{\pu}  y_{\pu}^{\dagger}}\Big)
	 \Big].
	 \end{align}

\subsection{SUSY breaking parameters}

\paragraph{Majorana gaugino masses}
\begin{align} 
	\beta_{M_1}^{(1)}
	&=\frac{66}{5} g_{1}^{2} M_1, \\ 
	\beta_{M_1}^{(2)}
	&=\frac{2}{25} g_{1}^{2}
	\Big[
		398 g_{1}^{2} M_1 +135 g_{2}^{2} M_1 +440 g_{3}^{2} M_1 +440 g_{3}^{2} M_3\,\ttag +135 g_{2}^{2} M_2 -70 M_1 \,\tr\Big({y_{\pd}  y_{\pd}^{\dagger}}\Big) \nonumber \\
	 &-90 M_1 \,\tr\Big({y_{\pe}  y_{\pe}^{\dagger}}\Big) 
	 -130 M_1 \,\tr\Big({y_{\pu}  y_{\pu}^{\dagger}}\Big) +70 \,\tr\Big({y_{\pd}^{\dagger}  a_{\pd}}\Big) +90 \,\tr\Big({y_{\pe}^{\dagger}  a_{\pe}}\Big) +130 \,\tr\Big({y_{\pu}^{\dagger}  a_{\pu}}\Big) \Big],\\ 
	\beta_{M_2}^{(1)}
	&=2 g_{2}^{2} M_2,\\ 
	\beta_{M_2}^{(2)}
	&= 
	\frac{2}{5} g_{2}^{2} 
	\Big[
		9 g_{1}^{2} M_1 +120 g_{3}^{2} M_3\,\ttag +9 g_{1}^{2} M_2 +250 g_{2}^{2} M_2 +120 g_{3}^{2} M_2 -30 M_2 \,\tr\Big({y_{\pd}  y_{\pd}^{\dagger}}\Big) \nonumber \\ 
		& -10 M_2 \,\tr\Big({y_{\pe}  y_{\pe}^{\dagger}}\Big)
	 -30 M_2 \,\tr\Big({y_{\pu}  y_{\pu}^{\dagger}}\Big) +30 \,\tr\Big({y_{\pd}^{\dagger}  a_{\pd}}\Big) +10 \,\tr\Big({y_{\pe}^{\dagger}  a_{\pe}}\Big) +30 \,\tr\Big({y_{\pu}^{\dagger}  a_{\pu}}\Big)
	 \Big],\\ 
	\beta_{M_3}^{(1)}
	&=0,\\ 
	\beta_{M_3}^{(2)}
	&= \frac{2}{5} g_{3}^{2} 
	\Big[
		11 g_{1}^{2} M_1 +11 g_{1}^{2} M_3 +45 g_{2}^{2} M_3 +680 g_{3}^{2} M_3 +45 g_{2}^{2} M_2 -20 M_3 \,\tr\Big({y_{\pd}  y_{\pd}^{\dagger}}\Big) \nonumber \\ 
		& -20 M_3 \,\tr\Big({y_{\pu}  y_{\pu}^{\dagger}}\Big) +20 \,\tr\Big({y_{\pd}^{\dagger}  a_{\pd}}\Big) +20 \,\tr\Big({y_{\pu}^{\dagger}  a_{\pu}}\Big) 
	 \Big]\ttag.
\end{align}

\paragraph{Dirac gluino mass}
\begin{align} 
	\beta_{m_{D3}}^{(1)}
	&=-6\, g_{3}^{2} m_{D3}\,\ttag,\\
	\beta_{m_{D3}}^{(2)}
	&=\frac15\,g_{3}^{2} m_{D3}\Big[
	11 g_{1}^{2} +45 g_{2}^{2}+520 g_{3}^2
	-20\,\tr\Big({y_{\pd}  y_{\pd}^{\dagger}}\Big)
	-20\,\tr\Big({y_{\pu}  y_{\pu}^{\dagger}}\Big)
	\Big]\ttag.
\end{align}

\paragraph{Trilinear Soft-Breaking Parameters}
{\allowdisplaybreaks \begin{align} 
	\beta_{a_{\pd}}^{(1)} 
	& = 4 {y_{\pd}  y_{\pd}^{\dagger}  a_{\pd}} +2 {y_{\pd}  y_{\pu}^{\dagger}  a_{\pu}} +5 {a_{\pd}  y_{\pd}^{\dagger}  y_{\pd}} +{a_{\pd}  y_{\pu}^{\dagger}  y_{\pu}}-\frac{7}{15} g_{1}^{2} a_{\pd} -3 g_{2}^{2} a_{\pd} + \frac{16}{3} \Big(\ttag-2\Big)g_{3}^{2} a_{\pd} \nonumber \\ 
	 &+3 a_{\pd} \,\tr\Big({y_{\pd}  y_{\pd}^{\dagger}}\Big) +a_{\pd} \,\tr\Big({y_{\pe}  y_{\pe}^{\dagger}}\Big) +y_{\pd} \Big[2 \,\tr\Big({y_{\pe}^{\dagger}  a_{\pe}}\Big)  + 6 g_{2}^{2} M_2  + 6 \,\tr\Big({y_{\pd}^{\dagger}  a_{\pd}}\Big) \nonumber \\
	 & + \frac{14}{15} g_{1}^{2} M_1  + \frac{32}{3} g_{3}^{2} M_3\,\ttag \Big],\\ 
	\beta_{a_{\pd}}^{(2)} 
	& = \frac{6}{5} g_{1}^{2} {y_{\pd}  y_{\pd}^{\dagger}  a_{\pd}} +6 g_{2}^{2} {y_{\pd}  y_{\pd}^{\dagger}  a_{\pd}} -\frac{8}{5} g_{1}^{2} M_1 {y_{\pd}  y_{\pu}^{\dagger}  y_{\pu}} +\frac{8}{5} g_{1}^{2} {y_{\pd}  y_{\pu}^{\dagger}  a_{\pu}} \nonumber \\ 
	 &+\frac{6}{5} g_{1}^{2} {a_{\pd}  y_{\pd}^{\dagger}  y_{\pd}} +12 g_{2}^{2} {a_{\pd}  y_{\pd}^{\dagger}  y_{\pd}} +\frac{4}{5} g_{1}^{2} {a_{\pd}  y_{\pu}^{\dagger}  y_{\pu}} -6 {y_{\pd}  y_{\pd}^{\dagger}  y_{\pd}  y_{\pd}^{\dagger}  a_{\pd}} \nonumber \\ 
	 &-8 {y_{\pd}  y_{\pd}^{\dagger}  a_{\pd}  y_{\pd}^{\dagger}  y_{\pd}} -2 {y_{\pd}  y_{\pu}^{\dagger}  y_{\pu}  y_{\pd}^{\dagger}  a_{\pd}} -4 {y_{\pd}  y_{\pu}^{\dagger}  y_{\pu}  y_{\pu}^{\dagger}  a_{\pu}} -4 {y_{\pd}  y_{\pu}^{\dagger}  a_{\pu}  y_{\pd}^{\dagger}  y_{\pd}} \nonumber \\ 
	 &-4 {y_{\pd}  y_{\pu}^{\dagger}  a_{\pu}  y_{\pu}^{\dagger}  y_{\pu}} -6 {a_{\pd}  y_{\pd}^{\dagger}  y_{\pd}  y_{\pd}^{\dagger}  y_{\pd}} -4 {a_{\pd}  y_{\pu}^{\dagger}  y_{\pu}  y_{\pd}^{\dagger}  y_{\pd}} -2 {a_{\pd}  y_{\pu}^{\dagger}  y_{\pu}  y_{\pu}^{\dagger}  y_{\pu}} \nonumber \\ 
	 &+\frac{287}{90} g_{1}^{4} a_{\pd} +g_{1}^{2} g_{2}^{2} a_{\pd} +\frac{15}{2} g_{2}^{4} a_{\pd} +\frac{8}{9} g_{1}^{2} g_{3}^{2} a_{\pd} +8 g_{2}^{2} g_{3}^{2} a_{\pd} +\frac{128}9 g_{3}^{4} a_{\pd} \nonumber \\ 
	 &-12 {y_{\pd}  y_{\pd}^{\dagger}  a_{\pd}} \,\tr\Big({y_{\pd}  y_{\pd}^{\dagger}}\Big) -15 {a_{\pd}  y_{\pd}^{\dagger}  y_{\pd}} \,\tr\Big({y_{\pd}  y_{\pd}^{\dagger}}\Big) -\frac{2}{5} g_{1}^{2} a_{\pd} \,\tr\Big({y_{\pd}  y_{\pd}^{\dagger}}\Big) \nonumber \\ 
	 &+16 g_{3}^{2} a_{\pd} \,\tr\Big({y_{\pd}  y_{\pd}^{\dagger}}\Big) -4 {y_{\pd}  y_{\pd}^{\dagger}  a_{\pd}} \,\tr\Big({y_{\pe}  y_{\pe}^{\dagger}}\Big) -5 {a_{\pd}  y_{\pd}^{\dagger}  y_{\pd}} \,\tr\Big({y_{\pe}  y_{\pe}^{\dagger}}\Big) \nonumber \\ 
	 &+\frac{6}{5} g_{1}^{2} a_{\pd} \,\tr\Big({y_{\pe}  y_{\pe}^{\dagger}}\Big) -6 {y_{\pd}  y_{\pu}^{\dagger}  a_{\pu}} \,\tr\Big({y_{\pu}  y_{\pu}^{\dagger}}\Big) -3 {a_{\pd}  y_{\pu}^{\dagger}  y_{\pu}} \,\tr\Big({y_{\pu}  y_{\pu}^{\dagger}}\Big) \nonumber \\ 
	 &-\frac{2}{5} {y_{\pd}  y_{\pd}^{\dagger}  y_{\pd}} \Big[15 \,\tr\Big({y_{\pe}^{\dagger}  a_{\pe}}\Big)  + 30 g_{2}^{2} M_2  + 45 \,\tr\Big({y_{\pd}^{\dagger}  a_{\pd}}\Big)  + 4 g_{1}^{2} M_1 \Big]-6 {y_{\pd}  y_{\pu}^{\dagger}  y_{\pu}} \,\tr\Big({y_{\pu}^{\dagger}  a_{\pu}}\Big) \nonumber \\ 
	 &-9 a_{\pd} \,\tr\Big({y_{\pd}  y_{\pd}^{\dagger}  y_{\pd}  y_{\pd}^{\dagger}}\Big) -3 a_{\pd} \,\tr\Big({y_{\pd}  y_{\pu}^{\dagger}  y_{\pu}  y_{\pd}^{\dagger}}\Big) -3 a_{\pd} \,\tr\Big({y_{\pe}  y_{\pe}^{\dagger}  y_{\pe}  y_{\pe}^{\dagger}}\Big) \nonumber \\ 
	 &-\frac{2}{45} y_{\pd} \Big[287 g_{1}^{4} M_1 +45 g_{1}^{2} g_{2}^{2} M_1 +40 g_{1}^{2} g_{3}^{2} M_1 +40 g_{1}^{2} g_{3}^{2} M_3\,\ttag +360 g_{2}^{2} g_{3}^{2} M_3\,\ttag \nonumber \\ 
	 & +1280 g_{3}^{4} M_3\,\ttag +45 g_{1}^{2} g_{2}^{2} M_2 +675 g_{2}^{4} M_2 +360 g_{2}^{2} g_{3}^{2} M_2 +18 \Big(40 g_{3}^{2} M_3\,\ttag  - g_{1}^{2} M_1 \Big)\,\tr\Big({y_{\pd}  y_{\pd}^{\dagger}}\Big) \nonumber \\ 
	 & +54 g_{1}^{2} M_1 \,\tr\Big({y_{\pe}  y_{\pe}^{\dagger}}\Big) +18 g_{1}^{2} \,\tr\Big({y_{\pd}^{\dagger}  a_{\pd}}\Big) -720 g_{3}^{2} \,\tr\Big({y_{\pd}^{\dagger}  a_{\pd}}\Big) -54 g_{1}^{2} \,\tr\Big({y_{\pe}^{\dagger}  a_{\pe}}\Big)
	 +810 \,\tr\Big({y_{\pd}  y_{\pd}^{\dagger}  a_{\pd}  y_{\pd}^{\dagger}}\Big) \nonumber \\ & +135 \,\tr\Big({y_{\pd}  y_{\pu}^{\dagger}  a_{\pu}  y_{\pd}^{\dagger}}\Big) +270 \,\tr\Big({y_{\pe}  y_{\pe}^{\dagger}  a_{\pe}  y_{\pe}^{\dagger}}\Big) +135 \,\tr\Big({y_{\pu}  y_{\pd}^{\dagger}  a_{\pd}  y_{\pu}^{\dagger}}\Big) \Big],\\ 
	\beta_{a_{\pe}}^{(1)} 
	& = 4 {y_{\pe}  y_{\pe}^{\dagger}  a_{\pe}} +5 {a_{\pe}  y_{\pe}^{\dagger}  y_{\pe}} -\frac{9}{5} g_{1}^{2} a_{\pe} -3 g_{2}^{2} a_{\pe} +3 a_{\pe} \,\tr\Big({y_{\pd}  y_{\pd}^{\dagger}}\Big) +a_{\pe} \,\tr\Big({y_{\pe}  y_{\pe}^{\dagger}}\Big) \nonumber \\ 
	 &+y_{\pe} \Big[2 \,\tr\Big({y_{\pe}^{\dagger}  a_{\pe}}\Big)  + 6 g_{2}^{2} M_2  + 6 \,\tr\Big({y_{\pd}^{\dagger}  a_{\pd}}\Big)  + \frac{18}{5} g_{1}^{2} M_1 \Big],\\ 
	\beta_{a_{\pe}}^{(2)} & =  
	+\frac{6}{5} g_{1}^{2} {y_{\pe}  y_{\pe}^{\dagger}  a_{\pe}} +6 g_{2}^{2} {y_{\pe}  y_{\pe}^{\dagger}  a_{\pe}} -\frac{6}{5} g_{1}^{2} {a_{\pe}  y_{\pe}^{\dagger}  y_{\pe}} +12 g_{2}^{2} {a_{\pe}  y_{\pe}^{\dagger}  y_{\pe}} \nonumber \\ 
	 &-6 {y_{\pe}  y_{\pe}^{\dagger}  y_{\pe}  y_{\pe}^{\dagger}  a_{\pe}} -8 {y_{\pe}  y_{\pe}^{\dagger}  a_{\pe}  y_{\pe}^{\dagger}  y_{\pe}} -6 {a_{\pe}  y_{\pe}^{\dagger}  y_{\pe}  y_{\pe}^{\dagger}  y_{\pe}} +\frac{27}{2} g_{1}^{4} a_{\pe} +\frac{9}{5} g_{1}^{2} g_{2}^{2} a_{\pe} +\frac{15}{2} g_{2}^{4} a_{\pe} \nonumber \\ 
	 &-12 {y_{\pe}  y_{\pe}^{\dagger}  a_{\pe}} \,\tr\Big({y_{\pd}  y_{\pd}^{\dagger}}\Big) -15 {a_{\pe}  y_{\pe}^{\dagger}  y_{\pe}} \,\tr\Big({y_{\pd}  y_{\pd}^{\dagger}}\Big) -\frac{2}{5} g_{1}^{2} a_{\pe} \,\tr\Big({y_{\pd}  y_{\pd}^{\dagger}}\Big) \nonumber \\ 
	 &+16 g_{3}^{2} a_{\pe} \,\tr\Big({y_{\pd}  y_{\pd}^{\dagger}}\Big) -4 {y_{\pe}  y_{\pe}^{\dagger}  a_{\pe}} \,\tr\Big({y_{\pe}  y_{\pe}^{\dagger}}\Big) -5 {a_{\pe}  y_{\pe}^{\dagger}  y_{\pe}} \,\tr\Big({y_{\pe}  y_{\pe}^{\dagger}}\Big) \nonumber \\ 
	 &+\frac{6}{5} g_{1}^{2} a_{\pe} \,\tr\Big({y_{\pe}  y_{\pe}^{\dagger}}\Big) -6 {y_{\pe}  y_{\pe}^{\dagger}  y_{\pe}} \Big[2 g_{2}^{2} M_2  + 3 \,\tr\Big({y_{\pd}^{\dagger}  a_{\pd}}\Big)  + \,\tr\Big({y_{\pe}^{\dagger}  a_{\pe}}\Big)\Big]-9 a_{\pe} \,\tr\Big({y_{\pd}  y_{\pd}^{\dagger}  y_{\pd}  y_{\pd}^{\dagger}}\Big) \nonumber \\ 
	 &-3 a_{\pe} \,\tr\Big({y_{\pd}  y_{\pu}^{\dagger}  y_{\pu}  y_{\pd}^{\dagger}}\Big) -3 a_{\pe} \,\tr\Big({y_{\pe}  y_{\pe}^{\dagger}  y_{\pe}  y_{\pe}^{\dagger}}\Big) \nonumber \\ 
	 &-\frac{2}{5} y_{\pe} \Big[135 g_{1}^{4} M_1 +9 g_{1}^{2} g_{2}^{2} M_1 +9 g_{1}^{2} g_{2}^{2} M_2 +75 g_{2}^{4} M_2 +\Big(-2 g_{1}^{2} M_1  + 80 g_{3}^{2} M_3\,\ttag \Big)\,\tr\Big({y_{\pd}  y_{\pd}^{\dagger}}\Big) \nonumber \\ 
	 &+6 g_{1}^{2} M_1 \,\tr\Big({y_{\pe}  y_{\pe}^{\dagger}}\Big) +2 g_{1}^{2} \,\tr\Big({y_{\pd}^{\dagger}  a_{\pd}}\Big) -80 g_{3}^{2} \,\tr\Big({y_{\pd}^{\dagger}  a_{\pd}}\Big) -6 g_{1}^{2} \,\tr\Big({y_{\pe}^{\dagger}  a_{\pe}}\Big) \nonumber \\ 
	 &+90 \,\tr\Big({y_{\pd}  y_{\pd}^{\dagger}  a_{\pd}  y_{\pd}^{\dagger}}\Big) +15 \,\tr\Big({y_{\pd}  y_{\pu}^{\dagger}  a_{\pu}  y_{\pd}^{\dagger}}\Big) +30 \,\tr\Big({y_{\pe}  y_{\pe}^{\dagger}  a_{\pe}  y_{\pe}^{\dagger}}\Big) +15 \,\tr\Big({y_{\pu}  y_{\pd}^{\dagger}  a_{\pd}  y_{\pu}^{\dagger}}\Big) \Big],\\ 
	\beta_{a_{\pu}}^{(1)} 
	& = 2 {y_{\pu}  y_{\pd}^{\dagger}  a_{\pd}} +4 {y_{\pu}  y_{\pu}^{\dagger}  a_{\pu}} +{a_{\pu}  y_{\pd}^{\dagger}  y_{\pd}}+5 {a_{\pu}  y_{\pu}^{\dagger}  y_{\pu}} -\frac{13}{15} g_{1}^{2} a_{\pu} -3 g_{2}^{2} a_{\pu} +\frac{16}{3}\Big(\ttag-2\Big) g_{3}^{2} a_{\pu} \nonumber \\ 
	 &+3 a_{\pu} \,\tr\Big({y_{\pu}  y_{\pu}^{\dagger}}\Big) +y_{\pu} \Big[6 g_{2}^{2} M_2  + 6 \,\tr\Big({y_{\pu}^{\dagger}  a_{\pu}}\Big)  + \frac{26}{15} g_{1}^{2} M_1  + \frac{32}{3} g_{3}^{2} M_3\,\ttag \Big],
	\label{eq:beta-au-1loop}\\ 
	\beta_{a_{\pu}}^{(2)} 
	& = \frac{4}{5} g_{1}^{2} {y_{\pu}  y_{\pd}^{\dagger}  a_{\pd}} -\frac{4}{5} g_{1}^{2} M_1 {y_{\pu}  y_{\pu}^{\dagger}  y_{\pu}} -12 g_{2}^{2} M_2 {y_{\pu}  y_{\pu}^{\dagger}  y_{\pu}} +\frac{6}{5} g_{1}^{2} {y_{\pu}  y_{\pu}^{\dagger}  a_{\pu}} \nonumber \\ 
	 &+6 g_{2}^{2} {y_{\pu}  y_{\pu}^{\dagger}  a_{\pu}} +\frac{2}{5} g_{1}^{2} {a_{\pu}  y_{\pd}^{\dagger}  y_{\pd}} +12 g_{2}^{2} {a_{\pu}  y_{\pu}^{\dagger}  y_{\pu}} -4 {y_{\pu}  y_{\pd}^{\dagger}  y_{\pd}  y_{\pd}^{\dagger}  a_{\pd}} \nonumber \\ 
	 &-2 {y_{\pu}  y_{\pd}^{\dagger}  y_{\pd}  y_{\pu}^{\dagger}  a_{\pu}} -4 {y_{\pu}  y_{\pd}^{\dagger}  a_{\pd}  y_{\pd}^{\dagger}  y_{\pd}} -4 {y_{\pu}  y_{\pd}^{\dagger}  a_{\pd}  y_{\pu}^{\dagger}  y_{\pu}} -6 {y_{\pu}  y_{\pu}^{\dagger}  y_{\pu}  y_{\pu}^{\dagger}  a_{\pu}} \nonumber \\ 
	 &-8 {y_{\pu}  y_{\pu}^{\dagger}  a_{\pu}  y_{\pu}^{\dagger}  y_{\pu}} -2 {a_{\pu}  y_{\pd}^{\dagger}  y_{\pd}  y_{\pd}^{\dagger}  y_{\pd}} -4 {a_{\pu}  y_{\pd}^{\dagger}  y_{\pd}  y_{\pu}^{\dagger}  y_{\pu}} -6 {a_{\pu}  y_{\pu}^{\dagger}  y_{\pu}  y_{\pu}^{\dagger}  y_{\pu}} +\frac{2743}{450} g_{1}^{4} a_{\pu} \nonumber \\ 
	 &+g_{1}^{2} g_{2}^{2} a_{\pu} +\frac{15}{2} g_{2}^{4} a_{\pu} +\frac{136}{45} g_{1}^{2} g_{3}^{2} a_{\pu} +8 g_{2}^{2} g_{3}^{2} a_{\pu} +\frac{128}9 g_{3}^{4} a_{\pu} -6 {y_{\pu}  y_{\pd}^{\dagger}  a_{\pd}} \,\tr\Big({y_{\pd}  y_{\pd}^{\dagger}}\Big) \nonumber \\ 
	 &-3 {a_{\pu}  y_{\pd}^{\dagger}  y_{\pd}} \,\tr\Big({y_{\pd}  y_{\pd}^{\dagger}}\Big) -2 {y_{\pu}  y_{\pd}^{\dagger}  a_{\pd}} \,\tr\Big({y_{\pe}  y_{\pe}^{\dagger}}\Big) - {a_{\pu}  y_{\pd}^{\dagger}  y_{\pd}} \,\tr\Big({y_{\pe}  y_{\pe}^{\dagger}}\Big) \nonumber \\ 
	 &-12 {y_{\pu}  y_{\pu}^{\dagger}  a_{\pu}} \,\tr\Big({y_{\pu}  y_{\pu}^{\dagger}}\Big) -15 {a_{\pu}  y_{\pu}^{\dagger}  y_{\pu}} \,\tr\Big({y_{\pu}  y_{\pu}^{\dagger}}\Big) +\frac{4}{5} g_{1}^{2} a_{\pu} \,\tr\Big({y_{\pu}  y_{\pu}^{\dagger}}\Big) \nonumber \\ 
	 &+16 g_{3}^{2} a_{\pu} \,\tr\Big({y_{\pu}  y_{\pu}^{\dagger}}\Big) -\frac{2}{5} {y_{\pu}  y_{\pd}^{\dagger}  y_{\pd}} \Big[15 \,\tr\Big({y_{\pd}^{\dagger}  a_{\pd}}\Big)  + 2 g_{1}^{2} M_1  + 5 \,\tr\Big({y_{\pe}^{\dagger}  a_{\pe}}\Big) \Big]\nonumber \\ 
	 &-18 {y_{\pu}  y_{\pu}^{\dagger}  y_{\pu}} \,\tr\Big({y_{\pu}^{\dagger}  a_{\pu}}\Big) -3 a_{\pu} \,\tr\Big({y_{\pd}  y_{\pu}^{\dagger}  y_{\pu}  y_{\pd}^{\dagger}}\Big) -9 a_{\pu} \,\tr\Big({y_{\pu}  y_{\pu}^{\dagger}  y_{\pu}  y_{\pu}^{\dagger}}\Big) \nonumber \\ 
	 &-\frac{2}{225} y_{\pu} \Big\{2743 g_{1}^{4} M_1 +225 g_{1}^{2} g_{2}^{2} M_1 +680 g_{1}^{2} g_{3}^{2} M_1 +680 g_{1}^{2} g_{3}^{2} M_3\,\ttag +1800 g_{2}^{2} g_{3}^{2} M_3\,\ttag \nonumber \\ 
	 & +6400 g_{3}^{4} M_3\,\ttag +225 g_{1}^{2} g_{2}^{2} M_2 +3375 g_{2}^{4} M_2 +1800 g_{2}^{2} g_{3}^{2} M_2 \nonumber \\ 
	 &-180 \Big(20 g_{3}^{2}  + g_{1}^{2}\Big)\,\tr\Big({y_{\pu}^{\dagger}  a_{\pu}}\Big) +675 \,\tr\Big({y_{\pd}  y_{\pu}^{\dagger}  a_{\pu}  y_{\pd}^{\dagger}}\Big) +675 \,\tr\Big({y_{\pu}  y_{\pd}^{\dagger}  a_{\pd}  y_{\pu}^{\dagger}}\Big) \nonumber \\ 
	 &+4050 \,\tr\Big[{y_{\pu}  y_{\pu}^{\dagger}  a_{\pu}  y_{\pu}^{\dagger}} +180 \Big(20 g_{3}^{2} M_3\,\ttag  + g_{1}^{2} M_1 \Big)\,\tr\Big({y_{\pu}  y_{\pu}^{\dagger}}\Big)\Big] \Big\}.
\end{align}}

\paragraph{Bilinear Soft-Breaking Parameters}
{\allowdisplaybreaks  \begin{align} 
	\beta_{B_{\mu}}^{(1)} 
	& = \frac{6}{5} g_{1}^{2} M_1 \mu +6 g_{2}^{2} M_2 \mu +B_{\mu} \Big[3 \,\tr\Big({y_{\pd}  y_{\pd}^{\dagger}}\Big) -3 g_{2}^{2} + 3 \,\tr\Big({y_{\pu}  y_{\pu}^{\dagger}}\Big)  -\frac{3}{5} g_{1}^{2}  + \,\tr\Big({y_{\pe}  y_{\pe}^{\dagger}}\Big)\Big]\nonumber \\ 
	 &+6 \mu \,\tr\Big({y_{\pd}^{\dagger}  a_{\pd}}\Big) +2 \mu \,\tr\Big({y_{\pe}^{\dagger}  a_{\pe}}\Big) +6 \mu \,\tr\Big({y_{\pu}^{\dagger}  a_{\pu}}\Big), \\ 
	\beta_{B_{\mu}}^{(2)} 
	& = B_{\mu} \Big[\frac{207}{50} g_{1}^{4} +\frac{9}{5} g_{1}^{2} g_{2}^{2} +\frac{15}{2} g_{2}^{4} +\frac{2}{5} \Big(g_{1}^{2}-40 g_{3}^{2}\Big)\,\tr\Big({y_{\pd}  y_{\pd}^{\dagger}}\Big) +\frac{6}{5} g_{1}^{2} \,\tr\Big({y_{\pe}  y_{\pe}^{\dagger}}\Big) +\frac{4}{5} g_{1}^{2} \,\tr\Big({y_{\pu}  y_{\pu}^{\dagger}}\Big) \nonumber \\ 
	 &+16 g_{3}^{2} \,\tr\Big({y_{\pu}  y_{\pu}^{\dagger}}\Big) -9 \,\tr\Big({y_{\pd}  y_{\pd}^{\dagger}  y_{\pd}  y_{\pd}^{\dagger}}\Big) -6 \,\tr\Big({y_{\pd}  y_{\pu}^{\dagger}  y_{\pu}  y_{\pd}^{\dagger}}\Big) -3 \,\tr\Big({y_{\pe}  y_{\pe}^{\dagger}  y_{\pe}  y_{\pe}^{\dagger}}\Big)  \nonumber \\ & -9 \,\tr\Big({y_{\pu}  y_{\pu}^{\dagger}  y_{\pu}  y_{\pu}^{\dagger}}\Big) \Big] 
	 -\frac{2}{25} \mu \Big[207 g_{1}^{4} M_1 +45 g_{1}^{2} g_{2}^{2} M_1 +45 g_{1}^{2} g_{2}^{2} M_2 +375 g_{2}^{4} M_2 \nonumber \\ 
	 &+30 g_{1}^{2} M_1 \,\tr\Big({y_{\pe}  y_{\pe}^{\dagger}}\Big)+10 \Big(g_{1}^{2} M_1-40 g_{3}^{2} M_3\,\ttag \Big)\,\tr\Big({y_{\pd}  y_{\pd}^{\dagger}}\Big) +20 g_{1}^{2} M_1 \,\tr\Big({y_{\pu}  y_{\pu}^{\dagger}}\Big) \nonumber \\ 
	 &+400 g_{3}^{2} M_3\,\ttag \,\tr\Big({y_{\pu}  y_{\pu}^{\dagger}}\Big) +10 g_{1}^{2} \,\tr\Big({y_{\pd}^{\dagger}  a_{\pd}}\Big) -400 g_{3}^{2} \,\tr\Big({y_{\pd}^{\dagger}  a_{\pd}}\Big) -30 g_{1}^{2} \,\tr\Big({y_{\pe}^{\dagger}  a_{\pe}}\Big) -20 g_{1}^{2} \,\tr\Big({y_{\pu}^{\dagger}  a_{\pu}}\Big) \nonumber \\ 
	 &+450 \,\tr\Big({y_{\pd}  y_{\pd}^{\dagger}  a_{\pd}  y_{\pd}^{\dagger}}\Big) +150 \,\tr\Big({y_{\pd}  y_{\pu}^{\dagger}  a_{\pu}  y_{\pd}^{\dagger}}\Big) +150 \,\tr\Big({y_{\pe}  y_{\pe}^{\dagger}  a_{\pe}  y_{\pe}^{\dagger}}\Big) +150 \,\tr\Big({y_{\pu}  y_{\pd}^{\dagger}  a_{\pd}  y_{\pu}^{\dagger}}\Big) \nonumber \\ 
	 &+450 \,\tr\Big({y_{\pu}  y_{\pu}^{\dagger}  a_{\pu}  y_{\pu}^{\dagger}}\Big)-400 g_{3}^{2} \,\tr\Big({y_{\pu}^{\dagger}  a_{\pu}}\Big) \Big],\\ 
	\beta_{B_3}^{(1)} 
	& = - 12 g_{3}^{2} B_3,\\ 
	\beta_{B_3}^{(2)} 
	& = 72 g_{3}^{4} B_3.
	\end{align}} 
	
\paragraph{Soft-Breaking Scalar Masses}
\begin{align} 
	\sigma_{1,1} 
	& = \sqrt{\frac{3}{5}} g_1 \Big[\mm\pHu2-2 \,\tr\Big({\mm\pu2}\Big)  - \,\tr\Big({\mm\pl2}\Big)  - \mm\pHd2  + \,\tr\Big({\mm\pd2}\Big) + \,\tr\Big({\mm\pe2}\Big) + \,\tr\Big({\mm\pq2}\Big)\Big],\\ 
	\sigma_{2,11} 
	& = \frac{1}{10} g_{1}^{2} \Big[2 \,\tr\Big({\mm\pd2}\Big)  + 3 \,\tr\Big({\mm\pl2}\Big)  + 3 \mm\pHd2  + 3 \mm\pHu2  + 6 \,\tr\Big({\mm\pe2}\Big)  + 8 \,\tr\Big({\mm\pu2}\Big)  + \,\tr\Big({\mm\pq2}\Big)\Big],\\ 
	\sigma_{3,1} 
	& = \frac{1}{20\sqrt{15}} g_1 \Big[9 g_{1}^{2} \mm\pHu2-9 g_{1}^{2} \mm\pHd2 -45 g_{2}^{2} \mm\pHd2  +45 g_{2}^{2} \mm\pHu2 +4 \Big(20 g_{3}^{2}  + g_{1}^{2}\Big)\,\tr\Big({\mm\pd2}\Big) \nonumber \\ 
	 &-9 g_{1}^{2} \,\tr\Big({\mm\pl2}\Big) -45 g_{2}^{2} \,\tr\Big({\mm\pl2}\Big) +g_{1}^{2} \,\tr\Big({\mm\pq2}\Big) +45 g_{2}^{2} \,\tr\Big({\mm\pq2}\Big) +80 g_{3}^{2} \,\tr\Big({\mm\pq2}\Big) -32 g_{1}^{2} \,\tr\Big({\mm\pu2}\Big) \nonumber \\ 
	 &-160 g_{3}^{2} \,\tr\Big({\mm\pq2}\Big) +90 \mm\pHd2 \,\tr\Big({y_{\pd}  y_{\pd}^{\dagger}}\Big) +30 \mm\pHd2 \,\tr\Big({y_{\pe}  y_{\pe}^{\dagger}}\Big) -90 \mm\pHu2 \,\tr\Big({y_{\pu}  y_{\pu}^{\dagger}}\Big) \nonumber \\ 
	 &-30 \,\tr\Big({y_{\pd}  \m\pq^{2 *}  y_{\pd}^{\dagger}}\Big) -60 \,\tr\Big({y_{\pe}  y_{\pe}^{\dagger}  \m\pe^{2 *}}\Big) +30 \,\tr\Big({y_{\pe}  \m\pl^{2 *}  y_{\pe}^{\dagger}}\Big) +120 \,\tr\Big({y_{\pu}  y_{\pu}^{\dagger}  \m\pu^{2 *}}\Big) \nonumber \\ 
	 &-30 \,\tr\Big({y_{\pu}  \m\pq^{2 *}  y_{\pu}^{\dagger}}\Big) +36 g_{1}^{2} \,\tr\Big({\mm\pe2}\Big)-60 \,\tr\Big({y_{\pd}  y_{\pd}^{\dagger}  \m\pd^{2 *}}\Big)\Big],\\ 
	\sigma_{2,2} & = \frac{1}{2} \Big[3 \,\tr\Big({\mm\pq2}\Big)  + \mm\pHd2 + \mm\pHu2 + \,\tr\Big({\mm\pl2}\Big)\Big],\\ 
	\sigma_{2,3} & = \frac{1}{2} \Big[2 \,\tr\Big({\mm\pq2}\Big)  + 3(1+\ttag) \mm32  + \,\tr\Big({\mm\pd2}\Big) + \,\tr\Big({\mm\pu2}\Big)\Big].
\end{align} 
{\allowdisplaybreaks  \begin{align} 
	\beta_{\mm\pq2}^{(1)} 
	& = -\frac{2}{15} g_{1}^{2} |M_1|^2 -\frac{32}{3} g_{3}^{2}  |M_3|^2\,\ttag -6 g_{2}^{2}  |M_2|^2 +2 \mm\pHd2 {y_{\pd}^{\dagger}  y_{\pd}} +2 \mm\pHu2 {y_{\pu}^{\dagger}  y_{\pu}} +2 {a_{\pd}^{\dagger}  a_{\pd}} \nonumber \\ 
	 &+2 {a_{\pu}^{\dagger}  a_{\pu}} +{\mm\pq2  y_{\pd}^{\dagger}  y_{\pd}}+{\mm\pq2  y_{\pu}^{\dagger}  y_{\pu}}+2 {y_{\pd}^{\dagger}  \mm\pd2  y_{\pd}} +{y_{\pd}^{\dagger}  y_{\pd}  \mm\pq2}+2 {y_{\pu}^{\dagger}  \mm\pu2  y_{\pu}} \nonumber \\ 
	 &+{y_{\pu}^{\dagger}  y_{\pu}  \mm\pq2}+\frac{1}{\sqrt{15}} g_1  \sigma_{1,1}, \\ 
	\beta_{\mm\pq2}^{(2)}
	& = \frac{2}{5} g_{1}^{2} g_{2}^{2} |M_2|^2 +33 g_{2}^{4}  |M_2|^2 +32 g_{2}^{2} g_{3}^{2}  |M_2|^2 \nonumber \\ 
	 &+\frac{16}{45} g_{3}^{2} \Big\{15 \Big[10 g_{3}^{2} M_3\,\ttag  + 3 g_{2}^{2} \Big(2 M_3\,\ttag  + M_2\Big)\Big] + g_{1}^{2} \Big[2 M_3\,\ttag  + M_1\Big]\Big\} M_3^*\,\ttag +\frac{1}{5} g_{1}^{2} g_{2}^{2} M_1  M_2^*  \nonumber \\ &+16 g_{2}^{2} g_{3}^{2} M_3  M_2^*\,\ttag
	 +\frac{4}{5} g_{1}^{2} \mm\pHd2 {y_{\pd}^{\dagger}  y_{\pd}} +\frac{8}{5} g_{1}^{2} \mm\pHu2 {y_{\pu}^{\dagger}  y_{\pu}} \nonumber \\ 
	 &+\frac{1}{225} g_{1}^{2} M_1^* \Big\lceil\Big\{5 \Big[16 g_{3}^{2} \Big(2 M_1  + M_3\,\ttag\Big) + 9 g_{2}^{2} \Big(2 M_1  + M_2\Big)\Big] + 597 g_{1}^{2} M_1 \Big\} \nonumber \\ 
	 &+180 \Big\{2 M_1 {y_{\pd}^{\dagger}  y_{\pd}}  -2 {y_{\pu}^{\dagger}  a_{\pu}}  + 4 M_1 {y_{\pu}^{\dagger}  y_{\pu}}  - {y_{\pd}^{\dagger}  a_{\pd}} \Big\}\Big\rceil\nonumber \\ 
	 &-\frac{4}{5} g_{1}^{2} M_1 {a_{\pd}^{\dagger}  y_{\pd}} +\frac{4}{5} g_{1}^{2} {a_{\pd}^{\dagger}  a_{\pd}} -\frac{8}{5} g_{1}^{2} M_1 {a_{\pu}^{\dagger}  y_{\pu}} +\frac{8}{5} g_{1}^{2} {a_{\pu}^{\dagger}  a_{\pu}} \nonumber \\ 
	 &+\frac{2}{5} g_{1}^{2} {\mm\pq2  y_{\pd}^{\dagger}  y_{\pd}} +\frac{4}{5} g_{1}^{2} {\mm\pq2  y_{\pu}^{\dagger}  y_{\pu}} +\frac{4}{5} g_{1}^{2} {y_{\pd}^{\dagger}  \mm\pd2  y_{\pd}} +\frac{2}{5} g_{1}^{2} {y_{\pd}^{\dagger}  y_{\pd}  \mm\pq2} \nonumber \\ 
	 &+\frac{8}{5} g_{1}^{2} {y_{\pu}^{\dagger}  \mm\pu2  y_{\pu}} +\frac{4}{5} g_{1}^{2} {y_{\pu}^{\dagger}  y_{\pu}  \mm\pq2} -8 \mm\pHd2 {y_{\pd}^{\dagger}  y_{\pd}  y_{\pd}^{\dagger}  y_{\pd}} -4 {y_{\pd}^{\dagger}  y_{\pd}  a_{\pd}^{\dagger}  a_{\pd}} \nonumber \\ 
	 &-4 {y_{\pd}^{\dagger}  a_{\pd}  a_{\pd}^{\dagger}  y_{\pd}} -8 \mm\pHu2 {y_{\pu}^{\dagger}  y_{\pu}  y_{\pu}^{\dagger}  y_{\pu}} -4 {y_{\pu}^{\dagger}  y_{\pu}  a_{\pu}^{\dagger}  a_{\pu}} -4 {y_{\pu}^{\dagger}  a_{\pu}  a_{\pu}^{\dagger}  y_{\pu}} \nonumber \\ 
	 &-4 {a_{\pd}^{\dagger}  y_{\pd}  y_{\pd}^{\dagger}  a_{\pd}} -4 {a_{\pd}^{\dagger}  a_{\pd}  y_{\pd}^{\dagger}  y_{\pd}} -4 {a_{\pu}^{\dagger}  y_{\pu}  y_{\pu}^{\dagger}  a_{\pu}} -4 {a_{\pu}^{\dagger}  a_{\pu}  y_{\pu}^{\dagger}  y_{\pu}} \nonumber \\ 
	 &-2 {\mm\pq2  y_{\pd}^{\dagger}  y_{\pd}  y_{\pd}^{\dagger}  y_{\pd}} -2 {\mm\pq2  y_{\pu}^{\dagger}  y_{\pu}  y_{\pu}^{\dagger}  y_{\pu}} -4 {y_{\pd}^{\dagger}  \mm\pd2  y_{\pd}  y_{\pd}^{\dagger}  y_{\pd}} -4 {y_{\pd}^{\dagger}  y_{\pd}  \mm\pq2  y_{\pd}^{\dagger}  y_{\pd}} \nonumber \\ 
	 &-4 {y_{\pd}^{\dagger}  y_{\pd}  y_{\pd}^{\dagger}  \mm\pd2  y_{\pd}} -2 {y_{\pd}^{\dagger}  y_{\pd}  y_{\pd}^{\dagger}  y_{\pd}  \mm\pq2} -4 {y_{\pu}^{\dagger}  \mm\pu2  y_{\pu}  y_{\pu}^{\dagger}  y_{\pu}} -4 {y_{\pu}^{\dagger}  y_{\pu}  \mm\pq2  y_{\pu}^{\dagger}  y_{\pu}} \nonumber \\ 
	 &-4 {y_{\pu}^{\dagger}  y_{\pu}  y_{\pu}^{\dagger}  \mm\pu2  y_{\pu}} -2 {y_{\pu}^{\dagger}  y_{\pu}  y_{\pu}^{\dagger}  y_{\pu}  \mm\pq2} +6 g_{2}^{4}  \sigma_{2,2} +\frac{32}{3} g_{3}^{4}  \sigma_{2,3} +\frac{2}{15} g_{1}^{2}  \sigma_{2,11} +4 \frac{1}{\sqrt{15}} g_1  \sigma_{3,1} \nonumber \\ 
	 &-12 \mm\pHd2 {y_{\pd}^{\dagger}  y_{\pd}} \,\tr\Big({y_{\pd}  y_{\pd}^{\dagger}}\Big) -6 {a_{\pd}^{\dagger}  a_{\pd}} \,\tr\Big({y_{\pd}  y_{\pd}^{\dagger}}\Big) -3 {\mm\pq2  y_{\pd}^{\dagger}  y_{\pd}} \,\tr\Big({y_{\pd}  y_{\pd}^{\dagger}}\Big) \nonumber \\ 
	 &-6 {y_{\pd}^{\dagger}  \mm\pd2  y_{\pd}} \,\tr\Big({y_{\pd}  y_{\pd}^{\dagger}}\Big) -3 {y_{\pd}^{\dagger}  y_{\pd}  \mm\pq2} \,\tr\Big({y_{\pd}  y_{\pd}^{\dagger}}\Big) -4 \mm\pHd2 {y_{\pd}^{\dagger}  y_{\pd}} \,\tr\Big({y_{\pe}  y_{\pe}^{\dagger}}\Big) \nonumber \\ 
	 &-2 {a_{\pd}^{\dagger}  a_{\pd}} \,\tr\Big({y_{\pe}  y_{\pe}^{\dagger}}\Big) - {\mm\pq2  y_{\pd}^{\dagger}  y_{\pd}} \,\tr\Big({y_{\pe}  y_{\pe}^{\dagger}}\Big) -2 {y_{\pd}^{\dagger}  \mm\pd2  y_{\pd}} \,\tr\Big({y_{\pe}  y_{\pe}^{\dagger}}\Big) \nonumber \\ 
	 &- {y_{\pd}^{\dagger}  y_{\pd}  \mm\pq2} \,\tr\Big({y_{\pe}  y_{\pe}^{\dagger}}\Big) -12 \mm\pHu2 {y_{\pu}^{\dagger}  y_{\pu}} \,\tr\Big({y_{\pu}  y_{\pu}^{\dagger}}\Big) -6 {a_{\pu}^{\dagger}  a_{\pu}} \,\tr\Big({y_{\pu}  y_{\pu}^{\dagger}}\Big) \nonumber \\ 
	 &-3 {\mm\pq2  y_{\pu}^{\dagger}  y_{\pu}} \,\tr\Big({y_{\pu}  y_{\pu}^{\dagger}}\Big) -6 {y_{\pu}^{\dagger}  \mm\pu2  y_{\pu}} \,\tr\Big({y_{\pu}  y_{\pu}^{\dagger}}\Big) -3 {y_{\pu}^{\dagger}  y_{\pu}  \mm\pq2} \,\tr\Big({y_{\pu}  y_{\pu}^{\dagger}}\Big) \nonumber \\ 
	 &-6 {a_{\pd}^{\dagger}  y_{\pd}} \,\tr\Big({y_{\pd}^{\dagger}  a_{\pd}}\Big) -2 {a_{\pd}^{\dagger}  y_{\pd}} \,\tr\Big({y_{\pe}^{\dagger}  a_{\pe}}\Big) -6 {a_{\pu}^{\dagger}  y_{\pu}} \,\tr\Big({y_{\pu}^{\dagger}  a_{\pu}}\Big) \nonumber \\ 
	 &-6 {y_{\pd}^{\dagger}  a_{\pd}} \,\tr\Big({a_{\pd}^*  y_{\pd}^{T}}\Big) -6 {y_{\pd}^{\dagger}  y_{\pd}} \,\tr\Big({a_{\pd}^*  a_{\pd}^{T}}\Big) -2 {y_{\pd}^{\dagger}  a_{\pd}} \,\tr\Big({a_{\pe}^*  y_{\pe}^{T}}\Big) \nonumber \\ 
	 &-2 {y_{\pd}^{\dagger}  y_{\pd}} \,\tr\Big({a_{\pe}^*  a_{\pe}^{T}}\Big) -6 {y_{\pu}^{\dagger}  a_{\pu}} \,\tr\Big({a_{\pu}^*  y_{\pu}^{T}}\Big) -6 {y_{\pu}^{\dagger}  y_{\pu}} \,\tr\Big({a_{\pu}^*  a_{\pu}^{T}}\Big) \nonumber \\ 
	 &-6 {y_{\pd}^{\dagger}  y_{\pd}} \,\tr\Big({\mm\pd2  y_{\pd}  y_{\pd}^{\dagger}}\Big) -2 {y_{\pd}^{\dagger}  y_{\pd}} \,\tr\Big({\mm\pe2  y_{\pe}  y_{\pe}^{\dagger}}\Big) -2 {y_{\pd}^{\dagger}  y_{\pd}} \,\tr\Big({\mm\pl2  y_{\pe}^{\dagger}  y_{\pe}}\Big) \nonumber \\ 
	 &-6 {y_{\pd}^{\dagger}  y_{\pd}} \,\tr\Big({\mm\pq2  y_{\pd}^{\dagger}  y_{\pd}}\Big) -6 {y_{\pu}^{\dagger}  y_{\pu}} \,\tr\Big({\mm\pq2  y_{\pu}^{\dagger}  y_{\pu}}\Big) -6 {y_{\pu}^{\dagger}  y_{\pu}} \,\tr\Big({\mm\pu2  y_{\pu}  y_{\pu}^{\dagger}}\Big), \\ 
	\beta_{\mm\pl2}^{(1)} 
	& = -\frac{6}{5} g_{1}^{2}  |M_1|^2 -6 g_{2}^{2}  |M_2|^2 +2 \mm\pHd2 {y_{\pe}^{\dagger}  y_{\pe}} +2 {a_{\pe}^{\dagger}  a_{\pe}} +{\mm\pl2  y_{\pe}^{\dagger}  y_{\pe}}+2 {y_{\pe}^{\dagger}  \mm\pe2  y_{\pe}} \nonumber \\ 
	 &+{y_{\pe}^{\dagger}  y_{\pe}  \mm\pl2}- \sqrt{\frac{3}{5}} g_1  \sigma_{1,1}, \\ 
	\beta_{\mm\pl2}^{(2)} 
	& = \frac{3}{5} g_{2}^{2} \Big[3 g_{1}^{2} \Big(2 M_2  + M_1\Big) + 55 g_{2}^{2} M_2 \Big] M_2^* +\frac{12}{5} g_{1}^{2} \mm\pHd2 {y_{\pe}^{\dagger}  y_{\pe}} \nonumber \\ 
	 &+\frac{3}{25} g_{1}^{2} M_1^* \Big\{-20 {y_{\pe}^{\dagger}  a_{\pe}}  + 3 \Big[5 g_{2}^{2} \Big(2 M_1  + M_2\Big) + 69 g_{1}^{2} M_1 \Big]  + 40 M_1 {y_{\pe}^{\dagger}  y_{\pe}} \Big\}-\frac{12}{5} g_{1}^{2} M_1 {a_{\pe}^{\dagger}  y_{\pe}} \nonumber \\ 
	 &+\frac{12}{5} g_{1}^{2} {a_{\pe}^{\dagger}  a_{\pe}} +\frac{6}{5} g_{1}^{2} {\mm\pl2  y_{\pe}^{\dagger}  y_{\pe}} +\frac{12}{5} g_{1}^{2} {y_{\pe}^{\dagger}  \mm\pe2  y_{\pe}} +\frac{6}{5} g_{1}^{2} {y_{\pe}^{\dagger}  y_{\pe}  \mm\pl2} \nonumber \\ 
	 &-8 \mm\pHd2 {y_{\pe}^{\dagger}  y_{\pe}  y_{\pe}^{\dagger}  y_{\pe}} -4 {y_{\pe}^{\dagger}  y_{\pe}  a_{\pe}^{\dagger}  a_{\pe}} -4 {y_{\pe}^{\dagger}  a_{\pe}  a_{\pe}^{\dagger}  y_{\pe}} -4 {a_{\pe}^{\dagger}  y_{\pe}  y_{\pe}^{\dagger}  a_{\pe}} \nonumber \\ 
	 &-4 {a_{\pe}^{\dagger}  a_{\pe}  y_{\pe}^{\dagger}  y_{\pe}} -2 {\mm\pl2  y_{\pe}^{\dagger}  y_{\pe}  y_{\pe}^{\dagger}  y_{\pe}} -4 {y_{\pe}^{\dagger}  \mm\pe2  y_{\pe}  y_{\pe}^{\dagger}  y_{\pe}} -4 {y_{\pe}^{\dagger}  y_{\pe}  \mm\pl2  y_{\pe}^{\dagger}  y_{\pe}} \nonumber \\ 
	 &-4 {y_{\pe}^{\dagger}  y_{\pe}  y_{\pe}^{\dagger}  \mm\pe2  y_{\pe}} -2 {y_{\pe}^{\dagger}  y_{\pe}  y_{\pe}^{\dagger}  y_{\pe}  \mm\pl2} +6 g_{2}^{4}  \sigma_{2,2} +\frac{6}{5} g_{1}^{2}  \sigma_{2,11} -4 \sqrt{\frac{3}{5}} g_1  \sigma_{3,1} \nonumber \\ 
	 &-12 \mm\pHd2 {y_{\pe}^{\dagger}  y_{\pe}} \,\tr\Big({y_{\pd}  y_{\pd}^{\dagger}}\Big) -6 {a_{\pe}^{\dagger}  a_{\pe}} \,\tr\Big({y_{\pd}  y_{\pd}^{\dagger}}\Big) -3 {\mm\pl2  y_{\pe}^{\dagger}  y_{\pe}} \,\tr\Big({y_{\pd}  y_{\pd}^{\dagger}}\Big) \nonumber \\ 
	 &-6 {y_{\pe}^{\dagger}  \mm\pe2  y_{\pe}} \,\tr\Big({y_{\pd}  y_{\pd}^{\dagger}}\Big) -3 {y_{\pe}^{\dagger}  y_{\pe}  \mm\pl2} \,\tr\Big({y_{\pd}  y_{\pd}^{\dagger}}\Big) -4 \mm\pHd2 {y_{\pe}^{\dagger}  y_{\pe}} \,\tr\Big({y_{\pe}  y_{\pe}^{\dagger}}\Big) \nonumber \\ 
	 &-2 {a_{\pe}^{\dagger}  a_{\pe}} \,\tr\Big({y_{\pe}  y_{\pe}^{\dagger}}\Big) - {\mm\pl2  y_{\pe}^{\dagger}  y_{\pe}} \,\tr\Big({y_{\pe}  y_{\pe}^{\dagger}}\Big) -2 {y_{\pe}^{\dagger}  \mm\pe2  y_{\pe}} \,\tr\Big({y_{\pe}  y_{\pe}^{\dagger}}\Big) \nonumber \\ 
	 &- {y_{\pe}^{\dagger}  y_{\pe}  \mm\pl2} \,\tr\Big({y_{\pe}  y_{\pe}^{\dagger}}\Big) -6 {a_{\pe}^{\dagger}  y_{\pe}} \,\tr\Big({y_{\pd}^{\dagger}  a_{\pd}}\Big) -2 {a_{\pe}^{\dagger}  y_{\pe}} \,\tr\Big({y_{\pe}^{\dagger}  a_{\pe}}\Big) \nonumber \\ 
	 &-6 {y_{\pe}^{\dagger}  a_{\pe}} \,\tr\Big({a_{\pd}^*  y_{\pd}^{T}}\Big) -6 {y_{\pe}^{\dagger}  y_{\pe}} \,\tr\Big({a_{\pd}^*  a_{\pd}^{T}}\Big) -2 {y_{\pe}^{\dagger}  a_{\pe}} \,\tr\Big({a_{\pe}^*  y_{\pe}^{T}}\Big) \nonumber \\ 
	 &-2 {y_{\pe}^{\dagger}  y_{\pe}} \,\tr\Big({a_{\pe}^*  a_{\pe}^{T}}\Big) -6 {y_{\pe}^{\dagger}  y_{\pe}} \,\tr\Big({\mm\pd2  y_{\pd}  y_{\pd}^{\dagger}}\Big) -2 {y_{\pe}^{\dagger}  y_{\pe}} \,\tr\Big({\mm\pe2  y_{\pe}  y_{\pe}^{\dagger}}\Big) \nonumber \\ 
	 &-2 {y_{\pe}^{\dagger}  y_{\pe}} \,\tr\Big({\mm\pl2  y_{\pe}^{\dagger}  y_{\pe}}\Big) -6 {y_{\pe}^{\dagger}  y_{\pe}} \,\tr\Big({\mm\pq2  y_{\pd}^{\dagger}  y_{\pd}}\Big), \\ 
	\beta_{\mm\pHd2}^{(1)} & =  
	-\frac{6}{5} g_{1}^{2} |M_1|^2 -6 g_{2}^{2} |M_2|^2 - \sqrt{\frac{3}{5}} g_1 \sigma_{1,1} +6 \mm\pHd2 \,\tr\Big({y_{\pd}  y_{\pd}^{\dagger}}\Big) +2 \mm\pHd2 \,\tr\Big({y_{\pe}  y_{\pe}^{\dagger}}\Big) +6 \,\tr\Big({a_{\pd}^*  a_{\pd}^{T}}\Big) \nonumber \\ 
	 &+2 \,\tr\Big({a_{\pe}^*  a_{\pe}^{T}}\Big) +6 \,\tr\Big({\mm\pd2  y_{\pd}  y_{\pd}^{\dagger}}\Big) +2 \,\tr\Big({\mm\pe2  y_{\pe}  y_{\pe}^{\dagger}}\Big) +2 \,\tr\Big({\mm\pl2  y_{\pe}^{\dagger}  y_{\pe}}\Big) +6 \,\tr\Big({\mm\pq2  y_{\pd}^{\dagger}  y_{\pd}}\Big), \\ 
	\beta_{\mm\pHd2}^{(2)} & =  
	\frac{1}{25} \Big\{15 g_{2}^{2} \Big[3 g_{1}^{2} \Big(2 M_2  + M_1\Big) + 55 g_{2}^{2} M_2 \Big]M_2^* +g_{1}^{2} M_1^* \Big[621 g_{1}^{2} M_1 +90 g_{2}^{2} M_1 +45 g_{2}^{2} M_2 \nonumber \\ & -40 M_1 \,\tr\Big({y_{\pd}  y_{\pd}^{\dagger}}\Big) +120 M_1 \,\tr\Big({y_{\pe}  y_{\pe}^{\dagger}}\Big) +20 \,\tr\Big({y_{\pd}^{\dagger}  a_{\pd}}\Big) -60 \,\tr\Big({y_{\pe}^{\dagger}  a_{\pe}}\Big) \Big]\nonumber \\ 
	 &+10 \Big[15 g_{2}^{4} \sigma_{2,2} +3 g_{1}^{2} \sigma_{2,11} -2 \sqrt{15} g_1 \sigma_{3,1} +\Big(160 g_{3}^{2} |M_3|^2\,\ttag  -2 g_{1}^{2} \mm\pHd2  + 80 g_{3}^{2} \mm\pHd2 \Big)\,\tr\Big({y_{\pd}  y_{\pd}^{\dagger}}\Big) \nonumber \\ 
	 &+6 g_{1}^{2} \mm\pHd2 \,\tr\Big({y_{\pe}  y_{\pe}^{\dagger}}\Big) -80 g_{3}^{2} M_3^* \,\tr\Big({y_{\pd}^{\dagger}  a_{\pd}}\Big)\,\ttag +2 g_{1}^{2} M_1 \,\tr\Big({a_{\pd}^*  y_{\pd}^{T}}\Big) -80 g_{3}^{2} M_3 \,\tr\Big({a_{\pd}^*  y_{\pd}^{T}}\Big)\,\ttag \nonumber \\ 
	 &-2 g_{1}^{2} \,\tr\Big({a_{\pd}^*  a_{\pd}^{T}}\Big) +80 g_{3}^{2} \,\tr\Big({a_{\pd}^*  a_{\pd}^{T}}\Big) -6 g_{1}^{2} M_1 \,\tr\Big({a_{\pe}^*  y_{\pe}^{T}}\Big) +6 g_{1}^{2} \,\tr\Big({a_{\pe}^*  a_{\pe}^{T}}\Big) \nonumber \\ 
	 &-2 g_{1}^{2} \,\tr\Big({\mm\pd2  y_{\pd}  y_{\pd}^{\dagger}}\Big) +80 g_{3}^{2} \,\tr\Big({\mm\pd2  y_{\pd}  y_{\pd}^{\dagger}}\Big) +6 g_{1}^{2} \,\tr\Big({\mm\pe2  y_{\pe}  y_{\pe}^{\dagger}}\Big) +6 g_{1}^{2} \,\tr\Big({\mm\pl2  y_{\pe}^{\dagger}  y_{\pe}}\Big) \nonumber \\ 
	 &-2 g_{1}^{2} \,\tr\Big({\mm\pq2  y_{\pd}^{\dagger}  y_{\pd}}\Big) +80 g_{3}^{2} \,\tr\Big({\mm\pq2  y_{\pd}^{\dagger}  y_{\pd}}\Big) -90 \mm\pHd2 \,\tr\Big({y_{\pd}  y_{\pd}^{\dagger}  y_{\pd}  y_{\pd}^{\dagger}}\Big) -90 \,\tr\Big({y_{\pd}  y_{\pd}^{\dagger}  a_{\pd}  a_{\pd}^{\dagger}}\Big) \nonumber \\ 
	 &-15 \mm\pHd2 \,\tr\Big({y_{\pd}  y_{\pu}^{\dagger}  y_{\pu}  y_{\pd}^{\dagger}}\Big) -15 \mm\pHu2 \,\tr\Big({y_{\pd}  y_{\pu}^{\dagger}  y_{\pu}  y_{\pd}^{\dagger}}\Big) -15 \,\tr\Big({y_{\pd}  y_{\pu}^{\dagger}  a_{\pu}  a_{\pd}^{\dagger}}\Big) \nonumber \\ 
	 &-90 \,\tr\Big({y_{\pd}  a_{\pd}^{\dagger}  a_{\pd}  y_{\pd}^{\dagger}}\Big) -15 \,\tr\Big({y_{\pd}  a_{\pu}^{\dagger}  a_{\pu}  y_{\pd}^{\dagger}}\Big) -30 \mm\pHd2 \,\tr\Big({y_{\pe}  y_{\pe}^{\dagger}  y_{\pe}  y_{\pe}^{\dagger}}\Big) -30 \,\tr\Big({y_{\pe}  y_{\pe}^{\dagger}  a_{\pe}  a_{\pe}^{\dagger}}\Big) \nonumber \\ 
	 &-30 \,\tr\Big({y_{\pe}  a_{\pe}^{\dagger}  a_{\pe}  y_{\pe}^{\dagger}}\Big) -15 \,\tr\Big({y_{\pu}  y_{\pd}^{\dagger}  a_{\pd}  a_{\pu}^{\dagger}}\Big) -15 \,\tr\Big({y_{\pu}  a_{\pd}^{\dagger}  a_{\pd}  y_{\pu}^{\dagger}}\Big) -90 \,\tr\Big({\mm\pd2  y_{\pd}  y_{\pd}^{\dagger}  y_{\pd}  y_{\pd}^{\dagger}}\Big) \nonumber \\ 
	 &-15 \,\tr\Big({\mm\pd2  y_{\pd}  y_{\pu}^{\dagger}  y_{\pu}  y_{\pd}^{\dagger}}\Big) -30 \,\tr\Big({\mm\pe2  y_{\pe}  y_{\pe}^{\dagger}  y_{\pe}  y_{\pe}^{\dagger}}\Big) -30 \,\tr\Big({\mm\pl2  y_{\pe}^{\dagger}  y_{\pe}  y_{\pe}^{\dagger}  y_{\pe}}\Big) 
	 \nonumber \\ &-90 \,\tr\Big({\mm\pq2  y_{\pd}^{\dagger}  y_{\pd}  y_{\pd}^{\dagger}  y_{\pd}}\Big) 
	 -15 \,\tr\Big({\mm\pq2  y_{\pd}^{\dagger}  y_{\pd}  y_{\pu}^{\dagger}  y_{\pu}}\Big) -15 \,\tr\Big({\mm\pq2  y_{\pu}^{\dagger}  y_{\pu}  y_{\pd}^{\dagger}  y_{\pd}}\Big) 
	 	 \nonumber \\ & -15 \,\tr\Big({\mm\pu2  y_{\pu}  y_{\pd}^{\dagger}  y_{\pd}  y_{\pu}^{\dagger}}\Big) \Big]\Big\},\\ 
	\beta_{\mm\pHu2}^{(1)} 
	& = -\frac{6}{5} g_{1}^{2} |M_1|^2 -6 g_{2}^{2} |M_2|^2 +\sqrt{\frac{3}{5}} g_1 \sigma_{1,1} +6 \mm\pHu2 \,\tr\Big({y_{\pu}  y_{\pu}^{\dagger}}\Big) +6 \,\tr\Big({a_{\pu}^*  a_{\pu}^{T}}\Big) +6 \,\tr\Big({\mm\pq2  y_{\pu}^{\dagger}  y_{\pu}}\Big) \nonumber \\ 
	 &+6 \,\tr\Big({\mm\pu2  y_{\pu}  y_{\pu}^{\dagger}}\Big), \\ 
	\beta_{\mm\pHu2}^{(2)} 
	& = \frac{3}{5} g_{2}^{2} \Big[3 g_{1}^{2} \Big(2 M_2  + M_1\Big) + 55 g_{2}^{2} M_2 \Big]M_2^* +6 g_{2}^{4} \sigma_{2,2} +\frac{6}{5} g_{1}^{2} \sigma_{2,11} +4 \sqrt{\frac{3}{5}} g_1 \sigma_{3,1} 
	\nonumber \\ & 
	+\frac{8}{5} g_{1}^{2} \mm\pHu2 \,\tr\Big({y_{\pu}  y_{\pu}^{\dagger}}\Big) +32 g_{3}^{2} \mm\pHu2 \,\tr\Big({y_{\pu}  y_{\pu}^{\dagger}}\Big) +64 g_{3}^{2} |M_3|^2 \,\tr\Big({y_{\pu}  y_{\pu}^{\dagger}}\Big) \,\ttag
	\nonumber \\ & 
	+\frac{1}{25} g_{1}^{2} M_1^* \Big[-40 \,\tr\Big({y_{\pu}^{\dagger}  a_{\pu}}\Big)  + 45 g_{2}^{2} M_2  + 621 g_{1}^{2} M_1  + 80 M_1 \,\tr\Big({y_{\pu}  y_{\pu}^{\dagger}}\Big)  + 90 g_{2}^{2} M_1 \Big]\nonumber \\ 
	 &-32 g_{3}^{2} M_3^* \,\tr\Big({y_{\pu}^{\dagger}  a_{\pu}}\Big)\,\ttag -\frac{8}{5} g_{1}^{2} M_1 \,\tr\Big({a_{\pu}^*  y_{\pu}^{T}}\Big) -32 g_{3}^{2} M_3 \,\tr\Big({a_{\pu}^*  y_{\pu}^{T}}\Big)\,\ttag +\frac{8}{5} g_{1}^{2} \,\tr\Big({a_{\pu}^*  a_{\pu}^{T}}\Big) \nonumber \\ 
	 &+32 g_{3}^{2} \,\tr\Big({a_{\pu}^*  a_{\pu}^{T}}\Big) +\frac{8}{5} g_{1}^{2} \,\tr\Big({\mm\pq2  y_{\pu}^{\dagger}  y_{\pu}}\Big) +32 g_{3}^{2} \,\tr\Big({\mm\pq2  y_{\pu}^{\dagger}  y_{\pu}}\Big) +\frac{8}{5} g_{1}^{2} \,\tr\Big({\mm\pu2  y_{\pu}  y_{\pu}^{\dagger}}\Big) \nonumber \\ 
	 &+32 g_{3}^{2} \,\tr\Big({\mm\pu2  y_{\pu}  y_{\pu}^{\dagger}}\Big) -6 \mm\pHd2 \,\tr\Big({y_{\pd}  y_{\pu}^{\dagger}  y_{\pu}  y_{\pd}^{\dagger}}\Big) -6 \mm\pHu2 \,\tr\Big({y_{\pd}  y_{\pu}^{\dagger}  y_{\pu}  y_{\pd}^{\dagger}}\Big) \nonumber \\ 
	 &-6 \,\tr\Big({y_{\pd}  y_{\pu}^{\dagger}  a_{\pu}  a_{\pd}^{\dagger}}\Big) -6 \,\tr\Big({y_{\pd}  a_{\pu}^{\dagger}  a_{\pu}  y_{\pd}^{\dagger}}\Big) -6 \,\tr\Big({y_{\pu}  y_{\pd}^{\dagger}  a_{\pd}  a_{\pu}^{\dagger}}\Big) -36 \mm\pHu2 \,\tr\Big({y_{\pu}  y_{\pu}^{\dagger}  y_{\pu}  y_{\pu}^{\dagger}}\Big) \nonumber \\ 
	 &-36 \,\tr\Big({y_{\pu}  y_{\pu}^{\dagger}  a_{\pu}  a_{\pu}^{\dagger}}\Big) -6 \,\tr\Big({y_{\pu}  a_{\pd}^{\dagger}  a_{\pd}  y_{\pu}^{\dagger}}\Big) -36 \,\tr\Big({y_{\pu}  a_{\pu}^{\dagger}  a_{\pu}  y_{\pu}^{\dagger}}\Big) \nonumber \\ 
	 &-6 \,\tr\Big({\mm\pd2  y_{\pd}  y_{\pu}^{\dagger}  y_{\pu}  y_{\pd}^{\dagger}}\Big) -6 \,\tr\Big({\mm\pq2  y_{\pd}^{\dagger}  y_{\pd}  y_{\pu}^{\dagger}  y_{\pu}}\Big) -6 \,\tr\Big({\mm\pq2  y_{\pu}^{\dagger}  y_{\pu}  y_{\pd}^{\dagger}  y_{\pd}}\Big) \nonumber \\ 
	 &-36 \,\tr\Big({\mm\pq2  y_{\pu}^{\dagger}  y_{\pu}  y_{\pu}^{\dagger}  y_{\pu}}\Big) -6 \,\tr\Big({\mm\pu2  y_{\pu}  y_{\pd}^{\dagger}  y_{\pd}  y_{\pu}^{\dagger}}\Big) -36 \,\tr\Big({\mm\pu2  y_{\pu}  y_{\pu}^{\dagger}  y_{\pu}  y_{\pu}^{\dagger}}\Big), \\ 
	\beta_{\mm\pd2}^{(1)} 
	& = -\frac{8}{15} g_{1}^{2}  |M_1|^2 -\frac{32}{3} g_{3}^{2}  |M_3|^2\,\ttag +4 \mm\pHd2 {y_{\pd}  y_{\pd}^{\dagger}} +4 {a_{\pd}  a_{\pd}^{\dagger}} +2 {\mm\pd2  y_{\pd}  y_{\pd}^{\dagger}} +4 {y_{\pd}  \mm\pq2  y_{\pd}^{\dagger}} \nonumber \\ 
	 &+2 {y_{\pd}  y_{\pd}^{\dagger}  \mm\pd2} +2 \frac{1}{\sqrt{15}} g_1  \sigma_{1,1}, \\ 
	\beta_{\mm\pd2}^{(2)} 
	& = \frac{32}{45} g_{3}^{2} \Big[2 g_{1}^{2} \Big(2 M_3+ M_1\Big) + 75 g_{3}^{2} M_3 \Big] M_3^*\,\ttag +\frac{4}{5} g_{1}^{2} \mm\pHd2 {y_{\pd}  y_{\pd}^{\dagger}} +12 g_{2}^{2} \mm\pHd2 {y_{\pd}  y_{\pd}^{\dagger}} \nonumber \\ 
	 &+24 g_{2}^{2} |M_2|^2 {y_{\pd}  y_{\pd}^{\dagger}} -\frac{4}{5} g_{1}^{2} M_1 {y_{\pd}  a_{\pd}^{\dagger}} -12 g_{2}^{2} M_2 {y_{\pd}  a_{\pd}^{\dagger}} \nonumber \\ 
	 &+\frac{4}{225} g_{1}^{2} M_1^* \Big\{2 \Big[303 g_{1}^{2} M_1  + 40 g_{3}^{2} \Big(2 M_1  + M_3\,\ttag\Big)\Big]  -45 {a_{\pd}  y_{\pd}^{\dagger}}  + 90 M_1 {y_{\pd}  y_{\pd}^{\dagger}} \Big\}-12 g_{2}^{2} M_2^* {a_{\pd}  y_{\pd}^{\dagger}} \nonumber \\ 
	 &+\frac{4}{5} g_{1}^{2} {a_{\pd}  a_{\pd}^{\dagger}} +12 g_{2}^{2} {a_{\pd}  a_{\pd}^{\dagger}} +\frac{2}{5} g_{1}^{2} {\mm\pd2  y_{\pd}  y_{\pd}^{\dagger}} +6 g_{2}^{2} {\mm\pd2  y_{\pd}  y_{\pd}^{\dagger}} \nonumber \\ 
	 &+\frac{4}{5} g_{1}^{2} {y_{\pd}  \mm\pq2  y_{\pd}^{\dagger}} +12 g_{2}^{2} {y_{\pd}  \mm\pq2  y_{\pd}^{\dagger}} +\frac{2}{5} g_{1}^{2} {y_{\pd}  y_{\pd}^{\dagger}  \mm\pd2} +6 g_{2}^{2} {y_{\pd}  y_{\pd}^{\dagger}  \mm\pd2} \nonumber \\ 
	 &-8 \mm\pHd2 {y_{\pd}  y_{\pd}^{\dagger}  y_{\pd}  y_{\pd}^{\dagger}} -4 {y_{\pd}  y_{\pd}^{\dagger}  a_{\pd}  a_{\pd}^{\dagger}} -4 \mm\pHd2 {y_{\pd}  y_{\pu}^{\dagger}  y_{\pu}  y_{\pd}^{\dagger}} \nonumber \\ 
	 &-4 \mm\pHu2 {y_{\pd}  y_{\pu}^{\dagger}  y_{\pu}  y_{\pd}^{\dagger}} -4 {y_{\pd}  y_{\pu}^{\dagger}  a_{\pu}  a_{\pd}^{\dagger}} -4 {y_{\pd}  a_{\pd}^{\dagger}  a_{\pd}  y_{\pd}^{\dagger}} -4 {y_{\pd}  a_{\pu}^{\dagger}  a_{\pu}  y_{\pd}^{\dagger}} \nonumber \\ 
	 &-4 {a_{\pd}  y_{\pd}^{\dagger}  y_{\pd}  a_{\pd}^{\dagger}} -4 {a_{\pd}  y_{\pu}^{\dagger}  y_{\pu}  a_{\pd}^{\dagger}} -4 {a_{\pd}  a_{\pd}^{\dagger}  y_{\pd}  y_{\pd}^{\dagger}} -4 {a_{\pd}  a_{\pu}^{\dagger}  y_{\pu}  y_{\pd}^{\dagger}} \nonumber \\ 
	 &-2 {\mm\pd2  y_{\pd}  y_{\pd}^{\dagger}  y_{\pd}  y_{\pd}^{\dagger}} -2 {\mm\pd2  y_{\pd}  y_{\pu}^{\dagger}  y_{\pu}  y_{\pd}^{\dagger}} -4 {y_{\pd}  \mm\pq2  y_{\pd}^{\dagger}  y_{\pd}  y_{\pd}^{\dagger}} -4 {y_{\pd}  \mm\pq2  y_{\pu}^{\dagger}  y_{\pu}  y_{\pd}^{\dagger}} \nonumber \\ 
	 &-4 {y_{\pd}  y_{\pd}^{\dagger}  \mm\pd2  y_{\pd}  y_{\pd}^{\dagger}} -4 {y_{\pd}  y_{\pd}^{\dagger}  y_{\pd}  \mm\pq2  y_{\pd}^{\dagger}} -2 {y_{\pd}  y_{\pd}^{\dagger}  y_{\pd}  y_{\pd}^{\dagger}  \mm\pd2} -4 {y_{\pd}  y_{\pu}^{\dagger}  \mm\pu2  y_{\pu}  y_{\pd}^{\dagger}} \nonumber \\ 
	 &-4 {y_{\pd}  y_{\pu}^{\dagger}  y_{\pu}  \mm\pq2  y_{\pd}^{\dagger}} -2 {y_{\pd}  y_{\pu}^{\dagger}  y_{\pu}  y_{\pd}^{\dagger}  \mm\pd2} +\frac{32}{3} g_{3}^{4}  \sigma_{2,3} +\frac{8}{15} g_{1}^{2}  \sigma_{2,11} +8 \frac{1}{\sqrt{15}} g_1  \sigma_{3,1} \nonumber \\ 
	 &-24 \mm\pHd2 {y_{\pd}  y_{\pd}^{\dagger}} \,\tr\Big({y_{\pd}  y_{\pd}^{\dagger}}\Big) -12 {a_{\pd}  a_{\pd}^{\dagger}} \,\tr\Big({y_{\pd}  y_{\pd}^{\dagger}}\Big) -6 {\mm\pd2  y_{\pd}  y_{\pd}^{\dagger}} \,\tr\Big({y_{\pd}  y_{\pd}^{\dagger}}\Big) \nonumber \\ 
	 &-12 {y_{\pd}  \mm\pq2  y_{\pd}^{\dagger}} \,\tr\Big({y_{\pd}  y_{\pd}^{\dagger}}\Big) -6 {y_{\pd}  y_{\pd}^{\dagger}  \mm\pd2} \,\tr\Big({y_{\pd}  y_{\pd}^{\dagger}}\Big) -8 \mm\pHd2 {y_{\pd}  y_{\pd}^{\dagger}} \,\tr\Big({y_{\pe}  y_{\pe}^{\dagger}}\Big) \nonumber \\ 
	 &-4 {a_{\pd}  a_{\pd}^{\dagger}} \,\tr\Big({y_{\pe}  y_{\pe}^{\dagger}}\Big) -2 {\mm\pd2  y_{\pd}  y_{\pd}^{\dagger}} \,\tr\Big({y_{\pe}  y_{\pe}^{\dagger}}\Big) -4 {y_{\pd}  \mm\pq2  y_{\pd}^{\dagger}} \,\tr\Big({y_{\pe}  y_{\pe}^{\dagger}}\Big) \nonumber \\ 
	 &-2 {y_{\pd}  y_{\pd}^{\dagger}  \mm\pd2} \,\tr\Big({y_{\pe}  y_{\pe}^{\dagger}}\Big) -12 {y_{\pd}  a_{\pd}^{\dagger}} \,\tr\Big({y_{\pd}^{\dagger}  a_{\pd}}\Big) -4 {y_{\pd}  a_{\pd}^{\dagger}} \,\tr\Big({y_{\pe}^{\dagger}  a_{\pe}}\Big) \nonumber \\ 
	 &-12 {a_{\pd}  y_{\pd}^{\dagger}} \,\tr\Big({a_{\pd}^*  y_{\pd}^{T}}\Big) -12 {y_{\pd}  y_{\pd}^{\dagger}} \,\tr\Big({a_{\pd}^*  a_{\pd}^{T}}\Big) -4 {a_{\pd}  y_{\pd}^{\dagger}} \,\tr\Big({a_{\pe}^*  y_{\pe}^{T}}\Big) \nonumber \\ 
	 &-4 {y_{\pd}  y_{\pd}^{\dagger}} \,\tr\Big({a_{\pe}^*  a_{\pe}^{T}}\Big) -12 {y_{\pd}  y_{\pd}^{\dagger}} \,\tr\Big({\mm\pd2  y_{\pd}  y_{\pd}^{\dagger}}\Big) -4 {y_{\pd}  y_{\pd}^{\dagger}} \,\tr\Big({\mm\pe2  y_{\pe}  y_{\pe}^{\dagger}}\Big) \nonumber \\ 
	 &-4 {y_{\pd}  y_{\pd}^{\dagger}} \,\tr\Big({\mm\pl2  y_{\pe}^{\dagger}  y_{\pe}}\Big) -12 {y_{\pd}  y_{\pd}^{\dagger}} \,\tr\Big({\mm\pq2  y_{\pd}^{\dagger}  y_{\pd}}\Big), \\ 
	\beta_{\mm\pu2}^{(1)} & =  
	-\frac{32}{15} g_{1}^{2}  |M_1|^2 -\frac{32}{3} g_{3}^{2}  |M_3|^2\,\ttag +4 \mm\pHu2 {y_{\pu}  y_{\pu}^{\dagger}} +4 {a_{\pu}  a_{\pu}^{\dagger}} +2 {\mm\pq2  y_{\pu}  y_{\pu}^{\dagger}} +4 {y_{\pu}  \mm\pq2  y_{\pu}^{\dagger}} \nonumber \\ 
	 &+2 {y_{\pu}  y_{\pu}^{\dagger}  \mm\pu2} -4 \frac{1}{\sqrt{15}} g_1  \sigma_{1,1}, \\ 
	\beta_{\mm\pu2}^{(2)}
	 & = \frac{32}{45} g_{3}^{2} \Big[75 g_{3}^{2} M_3\,\ttag  + 8 g_{1}^{2} \Big(2 M_3  + M_1\Big)\Big] M_3^*\,\ttag -\frac{4}{5} g_{1}^{2} \mm\pHu2 {y_{\pu}  y_{\pu}^{\dagger}} +12 g_{2}^{2} \mm\pHu2 {y_{\pu}  y_{\pu}^{\dagger}} \nonumber \\ 
	 &+24 g_{2}^{2} |M_2|^2 {y_{\pu}  y_{\pu}^{\dagger}} +\frac{4}{5} g_{1}^{2} M_1 {y_{\pu}  a_{\pu}^{\dagger}} -12 g_{2}^{2} M_2 {y_{\pu}  a_{\pu}^{\dagger}} -12 g_{2}^{2} M_2^* {a_{\pu}  y_{\pu}^{\dagger}} \nonumber \\ 
	 &+\frac{4}{225} g_{1}^{2} M_1^* \Big\{45 \Big[-2 M_1 {y_{\pu}  y_{\pu}^{\dagger}}  + {a_{\pu}  y_{\pu}^{\dagger}}\Big] + 8 \Big[321 g_{1}^{2} M_1  + 40 g_{3}^{2} \Big(2 M_1  + M_3\,\ttag\Big)\Big] \Big\}-\frac{4}{5} g_{1}^{2} {a_{\pu}  a_{\pu}^{\dagger}} \nonumber \\ 
	 &+12 g_{2}^{2} {a_{\pu}  a_{\pu}^{\dagger}} -\frac{2}{5} g_{1}^{2} {\mm\pu2  y_{\pu}  y_{\pu}^{\dagger}} +6 g_{2}^{2} {\mm\pu2  y_{\pu}  y_{\pu}^{\dagger}} -\frac{4}{5} g_{1}^{2} {y_{\pu}  \mm\pq2  y_{\pu}^{\dagger}} \nonumber \\ 
	 &+12 g_{2}^{2} {y_{\pu}  \mm\pq2  y_{\pu}^{\dagger}} -\frac{2}{5} g_{1}^{2} {y_{\pu}  y_{\pu}^{\dagger}  \mm\pu2} +6 g_{2}^{2} {y_{\pu}  y_{\pu}^{\dagger}  \mm\pu2} -4 \mm\pHd2 {y_{\pu}  y_{\pd}^{\dagger}  y_{\pd}  y_{\pu}^{\dagger}} \nonumber \\ 
	 &-4 \mm\pHu2 {y_{\pu}  y_{\pd}^{\dagger}  y_{\pd}  y_{\pu}^{\dagger}} -4 {y_{\pu}  y_{\pd}^{\dagger}  a_{\pd}  a_{\pu}^{\dagger}} -8 \mm\pHu2 {y_{\pu}  y_{\pu}^{\dagger}  y_{\pu}  y_{\pu}^{\dagger}} -4 {y_{\pu}  y_{\pu}^{\dagger}  a_{\pu}  a_{\pu}^{\dagger}} \nonumber \\ 
	 &-4 {y_{\pu}  a_{\pd}^{\dagger}  a_{\pd}  y_{\pu}^{\dagger}} -4 {y_{\pu}  a_{\pu}^{\dagger}  a_{\pu}  y_{\pu}^{\dagger}} -4 {a_{\pu}  y_{\pd}^{\dagger}  y_{\pd}  a_{\pu}^{\dagger}} -4 {a_{\pu}  y_{\pu}^{\dagger}  y_{\pu}  a_{\pu}^{\dagger}} \nonumber \\ 
	 &-4 {a_{\pu}  a_{\pd}^{\dagger}  y_{\pd}  y_{\pu}^{\dagger}} -4 {a_{\pu}  a_{\pu}^{\dagger}  y_{\pu}  y_{\pu}^{\dagger}} -2 {\mm\pu2  y_{\pu}  y_{\pd}^{\dagger}  y_{\pd}  y_{\pu}^{\dagger}} -2 {\mm\pu2  y_{\pu}  y_{\pu}^{\dagger}  y_{\pu}  y_{\pu}^{\dagger}} \nonumber \\ 
	 &-4 {y_{\pu}  \mm\pq2  y_{\pd}^{\dagger}  y_{\pd}  y_{\pu}^{\dagger}} -4 {y_{\pu}  \mm\pq2  y_{\pu}^{\dagger}  y_{\pu}  y_{\pu}^{\dagger}} -4 {y_{\pu}  y_{\pd}^{\dagger}  \mm\pd2  y_{\pd}  y_{\pu}^{\dagger}} \nonumber \\ 
	 &-4 {y_{\pu}  y_{\pd}^{\dagger}  y_{\pd}  \mm\pq2  y_{\pu}^{\dagger}} -2 {y_{\pu}  y_{\pd}^{\dagger}  y_{\pd}  y_{\pu}^{\dagger}  \mm\pu2} -4 {y_{\pu}  y_{\pu}^{\dagger}  \mm\pu2  y_{\pu}  y_{\pu}^{\dagger}} -4 {y_{\pu}  y_{\pu}^{\dagger}  y_{\pu}  \mm\pq2  y_{\pu}^{\dagger}} \nonumber \\ 
	 &-2 {y_{\pu}  y_{\pu}^{\dagger}  y_{\pu}  y_{\pu}^{\dagger}  \mm\pu2} +\frac{32}{3} g_{3}^{4}  \sigma_{2,3} +\frac{32}{15} g_{1}^{2}  \sigma_{2,11} -16 \frac{1}{\sqrt{15}} g_1  \sigma_{3,1} -24 \mm\pHu2 {y_{\pu}  y_{\pu}^{\dagger}} \,\tr\Big({y_{\pu}  y_{\pu}^{\dagger}}\Big) \nonumber \\ 
	 &-12 {a_{\pu}  a_{\pu}^{\dagger}} \,\tr\Big({y_{\pu}  y_{\pu}^{\dagger}}\Big) -6 {\mm\pu2  y_{\pu}  y_{\pu}^{\dagger}} \,\tr\Big({y_{\pu}  y_{\pu}^{\dagger}}\Big) -12 {y_{\pu}  \mm\pq2  y_{\pu}^{\dagger}} \,\tr\Big({y_{\pu}  y_{\pu}^{\dagger}}\Big) \nonumber \\ 
	 &-6 {y_{\pu}  y_{\pu}^{\dagger}  \mm\pu2} \,\tr\Big({y_{\pu}  y_{\pu}^{\dagger}}\Big) -12 {y_{\pu}  a_{\pu}^{\dagger}} \,\tr\Big({y_{\pu}^{\dagger}  a_{\pu}}\Big) -12 {a_{\pu}  y_{\pu}^{\dagger}} \,\tr\Big({a_{\pu}^*  y_{\pu}^{T}}\Big) \nonumber \\ 
	 &-12 {y_{\pu}  y_{\pu}^{\dagger}} \,\tr\Big({a_{\pu}^*  a_{\pu}^{T}}\Big) -12 {y_{\pu}  y_{\pu}^{\dagger}} \,\tr\Big({\mm\pq2  y_{\pu}^{\dagger}  y_{\pu}}\Big) -12 {y_{\pu}  y_{\pu}^{\dagger}} \,\tr\Big({\mm\pu2  y_{\pu}  y_{\pu}^{\dagger}}\Big), \\ 
	\beta_{\mm\pe2}^{(1)}
	 & = -\frac{24}{5} g_{1}^{2}  |M_1|^2 +2 \Big(2 \mm\pHd2 {y_{\pe}  y_{\pe}^{\dagger}}  + 2 {a_{\pe}  a_{\pe}^{\dagger}}  + 2 {y_{\pe}  \mm\pl2  y_{\pe}^{\dagger}}  + {\mm\pe2  y_{\pe}  y_{\pe}^{\dagger}} + {y_{\pe}  y_{\pe}^{\dagger}  \mm\pe2}\Big)\nonumber \\ 
	 &+2 \sqrt{\frac{3}{5}} g_1  \sigma_{1,1}, \\ 
	\beta_{\mm\pe2}^{(2)} 
	& = \frac{2}{25} \Big\lceil6 g_{1}^{2} M_1^* \Big\{234 g_{1}^{2} M_1   + 5 \Big[-2 M_1 {y_{\pe}  y_{\pe}^{\dagger}}  + {a_{\pe}  y_{\pe}^{\dagger}}\Big]\Big\}+20 g_1  \Big(3 g_1 \sigma_{2,11}  + \sqrt{15} \sigma_{3,1} \Big)\nonumber \\ 
	 &-5 \Big\{30 g_{2}^{2} M_2^* {a_{\pe}  y_{\pe}^{\dagger}} +6 g_{1}^{2} {a_{\pe}  a_{\pe}^{\dagger}} -30 g_{2}^{2} {a_{\pe}  a_{\pe}^{\dagger}} +3 g_{1}^{2} {\mm\pe2  y_{\pe}  y_{\pe}^{\dagger}} \nonumber \\ 
	 &-15 g_{2}^{2} {\mm\pe2  y_{\pe}  y_{\pe}^{\dagger}} +6 g_{1}^{2} {y_{\pe}  \mm\pl2  y_{\pe}^{\dagger}} -30 g_{2}^{2} {y_{\pe}  \mm\pl2  y_{\pe}^{\dagger}} +3 g_{1}^{2} {y_{\pe}  y_{\pe}^{\dagger}  \mm\pe2} \nonumber \\ 
	 &-15 g_{2}^{2} {y_{\pe}  y_{\pe}^{\dagger}  \mm\pe2} +20 \mm\pHd2 {y_{\pe}  y_{\pe}^{\dagger}  y_{\pe}  y_{\pe}^{\dagger}} +10 {y_{\pe}  y_{\pe}^{\dagger}  a_{\pe}  a_{\pe}^{\dagger}} +10 {y_{\pe}  a_{\pe}^{\dagger}  a_{\pe}  y_{\pe}^{\dagger}} \nonumber \\ 
	 &+10 {a_{\pe}  y_{\pe}^{\dagger}  y_{\pe}  a_{\pe}^{\dagger}} +10 {a_{\pe}  a_{\pe}^{\dagger}  y_{\pe}  y_{\pe}^{\dagger}} +5 {\mm\pe2  y_{\pe}  y_{\pe}^{\dagger}  y_{\pe}  y_{\pe}^{\dagger}} +10 {y_{\pe}  \mm\pl2  y_{\pe}^{\dagger}  y_{\pe}  y_{\pe}^{\dagger}} \nonumber \\ 
	 &+10 {y_{\pe}  y_{\pe}^{\dagger}  \mm\pe2  y_{\pe}  y_{\pe}^{\dagger}} +10 {y_{\pe}  y_{\pe}^{\dagger}  y_{\pe}  \mm\pl2  y_{\pe}^{\dagger}} +5 {y_{\pe}  y_{\pe}^{\dagger}  y_{\pe}  y_{\pe}^{\dagger}  \mm\pe2} +30 {a_{\pe}  a_{\pe}^{\dagger}} \,\tr\Big({y_{\pd}  y_{\pd}^{\dagger}}\Big) \nonumber \\ 
	 &+15 {\mm\pe2  y_{\pe}  y_{\pe}^{\dagger}} \,\tr\Big({y_{\pd}  y_{\pd}^{\dagger}}\Big) +30 {y_{\pe}  \mm\pl2  y_{\pe}^{\dagger}} \,\tr\Big({y_{\pd}  y_{\pd}^{\dagger}}\Big) +15 {y_{\pe}  y_{\pe}^{\dagger}  \mm\pe2} \,\tr\Big({y_{\pd}  y_{\pd}^{\dagger}}\Big) \nonumber \\ 
	 &+10 {a_{\pe}  a_{\pe}^{\dagger}} \,\tr\Big({y_{\pe}  y_{\pe}^{\dagger}}\Big) +5 {\mm\pe2  y_{\pe}  y_{\pe}^{\dagger}} \,\tr\Big({y_{\pe}  y_{\pe}^{\dagger}}\Big) +10 {y_{\pe}  \mm\pl2  y_{\pe}^{\dagger}} \,\tr\Big({y_{\pe}  y_{\pe}^{\dagger}}\Big) \nonumber \\ 
	 &+5 {y_{\pe}  y_{\pe}^{\dagger}  \mm\pe2} \,\tr\Big({y_{\pe}  y_{\pe}^{\dagger}}\Big) +{y_{\pe}  a_{\pe}^{\dagger}} \Big[10 \,\tr\Big({y_{\pe}^{\dagger}  a_{\pe}}\Big)  + 30 g_{2}^{2} M_2  + 30 \,\tr\Big({y_{\pd}^{\dagger}  a_{\pd}}\Big)  -6 g_{1}^{2} M_1 \Big]\nonumber \\ 
	 &+30 {a_{\pe}  y_{\pe}^{\dagger}} \,\tr\Big({a_{\pd}^*  y_{\pd}^{T}}\Big) +10 {a_{\pe}  y_{\pe}^{\dagger}} \,\tr\Big({a_{\pe}^*  y_{\pe}^{T}}\Big) 
	 +2 {y_{\pe}  y_{\pe}^{\dagger}} \Big[3 g_{1}^{2} \mm\pHd2 -15 g_{2}^{2} \mm\pHd2  -30 g_{2}^{2} |M_2|^2 \nonumber \\ & +30 \mm\pHd2 \,\tr\Big({y_{\pd}  y_{\pd}^{\dagger}}\Big) +10 \mm\pHd2 \,\tr\Big({y_{\pe}  y_{\pe}^{\dagger}}\Big) +15 \,\tr\Big({a_{\pd}^*  a_{\pd}^{T}}\Big) +5 \,\tr\Big({a_{\pe}^*  a_{\pe}^{T}}\Big) \nonumber \\ 
	 & +15 \,\tr\Big({\mm\pd2  y_{\pd}  y_{\pd}^{\dagger}}\Big) +5 \,\tr\Big({\mm\pe2  y_{\pe}  y_{\pe}^{\dagger}}\Big) +5 \,\tr\Big({\mm\pl2  y_{\pe}^{\dagger}  y_{\pe}}\Big) +15 \,\tr\Big({\mm\pq2  y_{\pd}^{\dagger}  y_{\pd}}\Big) \Big]\Big\}\Big\rceil,\\ 
	\beta_{\mm32}^{(1)} & =  
	-24 g_{3}^{2} |M_3|^2\,\ttag, \\ 
	\beta_{\mm32}^{(2)} & =  
	24 g_{3}^{4} \Big(15 |M_3|^2\,\ttag  + \sigma_{2,3}\Big).
\end{align}} 

\bibliographystyle{JHEP}
\bibliography{library}
\end{fmffile}
\end{document}